\documentclass[aps,preprint,nofootinbib,eqsecnum]{revtex4}
\usepackage{amsmath}
\usepackage{amsfonts}
\usepackage{amssymb}
\usepackage{graphicx}

\input{tcilatex}

\begin{document}
\begin{flushleft}
USCHEP/02-013
\hfill hep-th/0204260 \\
UT-02-08\hfill \\
\end{flushleft}
\title{Computing in String Field Theory Using the Moyal Star Product}
\author{Itzhak Bars}
\email{bars@usc.edu}
\affiliation{CIT-USC Center for Theoretical Physics, Department of Physics and Astronomy,
University of Southern California, Los Angeles, CA 90089-2535, USA}
\author{Yutaka Matsuo}
\email{matsuo@phys.s.u-tokyo.ac.jp}
\affiliation{Department of Physics, University of Tokyo, Hongo 7-3-1, Bunkyo-ku, Tokyo
113-0033, Japan \\
}

\begin{abstract}
Using the Moyal star product, we define open bosonic string field theory
carefully, with a cutoff, for any number of string oscillators and any
oscillator frequencies. Through detailed computations, such as Neumann
coefficients for all string vertices, we show that the Moyal star product is
all that is needed to give a precise definition of string field theory. The
formulation of the theory as well as the computation techniques are
considerably simpler in the Moyal formulation. After identifying a monoid
algebra as a fundamental mathematical structure in string field theory, we
use it as a tool to compute with ease the field configurations for wedge,
sliver, and generalized projectors, as well as all the string interaction
vertices for perturbative as well as monoid-type nonperturbative states.
Finally, in the context of VSFT we analyze the small fluctuations around any
D-brane vacuum. We show quite generally that to obtain nontrivial mass and
coupling, as well as a closed strings, there must be an associativity
anomaly. We identify the detailed source of the anomaly, but leave its study
for future work.
\end{abstract}

\maketitle
\tableofcontents

\begin{flushleft}
USCHEP/02-013
\hfill hep-th/0204260 \\
UT-02-08\hfill  \\
\end{flushleft}

\widetext {
\noindent CITUSC/02-013 \ \hfill\ \ hep-th/0204260 \\
UT-02-08 \ \hfill   \\}

\newpage

\section{Introduction}

Witten's formulation of open string field theory (SFT) \cite{Witten} is one
of the few tools available to discuss nonperturbative phenomena in string
theory. Recent discussions of tachyon condensation that were carried out in
the context of vacuum string field theory (VSFT) showed the relevance of
D-branes and spurred renewed interest in the overall string field theory
framework \cite{r-RSZ}-\cite{rsz3}. The main objective of the present paper
is to systematically develop the Moyal star formulation of open string field
theory (MSFT) \cite{witmoy} and show how to carry out computations in detail
in this simplifying framework.

SFT is usually considered in position space in terms of functionals $%
\psi\left( x^{\mu}\left( \sigma\right) ,\phi\left( \sigma\right) \right) $
that depend on the string coordinate $x^{\mu}\left( \sigma\right) $ and
bosonized ghosts $\phi\left( \sigma\right) $. For convenience we will rename
the ghost as a 27$^{th}$ dimension, $\phi\left( \sigma\right) =x_{27}\left(
\sigma\right) ,$ and allow the $\mu$ index to run over 27 instead of 26
dimensions. The string field may be rewritten in terms of the string modes $%
\psi\left( x_{0}^{\mu},x_{2n}^{\mu},x_{2n-1}^{\mu}\right) $ which are
defined by the expansion for open strings%
\begin{equation}
x^{\mu}(\sigma)=x_{0}^{\mu}+\sqrt{2}\sum_{n\geq1}^{\infty}x_{n}^{\mu}\cos(n%
\sigma).   \label{mode-expansion}
\end{equation}
A star product was defined by Witten among these fields \cite{Witten}. This
amounts to matching the right half of the first string to the left half of
the second string and integrating over the overlap. Computations in SFT
using this overlap have shown that Witten's star product does indeed lead to
the correct description of interactions among strings \cite{GJ}\cite%
{giddings}.

Recently, it has been shown that Witten's star product can be reformulated
as the standard Moyal star product \cite{witmoy}. This is obtained by
transforming to a \textquotedblleft half Fourier space" in which only the
odd modes $x_{2n-1}$ are Fourier transformed to momentum space $p_{2n-1}$
while the even modes $x_{2n}$ remain in position space. In order to
diagonalize the star product in mode space a linear transformation is
applied in mode space to the Fourier variables $p_{2n-1}=\frac{2}{\theta}%
\sum_{m>0}p_{2m}T_{2m,2n-1}$. Then the Moyal star is defined in the phase
space of even string modes $\left( x_{2n}^{\mu},p_{2n}^{\mu}\right) $ except
the midpoint, and the overall star is a product over the even modes. The
product is local at the midpoint. This reformulates string field theory as a
noncommutative field theory where noncommutativity is independent for each
even mode, thus establishing an explicit link between open string field
theory and noncommutative geometry in a form which is familiar in old \cite%
{r-Moyal} and recent literature \cite{r-DouglasNekrasov}. This result was
originally obtained in \cite{witmoy} by using the split string formalism
\cite{r-Bordes}, \cite{r-RSZ}-\cite{witmoy} as an intermediate step, and by
now it has been confirmed through a different approach that focuses on the
spectroscopy of the Neumann coefficients \cite{DLMZ}\cite{belov}\cite%
{arafeva}\footnote{%
The relation between \cite{witmoy} and \cite{DLMZ} amounts to a change of
basis through orthogonal transformations which diagonalize the matrix $%
\left( \kappa_{e}\right) ^{1/2}T\left( \kappa_{o}\right) ^{-1/2}$ (see
notation in text). We will come back to this transformation at several
points in this paper. In particular, the discussions related to Eqs.(\ref%
{ktk},\ref{gt},\ref{spectrum}) provide answers to some issues raised in \cite%
{DLMZ}. \label{similarity}}.

In this paper we develop methods of computation in MSFT systematically and
apply them to explicit examples. Section II gives details of the formulation
of MSFT, including a cutoff procedure, and provides a dictionary for
relating it to other formulations of open string field theory. MSFT is
initially defined more generally for any number of oscillators and any
oscillator frequencies. The contact with standard open string field theory
is made when the number of oscillators is infinite and when the frequencies
match the free string oscillator frequencies. The oscillator and Virasoro
algebra are constructed as special field configurations in Moyal space that
can be star multiplied on either the left or right side of general fields.
This provides the first representation of the Virasoro algebra in
noncommutative space, distinguished by the fact that its basic building
blocks are half as many oscillators (only even or only odd ones) as the
usual case. Section III introduces a noncommutative algebra for a special
class of string fields which are generating functions for correlators. This
closed algebra forms a monoid with an explicit structure which plays an
important role as a computational tool. In section-IV it is shown that the
monoid algebra is sufficient to compute explicitly and with ease all the
interaction vertices for any number of oscillators and any frequencies. The
simplicity of such computations is one of the payoffs of the reformulation
provided by MSFT. We reproduce and generalize many results that were
obtained through other methods and obtain new ones that are computed for the
first time. In particular, in section-V we obtain simple analytic
expressions for the Neumann coefficients, including zero modes, for all $n$%
-point string vertices, for any frequencies. The spectroscopy of these
coefficients for the case of $n$=3 and an infinite number of oscillators
agrees with the available results in the literature \cite{rsz3}. This helps
establish MSFT as a precise definition of string field theory. In section-VI
we analyze the small fluctuations in VSFT \cite{r-RSZ}. We show quite
generally that, to obtain nontrivial masses and couplings for the small
fluctuations, there has to be an associativity anomaly of the star product
in any formulation of VSFT. We identify the source that could explain the
anomaly in detail for an infinite number of oscillators, but leave the full
resolution of the problem to future work

\section{Moyal formulation of string field theory (MSFT)}

\subsection{Moyal star product}

\label{moyalsection}The position representation $\psi\left(
x_{0},x_{2n},x_{2n-1}\right) $ of a string field is related to the
oscillator representation of the field $|\psi\rangle$ by the Fock space
bra-ket product $\psi\left( x_{0},x_{2n},x_{2n-1}\right) =$ $\langle
x|\psi\rangle.$ The position state $\langle x|$ is constructed in Fock space
from string oscillators $\alpha_{n}$ with frequencies $\kappa_{n}$ \cite{GJ}%
\footnote{%
Compared to the conventional harmonic oscillators $a_{n}$ used in \cite{GJ}
we normalize our string oscillators as $\alpha_{n}=\sqrt{\kappa_{n}}a_{n}$
and $\alpha_{-n}=\sqrt{\kappa_{n}}a_{n}^{\dagger}$ for $n\geq1.$ These
satisfy the commutation rules $\left[ \alpha_{n},\alpha_{m}\right]
=\varepsilon\left( n\right) \kappa_{\left\vert n\right\vert }\delta_{n+m}$
where $\varepsilon\left( n\right) =sign\left( n\right) $ for all $n\neq0.$ %
\label{oscill}}
\begin{equation}
\langle x|=\langle x_{0}|~\exp\sum_{n\geq1}\left( \frac{1}{2\kappa_{n}}%
\alpha_{n}^{2}+i\frac{\sqrt{2}x_{n}}{l_{s}}\alpha_{n}-\frac{\kappa_{n}}{%
2l_{s}^{2}}x_{n}^{2}\right) \prod_{n\geq1}\left( \frac{\kappa_{n}}{\pi
l_{s}^{2}}\right) ^{d/4},   \label{x}
\end{equation}
where $l_{s}$ is the fundamental string length, $x_{0}$ is the center of
mass mode of the string. The state $\langle x_{0}|$ may also be written out
explicitly as above for any frequency $\kappa_{0}.$

In much of our formulation it is not necessary to specify the number of
oscillators$,$ or the frequencies $\kappa_{\left\vert n\right\vert }$ as a
function of $n$. We will take advantage of this to easily define a somewhat
generalized formulation of string field theory by allowing arbitrary
frequencies $\kappa_{\left\vert n\right\vert }$ and number of oscillators $2N
$. This flexibility will permit us to discuss a cutoff theory as defined in
\cite{r-BarsMatsuo} and described below in detail. This will be important to
obtain well defined quantities and resolve associativity anomalies in string
field theory. Thus, we will not indicate upper limits in sums or products
(such as Eq.(\ref{x})) to imply that such equations are valid in either the
cutoff theory (with upper limit $n=2N$) or the full theory (with upper limit
$n=\infty$). To make contact with the usual open string field theory at
various points we need to set $\kappa_{\left\vert n\right\vert
}\rightarrow\left\vert n\right\vert $ and $2N\rightarrow\infty.$ While most
structures exist for a large range of parameters $\kappa_{\left\vert
n\right\vert },N$, certain quantities, such as the Virasoro algebra, exist
only in this limit.

The MSFT formulation is obtained by performing a Fourier transform only on
the odd string position modes of $\langle x|$ or equivalently $\psi\left(
x_{0},x_{2n},x_{2n-1}\right) .$ We will use the notation $e=2n$ and $o=2n-1$
for even and odd integers ($e$ excludes zero). The Fourier image in the
Moyal basis $A(\bar{x},x_{e},p_{e})$ is given as follows \cite{witmoy}

\begin{equation}
A(\bar{x},x_{e},p_{e})=\det\left( 2T\right) ^{d/2}\left( \int dx_{o}^{\mu
}\right) \,e^{-\frac{2i}{\theta}\eta_{\mu\nu}\sum_{e,o>0}p_{e}^{%
\mu}T_{eo}x_{o}^{\nu}}\psi(x_{0},x_{e},x_{o}),\   \label{A}
\end{equation}
where $d$ is the number of dimensions (d=27 including the bosonized ghosts),
$\theta\,$\ is a parameter that has units of area in phase space, $T_{eo}$
is a special matrix given below which is intimately connected to split
strings, and $\bar{x}=x\left( \pi/2\right) $ is the string midpoint which
may be rewritten in terms of $x_{0},x_{e}$ via Eq.(\ref{mode-expansion})
\begin{equation}
x_{0}=\bar{x}+\sum_{e>0}x_{e}w_{e},   \label{mid}
\end{equation}
where $w_{e}$ is given below. For the center of mass state $\langle x_{0}|$
this change of variables may be written as a translation of $\langle\bar{x}|$%
\begin{equation}
\langle x_{0}|=\langle\bar{x}+\sum_{e>0}x_{e}w_{e}|=\langle\bar{x}%
|\exp\left( ip\cdot\sum_{e>0}x_{e}w_{e}\right) ,
\end{equation}
where $p^{\mu}$ is the center of mass momentum operator for the full string.
Then the Moyal image (\ref{A}) of the position space state $\langle x|$ is
obtained by applying the Fourier transform as well as the change of
variables (\ref{mid}) to Eq.(\ref{x}). The result is
\begin{equation}
\langle\bar{x},x_{e},p_{e}|=\langle\bar{x}|~e^{\sum\left( \frac{\alpha
_{e}^{2}}{2\kappa_{e}}-\frac{\alpha_{o}^{2}}{2\kappa_{o}}\right)
}e^{-\sum_{ij}\xi_{i}\left( M_{0}\right)
_{ij}\xi_{j}-\sum_{i}\xi_{i}\lambda_{i}}\det\left(
4\kappa_{e}^{1/2}T\kappa_{o}^{-1/2}\right) ^{d/2},   \label{p}
\end{equation}
where $x_{0}$ is written in terms of $\bar{x}$ as above, and we have defined
$\xi_{i}^{\mu},\lambda_{i}^{\mu},\left( M_{0}\right) _{ij}$ as follows%
\begin{equation}
\xi_{i}^{\mu}=\left( x_{2}^{\mu},x_{4}^{\mu},\cdots,p_{2}^{\mu},p_{4}^{\mu
},\cdots\right) ,   \label{xi}
\end{equation}%
\begin{equation}
\lambda^{\mu}=\left(
\begin{array}{c}
-\frac{i\sqrt{2}}{l_{s}}\alpha_{e}^{\mu}-ip^{\mu}w_{e} \\
-\frac{2\sqrt{2}l_{s}}{\theta}\sum_{o>0}T_{eo}\kappa_{o}^{-1}\alpha_{o}^{\mu}%
\end{array}
\right) ,\;   \label{lambda}
\end{equation}%
\begin{equation}
M_{0}=\left(
\begin{array}{cc}
\frac{1}{2l_{s}^{2}}\kappa_{e}\delta_{ee^{\prime}} & 0 \\
0 & \frac{2l_{s}^{2}}{\theta^{2}}Z_{ee^{\prime}}%
\end{array}
\right)
,\;Z_{ee^{\prime}}=\sum_{o>0}T_{eo}\kappa_{o}^{-1}T_{e^{\prime}o}.\;\;
\label{M0}
\end{equation}
Then one can directly relate $A(\bar{x},x_{e},p_{e})$ to the state $%
|\psi\rangle$ in the oscillator formalism%
\begin{equation}
A(\bar{x},x_{e},p_{e})=\langle\bar{x},x_{e},p_{e}|\psi\rangle.
\label{psitoA}
\end{equation}

As shown in \cite{witmoy} the Witten star product becomes the Moyal star
product in the phase space of each even mode\footnote{%
It is also possible to take a Fourier transform for the even $x_{2n}$ and
end up with a formulation in the odd phase space $\left(
x_{2n-1},p_{2n-1}\right) .$ \label{oddspace}} except the midpoint
\begin{equation}
\left( A\ast B\right) (\bar{x},x_{e},p_{e})=e^{\frac{3i}{2}\bar{x}%
_{27}}A\left( \bar{x},x_{e},p_{e}\right) \,\,e^{\frac{i\theta}{2}%
\eta_{\mu\nu }\sum_{n}\left( \frac{\overleftarrow{\partial}}{\partial
x_{e}^{\mu}}\frac{\overrightarrow{\partial}}{\partial p_{e}^{\nu}}-\frac{%
\overleftarrow {\partial}}{\partial p_{e}^{\nu}}\frac{\overrightarrow{%
\partial}}{\partial x_{e}^{\mu}}\right) }B\left( \bar{x},x_{e},p_{e}\right)
\label{moyalstar}
\end{equation}
The product is local at the midpoint in all dimensions, and there is a
midpoint insertion $e^{i3\bar{x}^{27}/2}$ in the 27$^{th}$ dimension which
is the bosonized ghost coordinate. It should be emphasized that all $x_{0}$
dependence should first be rewritten in terms of $\bar{x}$ and $x_{e}$ by
using Eq.(\ref{mid}) before the Moyal star is computed\footnote{%
An alternative approach that produces the same result is to keep the $x_{0}$
dependence intact, but replace the derivative $\partial/\partial x_{e}$ in
the Moyal star product in Eq.(\ref{moyalstar}) by $\partial/\partial
x_{e}\rightarrow \partial/\partial x_{e}+\left( \partial x_{0}/\partial
x_{e}\right) $ $\partial/\partial x_{0}=\partial/\partial x_{e}+w_{e}$ $%
\partial/\partial x_{0}$. \label{x0}}. This setup can be related to the star
product in the oscillator formalism by the following formula in the
oscillator language%
\begin{equation}
\left( A\ast B\right) (\bar{x},x_{e},p_{e})=\langle_{1}\bar{x}%
,x_{e},p_{e}|\langle_{2}A|\langle_{3}B||V_{123}\rangle.
\end{equation}
Indeed in section-5 we will show in detail how the Neumann coefficients $%
\left( V_{n}^{rs}\right) _{kl}$ in $n$-point vertices in the oscillator
formalism, including the zero modes $\left( V_{n}^{rs}\right) _{k0},\left(
V_{n}^{rs}\right) _{00},$ directly follow from the straightforward and
concise Moyal product in Eq.(\ref{moyalstar}). Thus, MSFT is a precise
definition of string field theory.

Note that the noncommutativity is independent for each mode. The Moyal
product has been diagonalized in string mode space by the insertion of the
matrix $T_{eo}$ in the definition of the Fourier transform. Therefore for
each independent even string mode, except the midpoint, we have the star
commutation rules (for simplicity, we suppress the midpoint ghost insertion)%
\begin{equation}
\left[ x_{e}^{\mu},p_{e^{\prime}}^{\nu}\right] _{* }=i\theta\eta^{\mu\nu
}\delta_{ee^{\prime}},\;\;\left[ x_{e}^{\mu},x_{e^{\prime}}^{\nu}\right] _{*
}=\left[ p_{e}^{\mu},p_{e^{\prime}}^{\nu}\right] _{* }=0.
\end{equation}
Taking all the modes together we have a noncommutative space of $2dN$
dimensions labelled by $\xi_{i}^{\mu}$, with commutation relations
\begin{equation}
\left[ \xi_{i}^{\mu},\xi_{j}^{\mu}\right] _{* }=\sigma_{ij}\eta^{\mu\nu
},\;\sigma=i\theta\left(
\begin{array}{cc}
0 & 1 \\
-1 & 0%
\end{array}
\right) .   \label{sigma}
\end{equation}
Here the blocks $1$ ($0$) represent the unit (zero) matrices in even mode
space. In terms of $\sigma_{ij}$ the star is given by%
\begin{equation}
* =e^{\frac{3i}{2}\bar{x}_{27}}\exp\left( \frac{\sigma_{ij}}{2}\eta_{\mu\nu }%
\frac{\overleftarrow{\partial}}{\partial\xi_{i}^{\mu}}\frac{\overrightarrow {%
\partial}}{\partial\xi_{j}^{\nu}}\right) \,.   \label{star}
\end{equation}
It must be emphasized that the noncommutativity associated with $\theta$ is
a device for formulating string interactions. The commutators above do not
follow from quantum mechanics, and $\theta$ has no relation to the Planck
quantum $\hbar$ although it has the same units. The parameter $\theta$ is
present already in \textit{classical} string field theory. When string field
theory is quantized $\hbar$ is introduced as an additional parameter as a
measure of the quantum noncommutativity of fields. The $\theta$
noncommutativity is in a spacetime with $2d$ dimensions $\left( x_{e}^{\mu
},p_{e}^{\mu}\right) $ for each even mode $e$. One may think of $\sigma
_{ij}\eta^{\mu\nu}$ as a giant constant \textquotedblleft magnetic field" in
the space of all the modes. String theory has sufficient gauge symmetry to
insure unitarity in such a noncommutative spacetime which includes timelike
components $\xi_{i}^{0}$.

We emphasize that as far as $\sigma_{ij}$ is concerned, the midpoint
positions $\bar{x}^{\mu}$ are commutative. If a constant background
antisymmetric $B_{\mu\nu}$ field is introduced then some components of the
midpoint $\bar
{x}^{\mu}$ will also become noncommutative. In this case our
formalism is easily generalized to accommodate the noncommutative midpoint.
In this paper, for simplicity, we will assume that the $B_{\mu\nu}$
background is zero.

\subsection{The matrices $T,R,w,v$}

The matrix $T$ in Eq.(\ref{A}) plays a crucial role in bringing the star
product to the diagonal form in mode space as in Eq.(\ref{moyalstar}), and
is an important bridge in making connections between MSFT and other
formulations of SFT. Therefore we list some of its properties here. It is
labelled $T_{eo}~$with even positive integers $\left( e\right) $ on its left
and odd positive integers $\left( o\right) $ on its right (zero is excluded
from $e$). Its inverse is $R_{oe},$ and there are two special vectors with
components $w_{e}$ and $v_{o},$ related to each other by $v_{o}=\sum
_{e>0}w_{e}T_{eo},$ that play a role in connection with the midpoint Eq.(\ref%
{mid}). There are also the frequencies $\kappa_{2n},\kappa_{2n-1}$ of even
and odd string oscillators that we designate as two diagonal matrices $%
\kappa_{e},\kappa_{o}$. These quantities satisfy the following matrix
relations \cite{r-BarsMatsuo}%
\begin{align}
R & =\left( \kappa_{o}\right) ^{-2}\bar{T}\left( \kappa_{e}\right)
^{2},\quad R=\bar{T}+v\bar{w},\quad v=\bar{T}w,\quad w=\bar{R}v,  \label{tr}
\\
TR & =1,\quad RT=1,\;\quad\bar{R}R=1+w\bar{w},\;\;\bar{T}T=1-v\bar {v},\;
\label{rtv} \\
T\bar{T} & =1-\frac{w\bar{w}}{1+\bar{w}w},\quad Tv=\frac{w}{1+\bar{w}w},\quad%
\bar{v}v=\frac{\bar{w}w}{1+\bar{w}w}  \label{tt} \\
Rw & =v(1+\bar{w}w),\quad R\bar{R}=1+v\bar{v}\left( 1+\bar{w}w\right) .
\label{Rw}
\end{align}
where a bar on top of a symbol means its transpose. The four equations in
Eq.(\ref{tr}) are defining equations in the sense that they determine $%
T,R,w,v$ as we will see in the next paragraph. All the other equations
follow from these defining equations; we listed all of them for later
convenience.

The first two equations in Eq.(\ref{tr}) are uniquely solved by the $N\times
N$ matrices
\begin{equation}
T_{eo}=\frac{w_{e}v_{o}\kappa_{o}^{2}}{\kappa_{e}^{2}-\kappa_{o}^{2}}%
,\;R_{oe}=\frac{w_{e}v_{o}\kappa_{e}^{2}}{\kappa_{e}^{2}-\kappa_{o}^{2}}.
\label{TRcut}
\end{equation}
Inserting these into the last two equations in Eq.(\ref{tr}) gives
\begin{equation}
\sum_{e>0}\frac{w_{e}^{2}}{\kappa_{e}^{2}-\kappa_{o}^{2}}=\frac{1}{\kappa
_{o}^{2}},\;\sum_{o>0}\frac{v_{o}^{2}}{\kappa_{e}^{2}-\kappa_{o}^{2}}=\frac {%
1}{\kappa_{e}^{2}}.   \label{wv}
\end{equation}
These determine $w_{e}^{2},v_{o}^{2}$ for each component in terms of the
frequencies $\kappa_{e},\kappa_{o}.$ We now state a theorem: the $%
w_{e}^{2},v_{o}^{2}$ that satisfy Eq.(\ref{wv}) obey the following
orthogonality relations%
\begin{equation}
\sum_{e>0}\frac{w_{e}^{2}\kappa_{e}^{2}}{\left(
\kappa_{e}^{2}-\kappa_{o}^{2}\right) \left(
\kappa_{e}^{2}-\kappa_{o^{\prime}}^{2}\right) }=\frac{\delta_{oo^{\prime}}}{%
v_{o}^{2}\kappa_{o}^{2}},\;\sum_{o>0}\frac {v_{o}^{2}\kappa_{o}^{2}}{\left(
\kappa_{e}^{2}-\kappa_{o}^{2}\right) \left(
\kappa_{e^{\prime}}^{2}-\kappa_{o}^{2}\right) }=\frac{\delta _{ee^{\prime}}}{%
w_{e}^{2}\kappa_{e}^{2}},   \label{theorem}
\end{equation}
for any given $\kappa_{e},\kappa_{o}.$We proved this theorem analytically
for $o\neq o^{\prime}$ and $e\neq e^{\prime},$ and checked it via computer
algebraically and numerically for $o=o^{\prime}$ and $e=e^{\prime}$. Then,
as a corollary of the theorem (\ref{theorem}), the $T,R$ of Eq.(\ref{TRcut})
are each other's inverses $TR=1=RT.$

These $T,R,w,v$ satisfy all of the remaining relations in (\ref{rtv}-\ref{Rw}%
) for any $\kappa_{e},\kappa_{o}$, and any set of signs for $w_{e},v_{o},$
at any $N$. Using the remaining freedom, the signs of $w_{e},v_{o}$ are
chosen as
\begin{equation}
sign\left( w_{e}\right) =\left( \sqrt{-1}\right) ^{-e+2},\;sign\left(
v_{o}\right) =\left( \sqrt{-1}\right) ^{o-1},   \label{signss}
\end{equation}
to agree with the large $N$ theory. Also, one may choose $\kappa_{e}=e$, $%
\kappa_{o}=o$ in the cutoff theory for all $N,$ although this is not
necessary.

At large $N$ when $\kappa_{e}\rightarrow e$, $\kappa_{o}\rightarrow o,$
these $T,R,w,v$ become precisely the infinite matrices that emerged in the
split string formalism given by
\begin{align}
T_{2n,2m-1} & =\frac{2\left( -1\right) ^{m+n+1}}{\pi}\left( \frac {1}{2m-1+2n%
}+\frac{1}{2m-1-2n}\right) ,  \label{Tinf} \\
R_{2m-1,2n} & =\frac{4n\left( -1\right) ^{n+m}}{\pi\left( 2m-1\right) }%
\left( \frac{1}{2m-1+2n}-\frac{1}{2m-1-2n}\right)  \label{Rinf} \\
w_{2n} & =\sqrt{2}\left( -1\right) ^{n+1},\;\;v_{2n-1}=\frac{2\sqrt{2}}{\pi}%
\frac{\left( -1\right) ^{n+1}}{2n-1},\;  \label{winf} \\
\kappa_{2n} & =2n,\;\;\kappa_{2n-1}=2n-1.   \label{kinf}
\end{align}
These infinite matrices satisfy all the relations in Eqs.(\ref{tr}-\ref{Rw}%
). We emphasize that here we obtained them directly from the defining
relations in Eq.(\ref{tr}).

We have shown that the relations Eq.(\ref{tr}) play a defining role in the
theory. The oscillator frequencies $\kappa_{e},\kappa_{o}$ and the set of
signs (\ref{signss}) are additional inputs in defining the cutoff or the
infinite theory through these relations. In the cutoff theory $\kappa
_{e},\kappa_{o}$ may be taken to be identical to Eq.(\ref{kinf}) at any $N$,
or some other convenient choice that tends to Eq.(\ref{kinf}) at large $N.$

\subsection{Cutoff procedure}

SFT needs some regularization in any formulation to give rigorous
mathematical meaning to some computations. As discussed in \cite%
{r-BarsMatsuo} the origin of the singular behavior in MSFT is due to the
even vector $w$ whose norm becomes infinite as the number of modes goes to
infinity $\bar{w}w\rightarrow2N\rightarrow\infty$ as seen from Eq.(\ref{winf}%
). Therefore in some computations we will use a finite number of modes to
regularize certain quantities before taking the large $N$ limit. This is
somewhat similar to level truncation, but our cutoff procedure is more
reliable in that all the relations in Eqs.(\ref{tr}-\ref{Rw}) remain valid
at all values of the cutoff $N$ and any set of frequencies $%
\kappa_{e},\kappa_{o}.$ It turns out that for certain delicate computations,
where naive level truncation could not be fully trusted, our cutoff method
gives unambiguous results consistent with gauge invariance.

However, there will remain some open issues on how to recover certain
non-perturbative effects in the context of vacuum string field theory, such
as closed strings, tachyon mass, D-brane tensions, as we will explain in
Section-VI. Such issues are all related to the existence of a zero
eigenvalue in the infinite theory, as explained below, and its relation to
an associativity anomaly in the star product in a very special way \cite%
{r-BarsMatsuo}. To sharpen these issues first we will examine the theory
through our cutoff procedure, and then we will see the precise point on
which to focus to be able to extract nonperturbative information from vacuum
string field theory.

In the large $N$ limit the infinite matrices $T,R,w,v$ as well as the star
product (\ref{moyalstar}) have an associativity anomaly that needs to be
treated delicately \cite{r-BarsMatsuo}. In particular from Eq.(\ref{tt})
note that $Tv\rightarrow0$ when $\bar{w}w\rightarrow\infty$, indicating a
zero mode which is the cause for the associativity anomaly. At finite $N$
there is no associativity anomaly because the zero mode is shifted away from
zero as seen by computing the determinant of $T$
\begin{equation}
\det\left( T\right) =\frac{1}{\sqrt{1+\bar{w}w}}=\frac{\det\kappa_{o}}{%
\det\kappa_{e}}=\dprod \limits_{n\geq1}\frac{\kappa_{2n-1}}{\kappa_{2n}},
\label{detT}
\end{equation}
The right hand side follows from Eqs.(\ref{tr}-\ref{Rw}) as a nontrivial
relation\footnote{%
Note that if we had first sent the cutoff to infinity, Eq.(\ref{rtv}) would
give $T\bar{T}=1$ with a determinant $\det\left( T\bar{T}\right) =1,$ while $%
\bar{T}T=1-v\bar{v}$ with a determinant $\det\left( \bar{T}T\right) =1-\bar{v%
}v=0.$ This is another example of the associativity anomaly which is
resolved uniquely and controlled with our cutoff method. The naive level
truncation would still be ambiguous in this case, and would not yield the
type of analytic relations given in Eq.(\ref{detT}) or many others.}. Thus,
all ambiguities are uniquely controlled by the cutoff, and one can proceed
with confidence using associativity both for the matrices $T,R,w,v$ and for
the star product in Eq.(\ref{moyalstar}).

Generically, to perform a computation in the cutoff theory, we do not need
an explicit form of the finite matrices $T,R,w,v,\kappa_{e},\kappa_{o}$
since in most cases it is sufficient to use the relations (\ref{tr}-\ref{Rw}%
) which are valid at any cutoff $N$, including infinity. The useful
information that the cutoff theory supplies is the behavior of a computation
as a function of $\bar{w}w$ (see e.g. Eq.(\ref{detT}) or Eqs.(\ref{tt},\ref%
{Rw})). As seen from Eq.(\ref{winf}) $\bar{w}w$ blows up as the cutoff is
removed at the rate $\bar{w}w\rightarrow2N\rightarrow\infty.$ Knowing the $%
\bar{w}w$ dependence of a quantity determines its dependence on the cutoff
at the very end of a calculation when $N\rightarrow\infty$. This is the
cutoff procedure that we will use when a computation is delicate; but
otherwise it is not important to use the cutoff, and the explicit infinite
version of $T,R,v,w,\kappa_{e}\kappa_{o}$ in Eqs.(\ref{Tinf}-\ref{kinf}) may
be used directly.

We provide here another basis for $T,R,w,v,\kappa_{e},\kappa_{o}$ which
sheds additional insight into the nature of the cutoff and the structure of
Eqs.(\ref{tr}-\ref{Rw}). In the new basis we see more clearly why the large $%
N$ limit is tricky and different than the naive level truncation. Let us
apply two orthogonal transformations $S_{e},S_{o}$ to the relations (\ref{tr}%
-\ref{Rw}) to write them in a basis in which the vectors $w_{e},v_{o}$ point
in a single direction. Then we find from Eqs.(\ref{tr}-\ref{Rw}) $\bar{w}%
_{e}=\left( 0,\cdots,0,w\right) \bar{S}_{e}$ and $\bar{v}_{o}=\left(
0,\cdots,0,w\left( 1+w^{2}\right) ^{-1/2}\right) \bar{S}_{o}$ while the
matrices $T,R$ become block diagonal%
\begin{equation}
T=S_{e}\left(
\begin{array}{cc}
t & 0 \\
0 & \frac{1}{\sqrt{1+w^{2}}}%
\end{array}
\right) \bar{S}_{o},\;R=S_{o}\left(
\begin{array}{cc}
\bar{t} & 0 \\
0 & \sqrt{1+w^{2}}%
\end{array}
\right) \bar{S}_{e},\;t\bar{t}=\bar{t}t=1,   \label{blockdiagonal1}
\end{equation}
such that the $\left( N-1\right) \times\left( N-1\right) $ matrix $t$ is
orthogonal (note that $t$ could be replaced by $1$ by absorbing it into a
redefinition of $S_{e}$ or $S_{o}$). Then all the relations are satisfied by
the new forms of $T,R,v,w$ except for the first relation in Eq.(\ref{tr})
which contains the information on the oscillator frequencies. This last
relation determines $T,R,w,v$ uniquely in terms of the frequencies $\kappa
_{e},\kappa_{o}$ as discussed in \cite{r-BarsMatsuo}. When the rank of the
matrices $N$ goes to infinity, the single parameter $w^{2}$ also grows at
the rate $w^{2}\rightarrow2N\rightarrow\infty.$ This combined limit is the
nontrivial aspect in our cutoff procedure. In particular note that $T$
develops a zero eigenvalue which is the cause of associativity anomalies
\cite{r-BarsMatsuo}. But in all computations where an anomaly occurs in the
infinite theory, it comes from ambiguous terms $\frac{\infty}{\infty}$ that
become uniquely evaluated in the form of $\frac{w}{w}$ in the cutoff theory,
thus resolving the anomaly. The block diagonal basis of this paragraph sheds
further light into our consistent cutoff procedure, but we prefer using the
original basis, along with the consistency equations (\ref{tr}-\ref{Rw}),
since the explicit forms of $T,R,v,w,\kappa_{e},\kappa_{o}$ are available in
the infinite limit in the original basis as in Eqs.(\ref{Tinf}-\ref{kinf}).

Other bases may also be considered. In particular, it is useful to study the
basis in which the matrix $\left( \kappa_{e}\right) ^{1/2}T\left(
\kappa_{o}\right) ^{-1/2},$ which occurs prominently in many expressions, is
diagonal. We will see later that in this basis the Neumann coefficients for
all $n$-point string vertices are diagonal. The orthogonal transformations
that diagonalize this matrix are denoted $V^{e},V^{o}$
\begin{equation}
\left( \kappa_{e}\right) ^{1/2}T\left( \kappa_{o}\right) ^{-1/2}=V^{e}\tau%
\bar{V}^{o},   \label{ktk}
\end{equation}
where $\tau$ is a $N\times N$ diagonal matrix with eigenvalues $\tau_{k}$
labelled by integers $k=0,1,\cdots,\left( N-1\right) .$ These orthogonal
transformations provide the map between the bases introduced in \cite{witmoy}
and \cite{DLMZ}. In the large $N$ limit, the matrix elements become
functions of a continuous parameter $\left( V^{e}\right) _{ek}\rightarrow
V_{e}\left( k\right) $, $\left( V^{o}\right) _{ok}\rightarrow V_{o}\left(
k\right) ,$ $\tau_{k}\rightarrow\tau\left( k\right) ,$ and Eq.(\ref{ktk})
takes the form
\begin{equation}
\sqrt{e}T_{eo}\frac{1}{\sqrt{o}}=\int_{0}^{\infty}dk~V_{e}\left( k\right)
V_{o}\left( k\right) ~\tau\left( k\right) .
\end{equation}
such that the eigenvalues $\tau_{k}$ become a continuous function as seen
from Eq.(6.7) in \cite{DLMZ}
\begin{equation}
\tau\left( k\right) =\tanh\left( \frac{\pi k}{4}\right) .   \label{tau}
\end{equation}
Furthermore, the functions $V_{e}\left( k\right) ,V_{o}\left( k\right) $ are
obtained from the following generating functions extracted from Eqs.(3.4) in
\cite{DLMZ}\footnote{%
Our normalization of $V_{e}\left( k\right) $ and $V_{o}\left( k\right) $ are
consistent with the orthonormality conditions of the matrices $V^{e},V^{o}.$
Our $V_{e}\left( k\right) ,V_{o}\left( k\right) $ are related by a factor of
$\sqrt{2}$ to the $v_{e}\left( k\right) ,v_{o}\left( k\right) $ of \cite%
{DLMZ}.} for $k\geq0$ ,%
\begin{equation}
\sum_{o}\frac{V_{o}\left( k\right) \left( \tan z\right) ^{o}}{\sqrt{o}}=%
\frac{\sinh\left( kz\right) }{\sqrt{k\sinh\left( \frac{\pi k}{2}\right) }}%
,\;\;\sum_{e}\frac{V_{e}\left( k\right) \left( \tan z\right) ^{e}}{\sqrt{e}}=%
\frac{\cosh\left( kz\right) -1}{\sqrt{k\sinh\left( \frac{\pi k}{2}\right) }}
\label{vevo}
\end{equation}
As seen from the expressions in Eqs.(\ref{tt},\ref{detT},\ref{blockdiagonal1}%
), a zero mode is expected in the large $N$ limit, and this is explicitly
seen in the expression of $\tau\left( k\right) $ at $k=0.$The associativity
anomalies caused by the zero mode in the large $N$ limit, as discussed in
\cite{r-BarsMatsuo} must occur also in the continuous basis of \cite{DLMZ}.
These are absent in the cutoff theory because the similarity transformations
$V^{e},V^{o}$ are well defined $N\times N$ finite matrices, the basis
labelled by $k$ is discrete, and the lowest eigenvalue among the $\tau_{k}$
is positive definite. Our regularized theory removes the associativity
anomaly by shifting this mode away from zero in any formalism, including in
the $k$ basis. Thus some issues raised in \cite{DLMZ} are avoided and
resolved in our regularized theory (in this connection see also Eq.(\ref{gt}%
) and the spectrum of the Virasoro operator $L_{0}$ in Eq.(\ref{spectrum})).

In the infinite theory the zero eigenvalue of the matrix $\left( \kappa
_{e}\right) ^{1/2}T\left( \kappa_{o}\right) ^{-1/2}$ is related to a number
of important nonperturbative issues. We will see in section-VI that certain
non-perturbative effects vanish when the star product is strictly
associative, as guaranteed by our cutoff procedure. We will suggest that to
recover such non-perturbative effects some nonassociativity will need to be
introduced in the \textit{definition} of the inverse of the matrix $\left(
\kappa _{e}\right) ^{1/2}T\left( \kappa_{o}\right) ^{-1/2}.$

\subsection{Identity, nothing state, reality, trace, integral, gauge
invariant action}

The reformulation of the star product greatly simplifies computations of
interacting string fields. Recall that the $x$-representation $\psi\left(
x_{0},x_{2n},x_{2n-1}\right) $ is related to the oscillator representation
by the Fock space bra-ket product $\psi\left( x_{0},x_{2n},x_{2n-1}\right)
=\langle x|\psi\rangle$ where $\langle x|$ is constructed from oscillators
in Eq.(\ref{x}) and $|\psi\rangle$ is a string state, while the MSFT field $%
A\left( \bar{x},x_{2n},p_{2n}\right) $ is the Fourier transform given in Eq.(%
\ref{A}). In this section we give a few simple illustrations of the
simplifications obtained in MSFT which help make string field theory more
manageable. More involved examples of simplified computations will appear in
later sections in this paper.

The first example is the identity field. If we ignore the midpoint ghost
insertion we can easily notice that the only identity of the Moyal product
is the number 1. Taking into account the ghost field insertion, the identity
field is the pure midpoint field, $I=\exp\left( -3i\bar{x}_{27}/2\right) .$
The midpoint phase is insensitive to the Moyal star, it is designed to
cancel the midpoint ghost insertion in the definition in Eq.(\ref{moyalstar}%
), so it really acts like the number 1. Therefore, it satisfies
\begin{equation}
I* A=A* I=A   \label{identity}
\end{equation}
for any string field $A\left( \bar{x},x_{2n},x_{2n-1}\right) .$ This result
for $I$ can be verified directly by taking the Fourier transform of the
identity field in the $x$-representation which is proportional to $\dprod
\limits_{n\geq1}\delta\left( x_{2n-1}\right) $.

A second example is the \textquotedblleft nothing state". In the $x$%
-representation the nothing state is a constant. The corresponding Moyal
field is a delta function for all even momenta $A_{nothing}\sim\dprod
\limits_{n\geq1}\delta\left( p_{2n}\right) .$ In computing the Moyal star
product of extremely localized states, such as the nothing state, one must
be aware of some nonperturbative properties in the powers of $\theta$ \cite%
{r-Bars2}.

A third example is the reality condition on string fields. In the oscillator
representation this is an awkward condition. In the $x$-representation it
becomes $\psi^{\ast}\left( x_{0},x_{2n},x_{2n-1}\right) =\psi\left(
x_{0},x_{2n},-x_{2n-1}\right) .$ In the Moyal basis it takes its simplest
form, namely the field is a real function under complex conjugation in the
usual sense $A^{\ast}\left( \bar{x},x_{2n},p_{2n}\right) =A\left( \bar
{x}%
,x_{2n},p_{2n}\right) .$

A fourth example is the integration which is needed to define c-number
quantities such as an action. In the original formulation of SFT integration
corresponds to folding a string on itself and integrating over the overlap
and midpoint. In MSFT this simplifies to phase space integrals which define
a \textquotedblleft trace" as in other applications of noncommutative
geometry, and a further integral over the midpoint with a ghost insertion%
\begin{align}
Tr\left( A^{\gamma}\left( \bar{x}\right) \right) & \equiv\int\dprod
\limits_{n\geq1,\mu}\frac{dx_{2n}^{\mu}dp_{2n}^{\mu}}{2\pi\theta}%
A^{\gamma}\left( \bar{x},x_{2n},p_{2n}\right) ,  \label{traceA} \\
\int Tr\left( A^{\gamma}\left( \bar{x}\right) \right) & =\int\left( d\bar{x}%
^{\mu}\right) ~e^{-i\gamma\bar{x}_{27}}Tr\left( A^{\gamma}\left( \bar{x}%
\right) \right) ,\;
\end{align}
where $\gamma$ is the ghost number of the field $A^{\gamma}.$ In particular,
the action takes the form%
\begin{equation}
S=\int\left( d\bar{x}^{\mu}\right) ~e^{-i\frac{3}{2}\bar{x}_{27}}~Tr\left(
\frac{1}{2}A\ast\left( \mathcal{Q}A\right) +\frac{1}{3}A\ast A\ast A\right)
.   \label{action}
\end{equation}
Here $A~$has ghost number $-1/2,$ the kinetic operator $\mathcal{Q}$ has
ghost number 1, and the star product has ghost number 3/2 due to the
insertion in Eq.(\ref{moyalstar}). Therefore the action density has ghost
number $\gamma =$3/2 which explains the midpoint ghost insertion in the last
integral.

The choice of the kinetic operator $\mathcal{Q}$ corresponds to a choice of
a vacuum (or D-branes). For the open string vacuum (space filling D$_{25}$
brane), $\mathcal{Q}$ is the usual BRST operator $\mathcal{Q}_{B}$
constructed from ghosts and Virasoro operators. Later in this paper we will
construct the Virasoro operators in Moyal space and will study some of their
properties. Since the Virasoro algebra is infinite dimensional, we cannot
achieve closure unless we take an infinite number of oscillators. Therefore
to construct the theory, and discuss its gauge invariance around the
perturbative vacuum, we must take $\kappa_{e}=\left\vert e\right\vert ,$ $%
\kappa_{o}=\left\vert o\right\vert $ and $N=\infty.$

For the conjectured nonperturbative closed string vacuum (no D-branes) $%
\mathcal{Q}$ is constructed purely from ghosts. In particular it is
suggested in \cite{GRSZ} that $\mathcal{Q}=\frac{1}{2i}(c(i)-c(-i)) $ is the
fermionic ghost at the midpoint. This version of the theory is called vacuum
string field theory (VSFT). In this case $\mathcal{Q}$ satisfies the usual
properties of an exterior derivative without recourse to the Virasoro
algebra. Then our action, which formally looks like a Chern-Simons action,
is gauge invariant under the gauge transformation
\begin{equation}
\delta A=\mathcal{Q}\Lambda+\Lambda* A-A *\Lambda.   \label{gaugetrans}
\end{equation}
for any number of oscillators 2$N$ and any frequencies $\kappa_{e},\kappa
_{o}.$ Of course, the gauge invariance requires an associative star product.
The setup of our theory guarantees associativity rigorously at any \textit{%
finite} $N$ .

We will see that the sliver state introduces a singularity at \textit{%
infinite} $N$. This is directly related to the zero eigenvalue of the
infinite matrix $\left( \kappa_{e}\right) ^{1/2}T\left( \kappa_{o}\right)
^{-1/2}$ we discussed in the previous section. In order to obtain closed
strings in VSFT, and have nontrivial physical results, an associativity
anomaly needs to be introduced through the \textit{definition} of the
inverse of this infinite matrix \cite{r-BarsMatsuo}. Inevitably, this
implies less gauge invariance which is introduced in a rather subtle way.
VSFT is not well defined until the singularity is universally defined and
the gauge invariance principle understood. To study these issues we will
proceed with full gauge invariance at finite $N$ and examine its
conclusions. By doing so, we identify the source of the possible
associativity anomaly in this paper but leave its resolution to later work.

\subsection{Perturbative vacuum and basis in Moyal formalism}

We now establish several concrete maps between the usual formulation and the
Moyal formulation. These will be useful to compare the Moyal formalism to
others. We keep in mind that the statements we make hold at any $\kappa
_{e},\kappa_{o},N$ as well as in the usual limit. The perturbative string
states are represented by some very special field configurations. In $x$%
-space a string field $\psi\left( x_{0},x_{2n},x_{2n-1}\right) $ in the
\textit{perturbative} Hilbert space may be expanded in terms of a complete
set of perturbative string fields $\psi_{\left( n_{1},n_{2},\cdots\right) }$%
\begin{equation}
\psi=\sum_{n_{1},n_{2},\cdots\geq0}\phi^{\left( n_{1},n_{2},\cdots\right)
}\left( x_{0}\right) \,\,\psi_{\left( n_{1},n_{2},\cdots\right) }\left(
x_{e},x_{o}\right) ,
\end{equation}
The $\phi^{\left( n_{1},n_{2},\cdots\right) }\left( x_{0}\right) $ are local
fields that represent the excited levels of the string in position space (as
functions of the center of mass mode of the string $x_{0}^{\mu}$), while the
field configurations $\psi_{\left( n_{1},n_{2},\cdots\right) }\left(
x_{2n},x_{2n-1}\right) $ represent the string excitations that are obtained
by applying creation operators on the ground state field $\psi_{0}\left(
x_{e},x_{o}\right) \sim\exp\left( -
\sum_{n\geq1}\kappa_{n}x_{n}^{2}/2l_{s}^{2}\right) .$ Up to a normalization,
this basis is given by%
\begin{equation}
\psi_{\left( n_{1},n_{2},\cdots\right) }\left( x_{e},x_{o}\right) \sim\left(
\prod_{i\geq1}\left( \alpha_{-i}\right) ^{n_{i}}\right) \,\psi_{0}\left(
x_{e},x_{o}\right) .   \label{basis}
\end{equation}
The oscillators $\alpha_{n}^{\mu}$ become represented by differential
operators for \textit{all positive and negative integers} $n$ (not zero)
acting on any $\psi\left( x_{0},x_{\left\vert 2n\right\vert },x_{\left\vert
2n-1\right\vert }\right) $
\begin{equation}
\langle x|\alpha_{n}|\psi\rangle=-\frac{i}{\sqrt{2}}\left( \epsilon(n) \frac{%
\kappa_{\left\vert n\right\vert }}{l_{s}}x_{\left\vert n\right\vert }+l_{s}\,%
\frac{\partial}{\partial x_{\left\vert n\right\vert }}\right) \psi\left(
x\right)   \label{aodd}
\end{equation}
where $\kappa_{\left\vert n\right\vert }$ is the oscillator frequency, $l_{s}
$ is the string length scale and $\epsilon\left( n\right) =\frac{n}{%
\left\vert n\right\vert }$ is the sign function. For $n=0,$ instead of an
oscillator we have a derivative
\begin{equation}
\langle x|\alpha_{0}|\psi\rangle=-il_{s}\frac{\partial\psi\left( x\right) }{%
\partial x_{0}}.   \label{azero}
\end{equation}
These differential operators satisfy the standard string oscillator
commutation relations when acting on $\psi$
\begin{equation}
\left[ \alpha_{n}^{\mu},\alpha_{n}^{\nu}\right] =\epsilon\left( n\right)
\kappa_{n}\delta_{n+m}\eta^{\mu\nu}.   \label{alphan}
\end{equation}
In Eq.(\ref{basis}) for simplicity of notation we have omitted the spacetime
indices on the $\alpha_{-i}^{\mu_{i}}$, and the corresponding spacetime
indices on the $\phi^{\left( n_{1},n_{2},\cdots\right) },$ but it is well
understood that they should be included since they represent the spin of the
field. For example, when all $n_{i}=0$, the field $\phi^{\left(
0,0,\cdots\right) }\left( x_{0}\right) \equiv t\left( x_{0}\right) $ is the
tachyon field with no spin, while the first excitation $\alpha_{-1}^{\mu
}\psi_{0}\left( x_{2n},x_{2n-1}\right) $ is associated with the vector field
$\left( \phi^{\left( 1,0,\cdots\right) }\left( x_{0}\right) \right)
_{\mu}\equiv A_{\mu}\left( x_{0}\right) $ with spin 1.

In the Moyal basis we have the corresponding expansion in terms of a
complete set of perturbative string fields
\begin{equation}
A\left( \bar{x},x_{2n},p_{2n}\right)
=\sum_{n_{1},n_{2},\cdots\geq0}\phi^{\left( n_{1},n_{2},\cdots\right)
}\left( x_{0}\right) \,\,A_{\left( n_{1},n_{2},\cdots\right) }\left(
x_{2n},p_{2n}\right) ,   \label{fullA}
\end{equation}
with the same set of local fields $\phi^{\left( n_{1},n_{2},\cdots\right)
}\left( x_{0}\right) .$ But now $x_{0}$ has to be rewritten in terms of $%
\bar{x}$ and $x_{2n}$ as in Eq.(\ref{mid}) before applying the Moyal
products as described following Eq.(\ref{moyalstar}). The complete set $%
A_{\left( n_{1},n_{2},\cdots\right) }\left( x_{2n},p_{2n}\right) $ is
related to the complete set $\psi_{\left( n_{1},n_{2},\cdots\right) }\left(
x_{2n},x_{2n-1}\right) $ via the Fourier transform in Eq.(\ref{A}). It will
be sufficient to construct the ground state in the Moyal basis $A_{0}\left(
x_{2n},p_{2n}\right) $ since all excited states will be obtained by applying
$\beta$ oscillators in Moyal space which will be defined below.

The ground state field for the string is independent of $x_{0}$ or $\bar{x}$
(except for the ghost part which contributes only in the selection rule) and
is given by
\begin{equation}
\psi_{0}\left( x\right) =\left( \prod_{n\geq1}\left( \frac{\kappa_{2n}}{\pi
l_{s}^{2}}\frac{\kappa_{2n-1}}{\pi l_{s}^{2}}\right) ^{d/4}\right)
\,\,e^{-\sum_{n\geq1}\left( \frac{\kappa_{2n}}{2l_{s}^{2}}x_{2n}\cdot x_{2n}+%
\frac{\kappa_{2n-1}}{2l_{s}^{2}}x_{2n-1}\cdot x_{2n-1}\right) }.
\end{equation}
As seen from (\ref{aodd}) it is annihilated by the positive oscillators $%
\langle x|\alpha_{n}|\psi_{0}\rangle=0$ for $n=1,2,$ $\cdots$. The state is
normalized so that
\begin{equation}
\left( \prod_{n\geq1}\prod_{\mu=0}^{d}\int dx_{2n}^{\mu}dx_{2n-1}^{\mu
}\right) \,\,\left\vert \psi_{0}\left( x_{2n},x_{2n-1}\right) \right\vert
^{2}=1.   \label{norm1}
\end{equation}
The ground state field in the Moyal basis is obtained through the Fourier
transform of $\psi_{0}$ as in Eq.(\ref{A}), with the result%
\begin{equation}
A_{0}=\prod_{n\geq1}\left( 2^{4}\frac{\kappa_{2n-1}}{\kappa_{2n}}\right)
^{d/4}\exp\left( -\sum_{n\geq1}\frac{\kappa_{2n}}{2l_{s}^{2}}x_{2n}\cdot
x_{2n}-\sum_{n,m\geq1}\frac{2l_{s}^{2}}{\theta^{2}}Z_{2n,2m}~p_{2n}~\cdot
p_{2m}\right) .   \label{ground}
\end{equation}
where the matrix $Z$ is given explicitly by%
\begin{equation}
Z_{2n,2m}=\sum_{k\geq1}T_{2n,2k-1}\frac{1}{\kappa_{2k-1}}T_{2m,2k-1}.
\label{Z}
\end{equation}
In the infinite cutoff limit we have (i.e. using Eq.(\ref{Tinf}-\ref{kinf}))%
\begin{align}
\lim_{N\rightarrow\infty}Z_{2n,2m} & =\frac{\left( -1\right) ^{m+n+1}}{%
\pi^{2}\left( n^{2}-m^{2}\right) }\left[ \psi(\frac{1}{2}+n)+\psi(\frac {1}{2%
}-n)-\psi(\frac{1}{2}+m)-\psi(\frac{1}{2}-m)\right]  \label{zpsi} \\
& =\frac{4\left( -1\right) ^{m+n+1}}{\pi^{2}\left( n^{2}-m^{2}\right) }\left[
\sum_{r=1}^{n}\frac{1}{2r-1}-\sum_{r=1}^{m}\frac{1}{2r-1}\right] \\
& =\frac{4}{\pi^{2}}\left(
\begin{array}{cccc}
1 & \frac{1}{9} & -\frac{1}{15} & \cdots \\
\frac{1}{9} & \frac{5}{9} & \frac{1}{25} & \cdots \\
-\frac{1}{15} & \frac{1}{25} & \frac{7\times37}{3^{3}5^{2}} & \cdots \\
\vdots & \vdots & \vdots & \ddots%
\end{array}
\right)   \label{Znum}
\end{align}
where $\psi(z)=\Gamma^{\prime}(z)/\Gamma(z)$ is the digamma function.

The norm of $A_{0}\left( x_{2n},p_{2n}\right) $ is determined through its
relation to $\psi_{0}$ in Eq.(\ref{A}) and is given by
\begin{equation}
\prod_{n\geq1}\prod_{\mu=0}^{d}\int\frac{dx_{2n}^{\mu}dp_{2n}^{\mu}}{%
2\pi\theta}\,\left\vert A_{0}\left( x_{2n},p_{2n}\right) \right\vert ^{2}=1.
\label{norm2}
\end{equation}
Note that this measure is consistent with Eqs.(\ref{A},\ref{norm1}) as well
as (\ref{traceA}). In computing this norm we needed to use
\begin{equation}
\det\left( Z\right) =\det\left( \bar{T}T\right) \left( \prod_{n\geq 1}\frac{1%
}{\kappa_{2n-1}}\right) =\dprod \limits_{n\geq1}\frac{\kappa_{2n-1}}{\left(
\kappa_{2n}\right) ^{2}}
\end{equation}
where the right hand side is unambiguously computed by using the relation
Eq.(\ref{detT}).

In summary, the normalized vacuum field in Moyal space is given by
\begin{equation}
A_{0}=\mathcal{N}_{0}~e^{-\bar{\xi}M_{0}\xi},\;Tr\left( A_{0}* A_{0}\right)
=1,\;\mathcal{N}_{0}=\left( \frac{\det\left( 16\kappa_{o}\right) }{%
\det\kappa_{e}}\right) ^{d/4},   \label{AN0}
\end{equation}
with $M_{0}$ defined in (\ref{M0}). Note that $Tr\left( A_{0}* A_{0}\right)
=Tr\left( A_{0}^{2}\right) =1$ because the Moyal star product between two
factors can be removed under integration. Here we have defined the norm $%
\mathcal{N}_{0},$ and the matrix $\left( M_{0}\right) _{ij}$ sandwiched
between the $\xi_{i}^{\mu}$ whose basis is given in Eq.(\ref{xi}) This form,
or Eq.(\ref{ground}), is valid for either the cutoff or the infinite version
of the theory.

It is useful to record at this juncture some of the technical properties of
the matrix
\begin{equation}
\Gamma\equiv Z\kappa_{e}=T\kappa_{o}^{-1}\bar{T}\kappa_{e}=T\kappa_{o}R%
\kappa_{e}^{-1}   \label{gamma}
\end{equation}
which will come up in the course of computing several quantities in later
sections, including the wedge fields $W_{n}\left( x,p\right) $ or sliver
field $\Xi\left( x,p\right) $ as functions of the vacuum field $A_{0}$. The
inverse of $\Gamma$ is given by
\begin{equation}
\Upsilon=\kappa_{e}^{-1}\bar{R}\kappa_{o}R=\kappa_{e}T\kappa_{o}^{-1}R.
\label{Ginv}
\end{equation}
Using Eq.(\ref{detT}) we compute its determinant
\begin{equation}
\det\Gamma=\left( \det\left( T\right) \right) ^{2}\frac{\det\kappa_{e}}{%
\det\kappa_{o}}=\det T=\frac{1}{\sqrt{1+\bar{w}w}}.   \label{detL}
\end{equation}
Furthermore, using Eqs.(\ref{tr}-\ref{Rw}) we compute%
\begin{equation}
\Gamma\bar{\Gamma}=T\kappa_{o}^{-1}\bar{T}\kappa_{e}^{2}T\kappa_{o}^{-1}\bar{%
T}=T\kappa_{o}RT\kappa_{o}^{-1}\bar{T}=T\bar{T}=1-\frac{w\bar{w}}{1+\bar{w}w}
\label{gbarg}
\end{equation}
and
\begin{equation}
\bar{\Gamma}\Gamma=\kappa_{e}T\kappa_{o}^{-1}\left( \bar{T}T\right)
\kappa_{o}^{-1}\bar{T}\kappa_{e}=\kappa_{e}T\kappa_{o}^{-1}\left( 1-v\bar {v}%
\right) \kappa_{o}^{-1}\bar{T}\kappa_{e}=1-u~\bar{u}   \label{ggbar}
\end{equation}
where we have defined $u$ and used its properties as follows
\begin{equation}
u\equiv\kappa_{e}T\kappa_{o}^{-1}v=\bar{\Gamma}w,\;\bar{u}u=\bar{v}v=\frac{%
\bar{w}w}{1+\bar{w}w},\;\Gamma u=w\left( 1+\bar{w}w\right) ^{-1}   \label{u}
\end{equation}
We see that $\bar{\Gamma}u$ approaches 0 as $N\rightarrow\infty$, so $u$ is
a vector that tends to become the zero mode of $\Gamma$ in the large $N$
limit.

A numerical estimate of the eigenvalues of $\Gamma$ can be obtained by using
the numerical matrix $Z$ given in Eq.(\ref{Znum}) and then using naive level
truncation (this is less accurate than using the exact cutoff version of $T$%
). Numerical computations show that almost all of the eigenvalues of $\Gamma$
are $1$ except for a very small number of them which deviate from 1. For
example for $N=50$, within 1\% error, forty six eigenvalues are $1.00,$ and
the last four eigenvalues are 0.99272, 0.95752, 0.79796, 0.35755. Thus, the
fifty eigenvalues of $\Gamma$ are approximately
\begin{equation}
eigen\left( \Gamma\right) \approx\left\{ \left( 1.00\right) ,\ldots,\left(
1.00\right) ,\left( 0.99\right) ,\left( 0.96\right) ,\left( 0.80\right)
,\left( 0.36\right) \right\} .   \label{eigen}
\end{equation}
The approach to a zero eigenvalue as $\bar{w}w\rightarrow2N\rightarrow\infty$
is expected since a zero mode was already identified in the large $N$ limit,
but it is interesting that this zero eigenvalue seems to be almost isolated
in the numerical estimates. To see this analytically we can bring $\bar{%
\Gamma}$ to block diagonal form by orthogonal transformations that map the
vectors $w,u$ to point in a single direction $\bar{w}=\left(
0,\cdots,0,w\right) \bar{S}_{e}$ and $\bar{u}=\left( 0,\cdots,0,w\left(
1+w^{2}\right) ^{-1/2}\right) \bar{S}_{e}^{\prime}.$ Then consistently with
Eqs.(\ref{gbarg},\ref{ggbar}) we derive%
\begin{equation}
\bar{\Gamma}=S_{e}^{\prime}\left(
\begin{array}{cc}
\gamma & 0 \\
0 & \frac{1}{\sqrt{1+w^{2}}}%
\end{array}
\right) \bar{S}_{e},\;\;\gamma\bar{\gamma}=\bar{\gamma}\gamma=1,   \label{g}
\end{equation}
where the $\left( N-1\right) \times\left( N-1\right) $ block $\gamma$ is
orthogonal. Given the numerical estimates in Eq.(\ref{eigen}) we see that
the eigenvalue that tends to zero at large $N$ is indeed isolated, and that
we may take $\gamma=1$ since it can be absorbed into a redefinition of $S_{e}
$ or $S_{e}^{\prime}.$

We also compare this result to the basis that diagonalizes $%
\kappa_{e}^{1/2}T\kappa_{o}^{-1/2}$ as defined in Eq.(\ref{ktk}). We see
that $\bar{\Gamma}=$ $\kappa_{e}T\kappa_{o}^{-1}\bar{T}=$ $%
\kappa_{e}^{1/2}\left( \kappa_{e}^{1/2}T\kappa_{o}^{-1/2}\right) \left(
\kappa_{o}^{-1/2}\bar
{T}\kappa_{e}^{1/2}\right) \kappa_{e}^{-1/2}$ takes
the form
\begin{equation}
\bar{\Gamma}=\left( \kappa_{e}\right) ^{1/2}V^{e}\left( \tau\right) ^{2}\bar{%
V}^{e}\left( \kappa_{e}\right) ^{-1/2}=S_{e}^{\prime}\left(
\begin{array}{cc}
\gamma & 0 \\
0 & \frac{1}{\sqrt{1+w^{2}}}%
\end{array}
\right) \bar{S}_{e}   \label{gt}
\end{equation}
This provides the relation between the eigenvalues $\tau_{k}$ and the
eigenvalues of $\bar{\Gamma}$. In particular, we see that $\left( \det
\tau\right) ^{2}=\left( 1+w^{2}\right) ^{-1/2}.$ Also, for an infinite
number of modes, writing $\tau\left( k\right) =\tanh\left( \pi k/4\right) ,$
we see its compatibility with the numerical computation in Eq.(\ref{eigen}).

\subsection{Oscillators as Differential operators in Moyal Space}

By taking the Fourier transform of the oscillators expressions given in Eq.(%
\ref{aodd},\ref{azero}) we construct the oscillators as differential
operators $\bar{\beta}_{0},\beta_{e}^{x},\beta_{o}^{p}$ (or $\bar{\beta}%
_{e}^{x},\bar{\beta}_{e}^{p}$) acting on any field $A\left( \bar
{x}%
,x_{\left\vert e\right\vert },p_{\left\vert e\right\vert }\right) $ in Moyal
space. The notation we are using in this section is as follows: $e$
indicates positive or negative even numbers excluding zero, \thinspace$o$
indicates positive or negative odd numbers. The result of the Fourier
transform is also obtained directly by using the properties of the
oscillator state $\langle\bar{x},x_{e},p_{e}|$
\begin{align}
& \langle\bar{x},x_{e},p_{e}\left\vert \alpha_{0}\right\vert \psi
\rangle\equiv\beta_{0}A=-il_{s}\frac{\partial A}{\partial\bar{x}}
\label{paralles} \\
& \langle\bar{x},x_{e},p_{e}\left\vert \alpha_{e}\right\vert \psi
\rangle\equiv\beta_{e}^{x}A=\left( \bar{\beta}_{e}^{x}-w_{e}^{\prime}%
\beta_{0}\right) A  \label{paralles1} \\
& \langle\bar{x},x_{e},p_{e}\left\vert \alpha_{o}\right\vert \psi
\rangle\equiv\beta_{o}^{p}A=\sum_{e\neq0}\left( \bar{\beta}_{e}^{p}A\right)
U_{-e,o}   \label{paralles2}
\end{align}
Note that we have distinguished in our notation between $\beta_{e}^{x}$
versus $\bar{\beta}_{e}^{x},$ and $\beta_{o}^{p}$ versus $\bar{\beta}%
_{e}^{p},$ where%
\begin{equation}
\bar{\beta}_{e}^{x}=-\frac{i}{\sqrt{2}}\left( \epsilon(e) \frac{\kappa
_{\left\vert e\right\vert }}{l_{s}}x_{\left\vert e\right\vert }+l_{s}\,
\frac{\partial}{\partial x_{\left\vert e\right\vert }}\,\right) ,\;\;\bar{%
\beta}_{e}^{p}=\sqrt{\frac{1}{2}}\left( \epsilon(e)\frac {%
\theta\kappa_{\left\vert e\right\vert }}{2l_{s}}\frac{\partial}{\partial
p_{\left\vert e\right\vert }}\,+\frac{2l_{s}}{\theta}\,p_{\left\vert
e\right\vert }^{\mu}\right) .   \label{betaxp}
\end{equation}
The $\bar{\beta}_{e}^{x},\bar{\beta}_{e}^{p}$ commute with each other and
satisfy oscillator commutation rules among themselves
\begin{equation}
\left[ \bar{\beta}_{e}^{x},\bar{\beta}_{e^{\prime}}^{x}\right]
=\varepsilon\left( e\right) \kappa_{\left\vert e\right\vert }\delta
_{e+e^{\prime}},\;\;\left[ \bar{\beta}_{e}^{p},\bar{\beta}_{e^{\prime}}^{p}%
\right] =\varepsilon\left( e\right) \kappa_{\left\vert e\right\vert
}\delta_{e+e^{\prime}},\;\;\left[ \bar{\beta}_{e}^{x},\bar{\beta}_{e^{\prime
}}^{p}\right] =0.
\end{equation}
The extra shift\footnote{%
We have taken into account that $\partial/\partial x_{\left\vert
e\right\vert }$, as part of $\alpha_{e}$, acts on $\psi\left(
x_{0},x_{\left\vert e\right\vert },x_{\left\vert o\right\vert }\right) $ at
\textit{fixed} $x_{0}$ whereas $\partial/\partial x_{\left\vert e\right\vert
}$, as part of $\bar{\beta}_{e}^{x}$ acts on $A\left( \bar{x},x_{\left\vert
e\right\vert },p_{\left\vert e\right\vert }\right) $ at \textit{fixed} $\bar{%
x}.$ The latter is required for compatibility with the definition of $%
\partial/\partial x_{\left\vert e\right\vert }$ that appears in the Moyal
star product in Eq.(\ref{moyalstar}). This difference in the definition of $%
\partial/\partial x_{\left\vert e\right\vert }$ is taken into account by
replacing $x_{0}=\bar{x}+\sum_{\left\vert e\right\vert >0}w_{\left\vert
e\right\vert }x_{\left\vert e\right\vert }$ in $\psi$ and also replacing $%
\partial\psi/\partial x_{\left\vert e\right\vert }\rightarrow\partial
\psi/\partial x_{\left\vert e\right\vert }-\left( \partial x_{0}/\partial
x_{\left\vert e\right\vert }\right) \left( \partial\psi/\partial\bar
{x}%
\right) $ before taking the Fourier transform. This is the reason for the
shift $\frac{w_{\left\vert e\right\vert }}{\sqrt{2}}\beta_{0}$ that appears
as the difference between $\beta_{e}^{x}$ and $\bar{\beta}_{e}^{x}.$ It
should be noted that $\beta_{e}^{x}$ commutes with any function of $x_{0}=%
\bar{x}+\sum w_{\left\vert e\right\vert }x_{\left\vert e\right\vert }$ as a
consequence of this structure $\beta_{e}^{x}f\left( \bar{x}+\sum
w_{\left\vert e\right\vert }x_{\left\vert e\right\vert }\right) =f\left(
\bar{x}+\sum w_{\left\vert e\right\vert }x_{\left\vert e\right\vert }\right)
\beta_{e}^{x};$ this is analogous to $\alpha_{e}$ commuting with any
function of $x_{0}.$} with $w_{e}^{\prime}= w_{\left\vert e\right\vert }/%
\sqrt{2}$ in Eq.(\ref{paralles1}) does not change the commutation relations
since $\beta_{0}$ commutes with $\bar{\beta}_{e}^{x}$.

The matrix $U$ and its inverse $U^{-1},$ with matrix elements $%
U_{-e,o},\left( U^{-1}\right) _{-o,e},$ are given by\footnote{%
An intermediate step in deriving Eq.(\ref{paralles2}) from Fourier
transforms is the form $U_{-e,o}=\left( \frac{1}{2}\epsilon\left( e\right)
\epsilon\left( o\right) \frac{\kappa_{\left\vert e\right\vert }}{%
\kappa_{\left\vert o\right\vert }}+\frac{1}{2} \right) T_{\left\vert
e\right\vert ,\left\vert o\right\vert }.$ After inserting the expression for
$T_{\left\vert e\right\vert ,\left\vert o\right\vert },R_{\left\vert
o\right\vert ,\left\vert e\right\vert }$ in Eq.(\ref{TRcut}), the simpler
form of $U$ and $U^{-1}$ follow. \label{UTf}}%
\begin{equation}
U_{-e,o}=\frac{w_{e}^{\prime}v_{o}^{\prime}\kappa_{o}^{\prime}}{\kappa
_{e}^{\prime}-\kappa_{o}^{\prime}},\;\;\left( U^{-1}\right) _{-o,e}=\frac{%
w_{e}^{\prime}v_{o}^{\prime}\kappa_{e}^{\prime}}{\kappa_{e}^{\prime
}-\kappa_{o}^{\prime}},   \label{UUinv}
\end{equation}
where we have extended the definition of the quantities $w_{e},v_{o},%
\kappa_{e},\kappa_{o}$ to both positive and negative values of $e,o$ as
follows%
\begin{equation}
v_{o}^{\prime}=\frac{ v_{\left\vert o\right\vert }}{\sqrt{2}}%
,\;\;w_{e}^{\prime}=\frac{w_{\left\vert e\right\vert }}{\sqrt{2}}%
,\;\kappa_{e}^{\prime }=\varepsilon\left( e\right) \kappa_{\left\vert
e\right\vert },\;\kappa _{o}^{\prime}=\varepsilon\left( o\right)
\kappa_{\left\vert o\right\vert }.   \label{extend}
\end{equation}
such that (after the sum over both positive and negative integers) $\bar
{v}%
^{\prime}v^{\prime}=\bar{v}v$ and $\bar{w}^{\prime}w^{\prime}=\bar{w}w.$
After the transformation with the matrix $U_{-e,o}$ one can verify easily
that $\left( \beta_{0},\beta_{e}^{x},\beta_{o}^{p}\right) $ are differential
operators that satisfy the oscillator commutation rules that are in one to
one correspondence with those of $\alpha_{n}$ given in Eq.(\ref{alphan})
\begin{equation}
\left[ \beta_{e}^{x},\beta_{e^{\prime}}^{x}\right] =\varepsilon\left(
e\right) \kappa_{\left\vert e\right\vert }\delta_{e+e^{\prime}},\;\;\left[
\beta_{o}^{p},\beta_{o^{\prime}}^{p}\right] =\varepsilon\left( o\right)
\kappa_{\left\vert o\right\vert }\delta_{o+o^{\prime}},\;\;\left[ \beta
_{e}^{x},\beta_{o}^{p}\right] =0,   \label{betadiffs}
\end{equation}
while $\beta_{0}$ commutes with all $\beta_{e}^{x},\beta_{o}^{p}.$

The $2N\times2N$ matrix $U$ plays the role of a Bogoliubov transformation (a
linear combination of positive and negative frequency oscillators) and
produces odd oscillators from the even ones, and vice versa. It satisfies
the following relations which follow from those of the $N\times N$ matrices $%
T,R,v,w$ in Eqs.(\ref{tr}-\ref{ktk})\footnote{%
Instead of deriving these relations from those of $T,R,v,w,$ it is also
possible to consider Eqs.(\ref{U1}) as the primary defining relations for $%
U,U^{-1},v_{o}^{\prime },w_{e}^{\prime}$ which determine $U,U^{-1}$ as in
Eqs.(\ref{UUinv}) and give $w^{\prime},v^{\prime}$ as the solutions of ${%
\kappa_{o}^{\prime}}^{-1}=\sum_{e\neq0}\left( w_{e}^{\prime}\right)
^{2}\left( \kappa_{e}^{\prime }-\kappa_{o}^{\prime}\right) ^{-1} ,$ ${%
\kappa_{e}^{\prime}}^{-1}=\sum _{o}\left( v_{o}^{\prime}\right) ^{2}\left(
\kappa_{e}^{\prime}-\kappa _{o}^{\prime}\right) ^{-1} $ for any $%
\kappa_{e}^{\prime},\kappa_{o}^{\prime}$. The rest of the relations in Eqs.(%
\ref{U2}-\ref{U4}) and those of $T,R,v,w$ follow from them.}%
\begin{align}
U^{-1} & =\kappa_{o}^{\prime-1}\bar{U}\kappa_{e}^{\prime},\quad U^{-1}=\bar{U%
}+v^{\prime}\bar{w}^{\prime},\quad v^{\prime}=\bar{U}w^{\prime},\quad
w^{\prime}=\bar{U}^{-1}v^{\prime},  \label{U1} \\
UU^{-1} & =1,\quad U^{-1}U=1,\;\;\bar{U}^{-1}U^{-1}=1+w^{\prime}\bar {w}%
^{\prime},\;\;\bar{U}U=1-v^{\prime}\bar{v}^{\prime},\;  \label{U2} \\
U\bar{U} & =1-\frac{w^{\prime}\bar{w}^{\prime}}{1+\bar{w}^{\prime}w^{\prime }%
},\quad Uv^{\prime}=\frac{w^{\prime}}{1+\bar{w}^{\prime}w^{\prime}},\quad
\bar{v}^{\prime}v^{\prime}=\frac{\bar{w}^{\prime}w^{\prime}}{1+\bar{w}%
^{\prime}w^{\prime}},  \label{U3} \\
U^{-1}w^{\prime} & =v^{\prime}(1+\bar{w}^{\prime}w^{\prime}),\quad U^{-1}%
\bar{U}^{-1}=1+v^{\prime}\bar{v}^{\prime}\left( 1+\bar{w}^{\prime
}w^{\prime}\right) .   \label{U4}
\end{align}
where $\bar{U}$ is the transpose of $U.$

In the large $N$ limit, after using Eqs.(\ref{Tinf}-\ref{kinf}), the
infinite matrix elements $U_{-e,o},$ $U_{-o,e}^{-1},\kappa_{e}^{\prime},%
\kappa _{o}^{\prime},v_{o}^{\prime},w_{o}^{\prime}$ get simplified to the
form
\begin{equation}
\,U_{-e,o}=\frac{2}{\pi}\frac{i^{o-e-1}}{o-e},\;\;U_{-o,e}^{-1}=\frac{2}{\pi
}\frac{e}{o}\frac{i^{o-e-1}}{o-e},\;w_{e}^{\prime}=i^{e+2},\;\;v_{o}^{^{%
\prime}}=\frac{2}{\pi}\frac{i^{o-1}}{o},\;\kappa_{e}^{\prime}=e,\;%
\kappa_{o}^{\prime}=o.
\end{equation}
where $i=\sqrt{-1}.$ The relations (\ref{U1}-\ref{U4}) can be verified
explicitly in the infinite cutoff limit by using the following identities
that are valid only for integers (recall that $U_{-e,o}$ or $U_{-o,e}^{-1}$
do not include $e=0$)%
\begin{gather}
\sum_{k=-\infty}^{\infty}\frac{1}{2m+2k-1}\frac{1}{2k-1}\frac{1}{2n-2k+1}=%
\frac{\pi^{2}}{8m}\delta_{2m+2n} \\
\sum_{k=-\infty}^{\infty}\frac{1}{2m+2k-1}\left( 2k\right) \frac{1}{2n-2k+1}%
=-\frac{\pi^{2}}{4}\left( 2n+1\right) \delta_{2m+2n} \\
\sum_{k=-\infty}^{\infty}\frac{1}{2m+2k-1}\frac{1}{2n-2k+1}=-\frac{\pi^{2}}{4%
}\delta_{2m+2n} \\
\sum_{k=-\infty}^{\infty}\frac{1}{2m+2k-1}=0,\;\;\sum_{k=-\infty}^{\infty }%
\frac{\left( -1\right) ^{k}}{2m+2k-1}=-\frac{\pi}{2}\left( -1\right) ^{m}.
\end{gather}
The conditional convergence of these sums cause anomalies in multiple sums
if a cutoff is not present. For example if $\bar{w}^{\prime}w^{\prime}=\infty
$ is first set in relations (\ref{U1}-\ref{U4}), and then one computes $%
U^{-1}Uv^{\prime}$ one finds two different answers: $\left( U^{-1}U\right)
v^{\prime}=v^{\prime}$ versus $U^{-1}\left( Uv^{\prime}\right) =0.$ This is
the source of many associativity anomalies in string field theory as
discussed in \cite{r-BarsMatsuo}. For this reason one must proceed carefully
with a cutoff $N$, and take the large $N$ limit only at the very end of a
calculation to obtain unique and unambiguous answers.

Note that in the infinite cutoff limit $U_{-e,o}$ is a function of only the
sum of its arguments. This is a useful property that is not necessarily
shared by the cutoff version of $U,$ and it leads to the derivation of
additional properties for the infinite $U,$ such as $U_{-o+e^{\prime%
\prime},e^{\prime}}=U_{-o,e^{\prime\prime}+e^{\prime}}$ and
\begin{equation}
\sum_{o}U_{e,o}^{-1}U_{-o+e^{\prime\prime},e^{\prime}}=\delta_{e+e^{\prime
}+e^{\prime\prime}},   \label{Ushifted}
\end{equation}
which follows from $U^{-1}U=1.$ This relation holds for any shift $%
e^{\prime\prime}$ in the infinite theory, but is only true for $e^{\prime
\prime}=0$ in the cutoff theory.

We now analyze the action of the oscillators on the vacuum state. By
construction, the vacuum field $A_{0}$ defined in the previous section
should be annihilated by the positive frequency oscillators $%
\beta_{\left\vert e\right\vert }^{x},\beta_{\left\vert o\right\vert }^{p}.$
It is instructive to verify this property directly by using the explicit $%
A_{0}$ given in Eq.(\ref{ground})
\begin{align}
& \beta_{e}^{x}A_{0}\left( x_{\left\vert e\right\vert },p_{\left\vert
e\right\vert }\right) =-\frac{i}{\sqrt{2}}\left( \epsilon(e)\frac {%
\kappa_{\left\vert e\right\vert }}{l_{s}}x_{\left\vert e\right\vert }+l_{s}%
\frac{\partial}{\partial x_{\left\vert e\right\vert }}\right) A_{0}+0
\label{bxA0} \\
& =\frac{i}{\sqrt{2}}\frac{\kappa_{\left\vert e\right\vert }}{l_{s}}%
x_{\left\vert e\right\vert }\left( 1-\epsilon\left( e\right) \right)
A_{0}=0\quad iff\quad e>0.
\end{align}
We see that the shift in Eq.(\ref{paralles1}) plays no role here because $%
A_{0}$ is independent of $x_{0}$ or $\bar{x}.$ Similarly we have%
\begin{align}
& \beta_{o}^{p}A_{0}\left( x_{\left\vert e\right\vert },p_{\left\vert
e\right\vert }\right) =\sum_{e\neq0}\left( \beta_{e}^{p}A_{0}\right) U_{-e,o}
\label{b-1} \\
& =\sum_{e\neq0}\sqrt{\frac{1}{2}}\left( \epsilon(e)\frac{\theta
\kappa_{\left\vert e\right\vert }}{2l_{s}}\frac{\partial}{\partial
p_{\left\vert e\right\vert }}\,+\frac{2l_{s}}{\theta} \,p_{\left\vert
e\right\vert }\right) A_{0}U_{-e,o} \\
& =\sum_{e\neq0}\sqrt{\frac{1}{2}}\left( -\epsilon(e)\frac{\theta
\kappa_{\left\vert e\right\vert }}{2l_{s}}\frac{4l_{s}^{2}}{\theta^{2}}%
\left( pZ\right) _{\left\vert e\right\vert }\,+\frac{2l_{s}}{\theta}%
\,p_{\left\vert e\right\vert }\right) A_{0}U_{-e,o} \\
& =\sqrt{\frac{1}{2}}\frac{2l_{s}}{\theta}A_{0}\sum_{e\neq0}\left(
-\epsilon(e)\left( pZ\right) _{\left\vert e\right\vert }\,\kappa_{\left\vert
e\right\vert }+p_{\left\vert e\right\vert }\right) U_{-e,o}, \\
& =\sqrt{\frac{1}{2}}\frac{2l_{s}}{\theta}\left(
\sum_{e>0}\,p_{e}T_{e,\left\vert o\right\vert }\right) \left(
1-\,\epsilon\left( o\right) \right) A_{0}=0\quad iff\quad o>0   \label{b-2}
\end{align}
where Eq.(\ref{b-2}) was computed by rewriting the sum only over positive
integers, and using the relations$^{\ref{UTf}}$ between $U$ and $T$.
\begin{align}
& \sum_{e>0}\,p_{e}^{\mu}\left(
\begin{array}{c}
-\sum_{e^{\prime}>0}Z_{e,e^{\prime}}\,\kappa_{e^{\prime}}\left(
U_{-e^{\prime},o}-U_{e^{\prime},o}\right) \\
+ \left( U_{-e,o}+U_{e,o}\right)%
\end{array}
\right)  \label{verify0} \\
& =\sum_{e>0}\,p_{e}^{\mu}\left(
\sum_{e^{\prime}>0}-Z_{e,e^{\prime}}R_{\left\vert o\right\vert
,e^{\prime}}\kappa_{o}\,\epsilon (o)+T_{e,\left\vert o\right\vert } \right)
\\
& =\left( 1-\,\epsilon\left( o\right) \right) \sum_{e>0}\,p_{e}^{\mu
}T_{e,\left\vert o\right\vert }
\end{align}
To prove the last line we have used the definition of $Z_{2k,2n}$ given in (%
\ref{Z}) and applied $RT=1.$ These properties of the vacuum field $A_{0}$
hold both in the cutoff and the infinite theory.

The perturbative string states can now be constructed by applying the
oscillators $\beta_{-\left\vert e\right\vert }^{x},$ $\beta_{-\left\vert
o\right\vert }^{p}$ on the perturbative vacuum $A_{0}$ of Eq.(\ref{AN0}).

\subsection{Oscillators as Fields in Moyal Space}

Instead of differential operators $\bar{\beta}_{e}^{x},\bar{\beta}_{e}^{p}$
or $\beta_{o}^{p}$, it is possible to construct the oscillators in terms of
star products among fields in Moyal space. To this end we define $\beta_{e}$
for $e\neq0$ as a function of $x_{\left\vert e\right\vert },p_{\left\vert
e\right\vert }$ in Moyal space
\begin{equation}
\beta_{e}=\sqrt{\frac{1}{2}}\left( -i\,\epsilon(e)\,\frac{\kappa_{\left\vert
e\right\vert }}{2l_{s}}x_{\left\vert e\right\vert }+\frac{l_{s}}{\theta }%
\,p_{\left\vert e\right\vert }\right) ,   \label{betae}
\end{equation}
These are functions in phase space, not differential operators. For $\beta
_{0}$ we have a differential operator $\beta_{0}=-il_{s}\partial_{\bar{x}}$
as before. Under the Moyal star product obeyed by fields they satisfy%
\begin{equation}
\left[ \beta_{e}^{\mu},\beta_{e^{\prime}}^{\nu}\right] _{\ast}=\frac{1}{2}%
\eta^{\mu\nu}\varepsilon\left( e\right) \kappa_{\left\vert e\right\vert
}\delta_{e+e^{\prime}}.   \label{betaev}
\end{equation}
while $\beta_{0}$ commutes with all others. Note the factor of $\frac{1}{2}$
compared to Eq.(\ref{betadiffs}). These can be star multiplied on either the
left or right side of string fields. When the star products are evaluated by
using explicitly the Moyal product we find the following two combinations
that act like differential operator representations of oscillators acting
only on $x_{\left\vert e\right\vert }$ or only on $p_{\left\vert
e\right\vert }$ dependence
\begin{align}
\beta_{e}\ast A-A\ast\beta_{-e} & =\bar{\beta}_{e}^{x}A, \\
\beta_{e}\ast A+A\ast\beta_{-e} & =\bar{\beta}_{e}^{p}A.
\end{align}
where the differential operators $\bar{\beta}_{e}^{x},\bar{\beta}_{e}^{p}$
are given in Eq.(\ref{betaxp}). This implies that we can construct the
effect of all oscillators $\alpha_{n}$ for $n\neq0$ in terms of only star
products by substituting the differential operators $\bar{\beta}_{e}^{x}A$, $%
\bar{\beta }_{e}^{p}A$ everywhere by the star product expressions given
above. In terms of them we can write%
\begin{align}
& \langle\bar{x},x_{e},p_{e}\left\vert \alpha_{0}\right\vert \psi
\rangle=\beta_{0}A\equiv-il_{s}\partial_{\bar{x}}A \\
& \langle\bar{x},x_{e},p_{e}\left\vert \alpha_{e}\right\vert \psi
\rangle=\left( \beta_{e}\ast A-A\ast\beta_{-e}\right)
-w_{e}^{\prime}\beta_{0}A  \label{osc-rules} \\
& \langle\bar{x},x_{e},p_{e}\left\vert \alpha_{o}\right\vert \psi
\rangle=\beta_{o}\ast A+A\ast\beta_{-o}
\end{align}
where $\beta_{o}$ is the Bogoliubov transform of $\beta_{e}$
\begin{equation}
\beta_{o}=\sum_{e\neq0}\beta_{e}U_{-e,o}.   \label{bodd}
\end{equation}
$\beta_{o}$ does not star commute with $\beta_{e},$
\begin{equation}
\left[ \beta_{-e},\beta_{o}\right] _{\ast}=\frac{1}{2}\eta^{\mu\nu
}\varepsilon\left( e\right) \kappa_{\left\vert e\right\vert }U_{-e,o}
\end{equation}
but $\beta_{o}$ satisfies the star commutator of oscillators with the odd
frequencies%
\begin{equation}
\left[ \beta_{o}^{\mu},\beta_{o^{\prime}}^{\nu}\right] _{\ast}=\frac{1}{2}%
\eta^{\mu\nu}\varepsilon\left( o\right) \kappa_{o}\delta_{o+o^{\prime}}.
\end{equation}
This is verified by using Eq.(\ref{betaev}) and the properties of $U$ given
in Eq.(\ref{U1}) below. Note again the factor of $\frac{1}{2}$ compared to
Eq.(\ref{betadiffs}). It is also possible to display the properties of $%
\beta_{o}$ by performing the transformation to the odd phase space$^{\ref%
{oddspace}}$, which we give here without proof.
\begin{align}
\beta_{o} & =\sum_{e\neq0}\sqrt{\frac{1}{2}}\left( -i\,\epsilon(e)\, \frac{%
\kappa_{\left\vert e\right\vert }}{2l_{s}}x_{\left\vert e\right\vert }+\frac{%
l_{s}}{\theta}p_{\left\vert e\right\vert }\right) U_{-e,o} \\
& =\sqrt{\frac{1}{2}}\left( -i\,\epsilon(o)\,\frac{\kappa_{o}}{2l_{s}}%
x_{\left\vert o\right\vert }+\frac{il_{s}}{\theta}p_{\left\vert o\right\vert
}\right)   \label{bodd2}
\end{align}

We have seen that the only fundamental oscillators in MSFT are the
oscillators in Moyal space, either the $\beta_{e}^{\mu}$ of Eq.(\ref{betae})
or the $\beta_{o}$ of Eqs.(\ref{bodd},\ref{bodd2}). Either set may be
considered as a special set of string fields.

\subsection{Differential Virasoro Operators in Moyal space}

We can now construct the Virasoro operators by using the correspondence we
have established between the $\alpha$ and $\beta$ oscillators. We find%
\begin{equation}
\langle\bar{x},x_{e},p_{e}|L_{e}^{\alpha}|\psi\rangle=L_{e}^{\beta
}A,\;\;\langle\bar{x},x_{e},p_{e}|L_{o}^{\alpha}|\psi\rangle=L_{o}^{\beta}A
\end{equation}
where we have the differential operators (sum over $e^{\prime}$ includes $%
e^{\prime}=0$)
\begin{align}
L_{e}^{\beta} & =\frac{1}{2}\sum_{e^{\prime}}:\beta_{-e^{\prime}}^{x}\cdot%
\beta_{e+e^{\prime}}^{x}:+\frac{1}{2}\sum_{o^{\prime}}:\beta_{-o^{\prime
}}^{p}\cdot\beta_{e+o^{\prime}}^{p}:  \label{Lev} \\
L_{o}^{\beta} & =\sum_{e^{\prime}}:\beta_{o-e^{\prime}}^{p}\cdot
\beta_{e^{\prime}}^{x}:   \label{Lod}
\end{align}
The normal ordering of the differential operator is the same as the $\alpha$
oscillators. Then it is evident that the positive ones annihilate the vacuum
field $A_{0}$ that satisfies Eqs.(\ref{bxA0},\ref{b-1}). These Virasoro
operators obviously satisfy the Virasoro algebra since the $%
\beta_{e}^{x},\beta_{o}^{p}$ have identical commutation relations to the $%
\alpha _{e},\alpha_{o}$ respectively. Closure of the Virasoro algebra is
possible only when the cutoff $N$ is sent to infinity\footnote{%
In mathematically rigorous sense, the closure is subtle even in the large $N$
limit. A sample computation of $[L_{n},L_{-n}]-2nL_{0}$ reveals that there
is roughly speaking $n$ terms located near $\beta_{\pm N}$ whose
coefficients diverge as $N\rightarrow\infty$. In this sense, we may at most
have the convergence of the operator algebra only in the weak topology.}.

We now introduce the cutoff to study certain issues related to anomalies. In
particular the cutoff version of the Virasoro operator $L_{0}$ which
determines the spectrum of the cutoff theory is%
\begin{equation}
L_{0}^{\beta }\left( N\right) =\frac{1}{2}\beta _{0}^{2}+\sum_{e\geq
2}^{2N}\beta _{-e}^{x}\cdot \beta _{e}^{x}+\sum_{o\geq 1}^{2N-1}\beta
_{-o}^{p}\cdot \beta _{o}^{p}.  \label{L0o}
\end{equation}%
The spectrum of this operator is the same as the one described in the
previous section, which is obtained by applying even and odd creation
operators on the vacuum state $A_{0}$ in the cutoff theory. The oscillator
frequencies $\kappa _{\left\vert e\right\vert },\kappa _{\left\vert
o\right\vert }$ in Eq.(\ref{betadiffs}) determine the spectrum, as usual.

Next, by using the 2$N\times2N$ Bogoliubov transformation $U$ we can express
the odd $\beta_{o}^{p}$ in terms of the even $\bar{\beta}_{e}^{p}.$ After
using Eq.(\ref{U3}) the Virasoro operator $L_{0}$ takes the form%
\begin{equation}
L_{0}^{\beta}\left( N\right) =\frac{1}{2}\beta_{0}^{2}+\sum_{e\geq2}^{2N}%
\left( \beta_{-e}^{x}\cdot\beta_{e}^{x}+\bar{\beta}_{-e}^{p}\cdot \bar{\beta}%
_{e}^{p}\right) -\frac{1}{4\left( 1+\bar{w}w\right) }\left(
\sum_{e>0}^{2N}w_{e}\left( \bar{\beta}_{e}^{p}+\bar{\beta}_{-e}^{p}\right)
\right) ^{2}.   \label{spectrum}
\end{equation}
where only even oscillators with only even frequencies $\kappa_{\left\vert
e\right\vert }$ appear. Note that there is no zero mode $\bar{\beta}_{0}^{p}$
since it does not exist in the formalism and therefore it is taken as zero.
Formally the second term vanishes when the cutoff is removed since $\bar
{w}%
w\rightarrow\infty~$in the infinite mode limit. It would appear then that
the spectrum is different than the original theory since now only $%
\kappa_{\left\vert e\right\vert }$ appears in the spectrum without any
information of the $\kappa_{o}.$ This is an anomaly related to the other
anomalous cases that we discussed before. To understand the problem let us
focus on the second term which plays a subtle role. First, in its absence, a
false perturbative vacuum state would be given by a function proportional to
$\exp\left( -\frac{1}{2l_{s}^{2}}\kappa_{\left\vert e\right\vert
}x_{\left\vert e\right\vert }^{2}-\frac{2l_{s}^{2}}{\theta^{2}\kappa
_{\left\vert e\right\vert }}p_{\left\vert e\right\vert }^{2}\right) $ as
assumed in \cite{DLMZ}. This cannot be the correct perturbative vacuum since
we have already determined that it is given by the field $A_{0}$ of Eq.(\ref%
{ground}) which is different. Indeed, due to the presence of the second term
in Eq.(\ref{spectrum}) the false vacuum is not an eigenfunction of the $%
L_{0}^{\beta}\left( N\right) $ of Eq.(\ref{spectrum}). The correct vacuum
state is the $A_{0}$ at any value of the cutoff, as is evident from the
Bogoliubov transformed $L_{0}^{\beta}\left( N\right) $ in Eq.(\ref{L0o}).

A related anomaly occurs in the commutation rules if the $L_{n}$ are
expressed in terms of only the even $\bar{\beta}_{e}^{p}$ oscillators. To
see this, we apply the Bogoliubov transformation $U$ to Eqs.(\ref{Lev},\ref%
{Lod}) to obtain\footnote{%
Strictly speaking these expressions are valid at $N=\infty$ because shifted
formulas, such as Eq.(\ref{Ushifted}) which are needed for this derivation,
are valid only at $N=\infty.$ So, we will be a bit sloppy in the following
argument, because we have introduced the cutoff $N$ in Eqs.(\ref{signs})
after the Bogoliubov transformation (by contrast Eq.(\ref{spectrum}) is
rigorous because the unshifted formulas are valid at any $N$).}
\begin{align}
L_{e}^{\beta}\left( N\right) & =\frac{1}{2}\sum_{e^{\prime}=-2N}^{2N}:\left(
\beta_{-e^{\prime}}^{x}\cdot\beta_{e+e^{\prime}}^{x}+\bar{\beta }%
_{-e^{\prime}}^{p}\cdot\bar{\beta}_{e+e^{\prime}}^{p}\right) :  \label{signs}
\\
& +\frac{w_{\vert e\vert}}{4\sqrt{2}\left( 1+\bar{w}w\right) }\left(
\sum_{e^{\prime}>0}^{2N}w_{\left\vert e^{\prime}\right\vert }\left( \bar{%
\beta}_{e^{\prime}}^{p}+\bar{\beta}_{-e^{\prime}}^{p}\right) \right) ^{2},
\notag \\
L_{o}^{\beta}\left( N\right) &
=\sum_{e^{\prime},e^{\prime\prime}=-2N}^{2N}U_{-e^{\prime\prime},o-e^{%
\prime}}\bar{\beta}_{e^{\prime\prime}}^{p}\cdot\beta_{e^{\prime}}^{x}.
\end{align}
As usual, closure of the Virasoro algebra is not possible unless $%
N\rightarrow\infty$. At infinite $N$ the second term in $L_{e}^{\beta}\left(
N\right) $ is formally zero. Using only the first term in $L_{e}^{\beta
}\left( N\right) $ at infinite $N$, we find that closure works in the
commutators $\left[ L_{e},L_{e}\right] $ and $\left[ L_{o},L_{o}\right] $
but it does not work in the commutator $\left[ L_{e},L_{o}\right] .$ Again
this is an anomaly because closure was guaranteed before the Bogoliubov
transformation was applied. The subtle point involves the second term in Eq.(%
\ref{signs}) which is formally zero. In fact, some of its commutators yield
finite results if first the commutator is evaluated and then the large $N$
limit is applied. Indeed, if the large $N$ limit is taken after all
commutators are performed, then all commutators of the Virasoro algebra
close correctly. We emphasize again that the closure of the algebra was
evident from the beginning by using the version with $\beta_{o}^{p}.$ The
lesson learned is that it is important to use the cutoff theory.

\subsection{Virasoro Fields in Moyal Space}

Next we would like to point out a fundamental structure for the Virasoro
fields in Moyal space, and in this process build a new representation of the
Virasoro algebra. We will see that the Virasoro fields can be star
multiplied either on the left side or the right side of an arbitrary field $A
$ in Moyal space. These generate independent left/right Virasoro
transformations. A special combination of the left and right star products
generate the usual differential form of the Virasoro operators that we
discussed in the previous section.

First we define the following Virasoro fields $\mathcal{L}_{e}$ and $%
\mathcal{L}_{o}$ that are functions of $x_{\left\vert e\right\vert
},p_{\left\vert e\right\vert }$
\begin{align}
\mathcal{L}_{e} & =\sum_{o^{\prime}}\beta_{-o^{\prime}}\cdot\beta
_{e+o^{\prime}}  \label{Leven} \\
\mathcal{L}_{o} & =\sum_{o^{\prime},o^{\prime\prime}}U_{o+o^{\prime\prime
},o^{\prime}}~\beta_{-o^{\prime\prime}}\cdot\beta_{-o^{\prime}}\;
\end{align}
Note the factor of 1/2 is absent compared to Eq.(\ref{Lev}). We chose to use
the odd Moyal oscillators $\beta_{o}$ of Eq.(\ref{bodd}) as the building
blocks of all the Virasoro fields $\mathcal{L}_{e},\mathcal{L}_{o}.$ These
can be rewritten in terms of the $\beta_{e}$ by using Eq.(\ref{betae}) or
its inverse. However, to avoid anomalies of the type discussed in Eqs.(\ref%
{spectrum},\ref{signs}) the use of the $\beta_{o}$ as the building blocks
are more convenient. Evidently, the Virasoro algebra can close only in the
infinite cutoff limit.

These $\mathcal{L}_{e},\mathcal{L}_{o}$ are fields in Moyal space, not
differential operators. No normal ordering is needed in Moyal space. When we
evaluate their star products with any field $A\left( \bar{x},x_{\left\vert
e\right\vert },p_{\left\vert e\right\vert }\right) $ in the following
combinations, they produce the Virasoro differential operators $L_{e}^{\beta
}A,$ $L_{o}^{\beta}A$ that we discussed in the previous section\footnote{%
Here, for simplicity, we have suppressed the shift proportional to $%
-w_{e}^{\prime }\beta_{0}$ that appeared in Eq.(\ref{osc-rules}). Therefore,
strictly speaking the formulas for $\mathcal{L}_{e},\mathcal{L}_{o}$ are
valid for fields independent of the midpoint $\bar{x}.$ However it is
straightforward to generalize $\mathcal{L}_{e},\mathcal{L}_{o}$ by including
$\beta_{0}$ to obtain the full $L_{e}^{\beta},L_{o}^{\beta}$ of the previous
section.}%
\begin{align}
L_{e}^{\alpha}\psi\; & \leftrightarrow\;L_{e}^{\beta}A=\mathcal{L}_{e}\ast
A+A\ast\mathcal{L}_{-e},  \label{LeA} \\
L_{o}^{\alpha}\psi\; & \leftrightarrow\;L_{o}^{\beta}A=\mathcal{L}_{o}\ast
A-A\ast\mathcal{L}_{-o}.   \label{LoA}
\end{align}

By using the properties of $U$ and the fundamental star commutator $\left[
\beta_{o},\beta_{o^{\prime}}\right] _{* }=\frac{o}{2}\delta_{o+o^{\prime}}$
(note the 1/2) it can be shown that the Virasoro fields $\mathcal{L}_{e},%
\mathcal{L}_{o}$ satisfy the following star commutation rules with the
oscillator fields $\beta_{e},\beta_{o}$%
\begin{equation}
\left[ \mathcal{L}_{e},\beta_{o}\right] _{* }=-o\beta_{e+o},\;\left[
\mathcal{L}_{e},\beta_{e^{\prime}}\right] _{* }=-e^{\prime}\beta
_{e+e^{\prime}},\;\left[ \mathcal{L}_{o},\beta_{o^{\prime}}\right] _{*
}=-o^{\prime}\beta_{o^{\prime}+o},\;\left[ \mathcal{L}_{o},\beta_{e}\right]
_{* }=-e\beta_{e+o},
\end{equation}
>From these we can show that the Virasoro fields satisfy the Virasoro
algebra under star commutators%
\begin{align}
\left[ \mathcal{L}_{e},\mathcal{L}_{e^{\prime}}\right] _{* } & =\left(
e-e^{\prime}\right) \mathcal{L}_{e+e^{\prime}}+\delta_{e+e^{\prime}}\frac{%
a_{e}}{2},  \label{vir1} \\
\left[ \mathcal{L}_{o},\mathcal{L}_{o^{\prime}}\right] _{* } & =\left(
o-o^{\prime}\right) \mathcal{L}_{o+o^{\prime}}+\delta_{o+o^{\prime}}\frac{%
a_{o}}{2},  \label{vir2} \\
\left[ \mathcal{L}_{o},\mathcal{L}_{e}\right] _{* } & =\left( o-e\right)
\mathcal{L}_{o+e}.   \label{vir3}
\end{align}
The anomalies are half of the usual anomalies $a_{e},a_{o}$ that appear for
the differential operators $L_{e}^{\beta},L_{o}^{\beta}.$ Indeed using the
algebra that we have just obtained we can then show its consistency with the
usual Virasoro algebra obeyed by the differential operators. For example,
using the correspondence in Eq.(\ref{LeA}) we replace the differential
operators with Moyal star products%
\begin{align}
\left[ L_{e}^{\beta},L_{e^{\prime}}^{\beta}\right] A & =L_{e}^{\beta }\left(
L_{e^{\prime}}^{\beta}A\right) -\left( e\leftrightarrow e^{\prime }\right) \\
& =\left( \mathcal{L}_{e}* \left( L_{e^{\prime}}^{\beta}A\right) +\left(
L_{e^{\prime}}^{\beta}A\right) * \mathcal{L}_{-e}\right) -\left(
e\leftrightarrow e^{\prime}\right)
\end{align}
After inserting $L_{e^{\prime}}^{\beta}A$ again from Eq.(\ref{LeA}) this
expression reduces to star commutators of the Moyal fields that can be
evaluated through Eq.(\ref{vir1})
\begin{align}
\left[ L_{e}^{\beta},L_{e^{\prime}}^{\beta}\right] A & =\left[ \mathcal{L}%
_{e},\mathcal{L}_{e^{\prime}}\right] _{* }* A+A* \left[ \mathcal{L}%
_{-e^{\prime}},\mathcal{L}_{-e}\right] _{* } \\
& =\left( e-e^{\prime}\right) \mathcal{L}_{e+e^{\prime}}* A+\delta
_{e+e^{\prime}}\frac{a_{e}}{2}A \\
& +\left( -e^{\prime}+e\right) A* \mathcal{L}_{-e^{\prime}-e}+\delta_{-e^{%
\prime}-e}\frac{a_{-e^{\prime}}}{2}A \\
& =\left( e-e^{\prime}\right) L_{e+e^{\prime}}^{\beta}A+\delta
_{e+e^{\prime}}a_{e}A
\end{align}
This reproduces the correct closure and anomaly for the differential $%
L_{e}^{\beta}$ consistently with the closure of $\mathcal{L}_{e}$ and their
half-anomaly as above. The consistency of the other commutators $\left[
L_{o}^{\beta},L_{o^{\prime}}^{\beta}\right] A,~\left[ L_{e}^{\beta},L_{o}^{%
\beta}\right] A$ can be verified in the same way.

We have seen that the fundamental Virasoro operation consists of independent
left/right star products involving the Virasoro fields $\mathcal{L}_{e},%
\mathcal{L}_{o}$. Therefore, let us consider finite left/right
transformations in the form%
\begin{align}
A & \rightarrow u_{l}* A* u_{r},\; \\
u_{l,r} & =\exp_{* }\left( i\varepsilon_{l,r}^{e}\mathcal{L}%
_{e}+i\varepsilon_{l,r}^{o}\mathcal{L}_{o}\right)   \label{ulr}
\end{align}
Due to Eqs.(\ref{vir1}-\ref{vir3}) these close to form two Virasoro groups,
one on the left side, the other on the right side. To obtain the usual
Virasoro transformations consistently with Eqs.(\ref{LeA},\ref{LoA}) we need
to take the subgroup generated by $\varepsilon_{l}^{e}=\varepsilon_{r}^{e}$
and $\varepsilon_{l}^{o}=-\varepsilon_{r}^{o}$ in the form%
\begin{equation}
A\rightarrow e^{i\varepsilon^{e}\mathcal{L}_{e}+i\varepsilon^{o}\mathcal{L}%
_{o}}* A* e^{i\varepsilon^{e}\mathcal{L}_{e}-i\varepsilon ^{o}\mathcal{L}%
_{o}}
\end{equation}
So far the parameters $\varepsilon^{e,o}$ are complex. If $A\left(
x,p\right) $ is a real string field we must also require that both sides of
this equation are hermitian, using $A^{\dagger}=A,~\mathcal{L}_{n}^{\dagger
}=\mathcal{L}_{-n}.$ Then the allowed Virasoro transformation on real fields
is restricted to complex parameters that satisfy $\left( \varepsilon
^{e}\right) ^{\ast}=-\varepsilon^{-e}$ and $\left( \varepsilon^{o}\right)
^{\ast}=+\varepsilon^{-o}.$

Note that we have used only $\beta_{o}^{\mu}$ as the fundamental structure
and yet we built both $\mathcal{L}_{e}$ and $\mathcal{L}_{o}.$ This uses
half as many oscillators as the standard representation of the Virasoro
algebra, and therefore it is a new representation. Also, it seems to be the
first representation of the Virasoro algebra in the context of
noncommutative field theory.

\section{Monoid algebra in Noncommutative Geometry}

In this section we will introduce an algebra among a subset of string fields
that form a monoid. The mathematical structure of the monoid becomes a tool
for performing computations in string field theory.

\subsection{Generating Functions}

We start from the phase space of $N$ even modes of a bosonic string $\xi
_{i}^{\mu}=\left( x_{2}^{\mu},x_{4}^{\mu},\cdots,p_{2}^{\mu},p_{4}^{\mu
},\cdots\right) $ with $\mu=1,\cdots,d$ denoting the number of dimensions$.$
Eventually we will send $N$ to infinity, but at first all structures are
defined at finite $N.$ The commutators between $x_{2n},p_{2n}$ under the
Moyal star product define noncommutative coordinates in $2Nd$ dimensions as
in Eq.(\ref{sigma}) $[\xi_{i}^{\mu},\xi_{j}^{\nu}]_{\ast}=\eta^{\mu\nu}%
\sigma_{ij}.$

By a linear coordinate redefinition one may simplify any general skew
symmetric $\sigma$ to the canonical form given in Eq.(\ref{sigma}).
Therefore, unless we specify, our general formulas below are written for the
general\footnote{%
Our formulas are easily further generalized to any $\theta^{IJ}$ that is not
necessarily of the form $\sigma^{ij}\eta^{\mu\nu}$. Although we do not
discuss this in detail in this paper, such configurations are relevant for
strings in $B_{\mu\nu}$ backgrounds. For this generalization we use $%
I=\left( i\mu\right) $ and replace everywhere formally $\sigma^{ij}$ by $%
\theta^{IJ}$ and substitute $d\rightarrow1.$} skew symmetric purely
imaginary $\sigma$. The Moyal star product is then the one given by Eq.(\ref%
{star}). For the sake of simplicity of presentation we will suppress the
spacetime index $\mu,$ but will always assume its presence, and will take it
into account in all computations. Similarly, we will suppress the midpoint
insertion and establish it in computations when needed. This product defines
a commutative ring of functions $\mathcal{A}$ on $\mathbf{R}^{2Nd}$. The
integration of functions in phase space is interpreted as the trace of the
algebra $\mathcal{A}$, as in Eq.(\ref{traceA}).

In many computations a certain class of functions will play a primary role.
These are generating functions that are gaussians with shifts of the form $A=%
\mathcal{N}\mathcal{\,\,}\exp\left( -\eta_{m\nu}\left( x^{\mu}ax^{\nu
}+x^{\mu}bp^{\nu}+p^{\mu}c^{T}x^{\nu}+p^{\mu}cx^{\nu}\right) -\left( x^{\mu
}\lambda_{\mu}^{x}+p^{\mu}\lambda_{\mu}^{p}\right) \right) .$ In brief
notation, we write
\begin{equation}
A_{\mathcal{N},M,\lambda}=\mathcal{N}e^{-\bar{\xi}M\xi-\bar{\xi}\lambda}
\label{generating}
\end{equation}
where $M_{ij}$ is a $2N\times2N$ symmetric constant matrix, $\lambda^{\mu}$
is a $2N$-component constant spacetime vector, and $\mathcal{N}$ is an
overall normalization. The normalization may be related to the trace
\begin{equation}
Tr\left( A_{\mathcal{N},M,\lambda}\right) =\frac{\mathcal{N}e^{\frac{1}{4}%
\bar{\lambda}M^{-1}\lambda}}{\left( \det(2M\sigma)\right) ^{d/2}}.
\label{trA}
\end{equation}
The trace is computed under the assumption that the phase space integral in
Eq.(\ref{traceA}) converges, which implies a positive definite matrix $M.$
We will also be interested in more general complex $M$ for which the
integral is not necessarily well defined. For example, the identity field
has infinite trace, the Virasoro transformation of Eqs.(\ref{ulr}) do not
have a well defined trace because of the $i$ in the exponent, etc. So, we
wish to include in our set all possible complex $M$'s since there are such
string field configurations that are relevant. Whenever we compute traces of
gaussians we will use Eq.(\ref{trA}) under the assumption that it is well
defined.

The motivation for considering such gaussians comes from examining the
perturbative and nonperturbative sectors of the theory. We have seen that
the vacuum state $A_{0}$ of Eq.(\ref{AN0}) is of the form of Eq.(\ref%
{generating}) with a special $M_{0}$. All perturbative states have the form
of polynomials that multiply the gaussian $A_{0}$. Such polynomials can be
obtained by differentiating a generating function of the form $A_{\mathcal{N}%
,M,\lambda }\left( \xi\right) $ with respect to the parameters $\lambda.$
Non-perturbative states such as the sliver and many other non-perturbative
vacua are represented by fields of the form $A_{\mathcal{N},M,\lambda}\left(
\xi\right) .$ Furthermore, the Virasoro group that we identified in the
previous section also has the same structure.

More generally, any field $A\left( \bar{x},x_{e},p_{e}\right) $ can be
written as a superposition of gaussians of the form (\ref{generating}). This
is seen by writing $A\left( \bar{x},x_{e},p_{e}\right) =\langle\bar{x}%
,x_{e},p_{e}|\psi\rangle$ where $\langle\bar{x},x_{e},p_{e}|$ is the state
of Eq.(\ref{p}). In the coherent state basis where the oscillators $\alpha
_{n}^{\mu}$ are diagonal, we see from Eq.(\ref{p}) this becomes a
superposition of shifted gaussians
\begin{equation}
A\left( \bar{x},x_{e},p_{e}\right) =\int\left( d\lambda\right) ~e^{-\bar{\xi}%
M_{0}\xi-\bar{\xi}\lambda}~\psi\left( \bar{x},\lambda\right)   \label{Asum}
\end{equation}
where $\psi\left( \bar{x},\lambda\right) $ includes the measure of
integration and normalization. Thus we see that the structure $\mathcal{N}%
e^{-\bar{\xi}M_{0}\xi-\bar{\xi}\lambda},$ where $M_{0}$ of Eq.(\ref{M0})
appears, is a generating function for computations involving any set of
perturbative string fields.

For purely perturbative computations it is sufficient to consider the
restricted set $A_{\mathcal{N},M_{0},\lambda}=\mathcal{N}e^{-\bar{\xi}%
M_{0}\xi-\bar{\xi}\lambda}$ with different $\lambda^{\prime}$s but a fixed $%
M_{0}.$ But to consider nonperturbative sectors which correspond to D-brane
lumps described by gaussians with different $M$'s, and to compute
correlators that involve several D$_{p}$-brane sectors with different $p$'s,
we need to consider generating functions $A_{\mathcal{N},M,\lambda}\left(
\xi\right) $ with all possible $M,\lambda,\mathcal{N}.$

It must also be mentioned that there is a one-to-one correspondence between
the gaussians (\ref{generating}) in Moyal space and coherent states built on
a vacuum of squeezed states in the oscillator formalism. Squeezed states in
the oscillator formalism of \cite{GJ} are defined by $\exp\left( -\frac{1}{2}%
a^{\dagger}\mathcal{S}a^{\dagger}\right) |0\rangle.$ In the $x$%
-representation they are given by gaussians $\psi\left( x\right) \sim
\exp\left( -xLx\right) $ where
\begin{equation*}
L=\frac{1}{2l_{s}^{2}}\sqrt{\kappa}\frac{1-\mathcal{S}}{1+\mathcal{S}}\sqrt{%
\kappa}\equiv\left(
\begin{array}{cc}
L_{e} & \mathcal{L} \\
\mathcal{\bar{L}} & L_{o}%
\end{array}
\right) .
\end{equation*}
with $\kappa=diag\left( \kappa_{n}\right) $. By applying the Fourier
transform of Eq.(\ref{A}) on this form one obtains a gaussian $A\sim
\exp\left( -\xi M\xi\right) $ with
\begin{equation}
M=\left(
\begin{array}{cc}
L_{e}-4\mathcal{L}\left( L_{o}\right) ^{-1}\mathcal{\bar{L}} & \;\;\frac {4i%
}{\theta}\mathcal{L}\left( L_{o}\right) ^{-1}\bar{T} \\
\frac{4i}{\theta}T\left( L_{o}\right) ^{-1}\mathcal{\bar{L}} & \;\;\frac{%
4l_{s}^{2}}{\theta^{2}}T\left( L_{o}\right) ^{-1}\bar{T}%
\end{array}
\right)
\end{equation}
We see that the general symmetric $M$ is related to a general symmetric $L$
or equivalently to a general symmetric $S.$ If the matrix $S$ is block
diagonal in the even/odd mode space, one obtains $\mathcal{L}=0,$ which
simplifies these relations to the form
\begin{equation}
\mathcal{S}=\left(
\begin{array}{cc}
\mathcal{S}_{e} & 0 \\
0 & \mathcal{S}_{o}%
\end{array}
\right) \;\longleftrightarrow\;M=\left(
\begin{array}{cc}
\frac{1}{2l_{s}^{2}}\kappa_{e}^{1/2}\frac{1-\mathcal{S}_{e}}{1+\mathcal{S}%
_{e}}\kappa_{e}^{1/2} & 0 \\
0 & \frac{2l_{s}^{2}}{\theta^{2}}T\kappa_{o}^{-1/2}\frac{1+\mathcal{S}_{o}}{%
1-\mathcal{S}_{o}}\kappa_{o}^{-1/2}\bar{T}%
\end{array}
\right) \;   \label{squeezedtoA1}
\end{equation}

Similarly, the generating function $A_{\mathcal{N},M,\lambda}=\mathcal{N}e^{-%
\bar{\xi}M\xi-\bar{\xi}\lambda}$ with nonzero $\lambda$ is related to a
shifted squeezed state $\exp\left( -\frac{1}{2}a^{\dagger}\mathcal{S}%
a^{\dagger}-ha^{\dagger}\right) |p\rangle$, with momentum $p^{\mu}.$ Then,
for block diagonal $\mathcal{S}$, the vectors $\lambda$ and $h,p$ are
related by using Eqs.(\ref{psitoA},\ref{lambda})%
\begin{equation}
\lambda^{\mu}=\left(
\begin{array}{c}
\frac{i\sqrt{2}}{l_{s}}\sqrt{\kappa_{e}}\frac{1}{1+\mathcal{S}_{e}}%
h_{e}^{\mu }-ip^{\mu}w_{e} \\
\frac{2\sqrt{2}l_{s}}{\theta}\sum_{o>0}T_{eo}\kappa_{o}^{-1/2}\frac {1}{1-%
\mathcal{S}_{o}}h_{o}^{\mu}%
\end{array}
\right)   \label{squizzedtoA2}
\end{equation}

\subsection{Monoid}

In the following, we will focus on the shifted gaussian type generating
functions of Eq.(\ref{generating}). Generally we will allow $%
M_{ij},\lambda_{i},\mathcal{N}$ to be complex numbers. Applying the star
product on any two gaussians closes into a third gaussian of the same form
(suppressing the midpoint insertion)%
\begin{equation}
\left( \mathcal{N}_{1}e^{-\xi M_{1}\xi-\xi\lambda_{1}}\right) * \left(
\mathcal{N}_{2}e^{-\xi M_{2}\xi-\xi\lambda_{2}}\right) =\mathcal{N}%
_{12}e^{-\xi M_{12}\xi-\xi\lambda_{12}}.   \label{a12}
\end{equation}
Therefore these elements form a closed algebra under the Moyal star
multiplication,
\begin{equation}
A_{\mathcal{N}_{1},M_{1},\lambda_{1}}* A_{\mathcal{N}_{2},M_{2},%
\lambda_{2}}=A_{\mathcal{N}_{12},M_{12},\lambda_{12}}.
\end{equation}
The quantities $\mathcal{N}_{12},M_{12,}\lambda_{12}$ were computed in \cite%
{witmoy}, and those details are included in the appendix. It is convenient
to define%
\begin{equation}
m_{1}=M_{1}\sigma,\;m_{2}=M_{2}\sigma,\;m_{12}=M_{12}\sigma,
\end{equation}
where $\sigma$ is the antisymmetric noncommutativity matrix given Eq.(\ref%
{sigma}). Actually our formulas below hold for any general noncommutativity
matrix $\sigma$. Note that symmetric matrices $M_{1},M_{2},M_{12}$ imply
that under transposition the $m_{i},m_{12}$ satisfy
\begin{equation}
\bar{m}_{i}=-\sigma m_{i}\sigma^{-1},\;\bar{m}_{12}=-\sigma
m_{12}\sigma^{-1}.
\end{equation}
Then the result for $m_{12},\lambda_{12},\mathcal{N}_{12}$ given in the
Appendix is written more simply in the form
\begin{align}
m_{12} & =\left( m_{1}+m_{2}m_{1}\right) \left( 1+m_{2}m_{1}\right)
^{-1}+\left( m_{2}-m_{1}m_{2}\right) \left( 1+m_{1}m_{2}\right) ^{-1},
\label{m12} \\
\lambda_{12} & =\left( 1-m_{1}\right) \left( 1+m_{2}m_{1}\right)
^{-1}\lambda_{2}+\left( 1+m_{2}\right) \left( 1+m_{1}m_{2}\right)
^{-1}\lambda_{1}  \label{lambda12} \\
\mathcal{N}_{12} & =\frac{\mathcal{N}_{1}\mathcal{N}_{2}}{\det\left(
1+m_{2}m_{1}\right) ^{d/2}}e^{\frac{1}{4}\left( \left( \bar{\lambda}_{1}+%
\bar{\lambda}_{2}\right) \sigma\left( m_{1}+m_{2}\right) ^{-1}\left(
\lambda_{1}+\lambda_{2}\right) -\bar{\lambda}_{12}\sigma\left( m_{12}\right)
^{-1}\lambda_{12}\right) }   \label{n12}
\end{align}
One can show%
\begin{equation}
\det(m_{1}+m_{2})=\det m_{12}\det\left( 1+m_{2}m_{1}\right) ,
\end{equation}
and%
\begin{equation}
\left( \bar{\lambda}_{1}+\bar{\lambda}_{2}\right) \sigma\left(
m_{1}+m_{2}\right) ^{-1}\left( \lambda_{1}+\lambda_{2}\right) -\bar{\lambda }%
_{12}\sigma\left( m_{12}\right) ^{-1}\lambda_{12}=\bar{\lambda}_{a}\sigma
K_{ab}\lambda_{b}
\end{equation}
where
\begin{align}
K_{11} & =\left( m_{2}^{-1}+m_{1}\right) ^{-1},\quad K_{12}=\left(
1+m_{2}m_{1}\right) ^{-1}, \\
K_{21} & =-\left( 1+m_{1}m_{2}\right) ^{-1},\quad K_{22}=\left(
m_{2}+m_{1}^{-1}\right) ^{-1}.
\end{align}
Other useful forms of $m_{12},\lambda_{12},\mathcal{N}_{12}$ are included in
the appendix.

If we ignore the midpoint insertion, the identity element $I$ discussed in
Eq.(\ref{identity}) can be thought of as an element of the monoid with $%
\mathcal{N}=1,$ $M=0,$ $\lambda=0,$ since $A_{1,0,0}=1$ is the natural
number one. Indeed inserting these values in the formulas above we verify
that $A_{1,0,0}=1$ is the identity element in the monoid. Furthermore, using
the formulas above we see that for a generic $A_{\mathcal{N},M,\lambda}$
there is an inverse $A_{\tilde{N},\tilde{M},\tilde{\lambda}}$ under the star
product, $A_{\mathcal{N},M,\lambda}* A_{\tilde{N},\tilde{M},\tilde{\lambda}%
}=1=A_{\tilde{N},\tilde{M},\tilde{\lambda}}* A_{\mathcal{N},M,\lambda}$,
where%
\begin{equation}
\tilde{M}=-M,\quad\tilde{\lambda}=\frac{m+1}{m-1}\lambda,\quad\mathcal{%
\tilde {N}}=\frac{1}{\mathcal{N}}\left( \det\left( 1-m^{2}\right) \right)
^{d/2}e^{\bar{\lambda}\sigma m\left( 1-m^{2}\right) ^{-2}\lambda}.
\end{equation}
Evidently, the inverse does not exist when $m^{2}$ has eigenvalues $1.$ In
particular, the vacuum field $A_{0}$ of Eq.(\ref{AN0}) which is an element
of the monoid, has an inverse. We note that when the inverse exists, it is
not normalizable under Tr$\left( A^{2}\right) $ since $\tilde{M}=-M$ is
negative definite when $M$ is positive definite; however this does not
prevent us from using the properties of the monoid under star products.

Thus, the algebra generated by the set of functions $A_{\mathcal{N}%
,M,\lambda }$ has the following properties:

\begin{itemize}
\item The algebra is closed under star products.

\item The product is associative.

\item It has a identity given by the number $1$ (suppressing the midpoint
insertion in $I$)

\item While the generic element has an inverse, not every element has an
inverse.
\end{itemize}

The structure is almost a group, but not quite because not every element in
the set has an inverse. This kind of algebraic structure is called a \emph{%
unitary semigroup} or \emph{monoid} in the mathematical literature. In our
case we have a monoid with special properties which we identify as a
fundamental algebra in open string field theory. The general formulas above
give the structure of the monoid. They will form the basis for all the
computations we will present in the rest of the paper.

Note that there are subsets of complex $M_{ij},\lambda_{i},\mathcal{N}$ for
which the submonoid has an inverse for every element. For such subsets the
monoid becomes a genuine infinite dimensional group. In particular, the
exponentiated Virasoro transformations of Eq.(\ref{ulr}), acting on string
fields from either left or right, form an infinite dimensional subgroup of
exactly this type.

As far as we know these types of structures have not been investigated in
the mathematical literature or in the context of noncommutative geometry.

\section{Computations in MSFT Using the Monoid}

\subsection{Powers and traces with same $m$ and $\protect\lambda$}

>From Eq.(\ref{a12}) we see that the $n$'th star-power of a generating
function is also a generating function of the same form%
\begin{equation}
\left( \mathcal{N}e^{-\bar{\xi}M\xi-\bar{\xi}\lambda}\right) _{\ast}^{n}=%
\mathcal{N}^{\left( n\right) }e^{-\bar{\xi}M^{\left( n\right) }\xi-\bar{\xi}%
\lambda^{\left( n\right) }}.   \label{powers}
\end{equation}
Multiplying one more time on both sides of this equation gives an iteration
according to Eqs.(\ref{m12}-\ref{n12})
\begin{align}
m^{\left( n+1\right) } & =\left( m+m^{\left( n\right) }m\right) \left(
1+m^{\left( n\right) }m\right) ^{-1}+\left( m^{\left( n\right) }-mm^{\left(
n\right) }\right) \left( 1+mm^{\left( n\right) }\right) ^{-1},  \label{I1} \\
\lambda_{\mu}^{\left( n+1\right) } & =\left( 1+m^{\left( n\right) }\right)
\left( 1+mm^{\left( n\right) }\right) ^{-1}\lambda_{\mu}+\left( 1-m\right)
\left( 1+m^{\left( n\right) }m\right) ^{-1}\lambda_{\mu }^{\left( n\right) },
\label{I2} \\
\mathcal{N}^{\left( n+1\right) } & =\frac{\mathcal{N}^{\left( n\right) }%
\mathcal{N}}{\det\left( 1+mm^{\left( n\right) }\right) ^{d/2}}\,e^{\frac{1}{4%
}\left( \bar{\lambda}+\bar{\lambda}^{\left( n\right) }\right) \sigma\left(
m+m^{\left( n\right) }\right) ^{-1}\left( \lambda+\lambda^{\left( n\right)
}\right) -\frac{1}{4}\bar{\lambda }^{\left( n+1\right) }\sigma\left(
m^{\left( n+1\right) }\right) ^{-1}\lambda^{\left( n+1\right) }.}
\label{I3}
\end{align}
If we apply a similarity transformation that diagonalizes $m,$ and perform
the iteration of Eq.(\ref{I1}) in the diagonal basis, we easily see that $%
m^{\left( n\right) }$ and $m^{\left( n+1\right) }$ must also be diagonal in
the same basis. From this we conclude that $m$ commutes with $m^{\left(
n\right) }$. Using their commutativity we simplify these formulas as follows
\begin{align}
m^{\left( n+1\right) } & =\left( m+m^{\left( n\right) }\right) \left(
1+mm^{\left( n\right) }\right) ^{-1}  \label{I11} \\
\lambda^{\left( n+1\right) } & =\left( 1+mm^{\left( n\right) }\right) ^{-1}
\left[ \left( \lambda+\lambda^{\left( n\right) }\right) +m^{\left( n\right)
}\lambda-m\lambda^{\left( n\right) }\right]   \label{I22}
\end{align}
The explicit solution of the iteration is then given by%
\begin{align}
m^{\left( n\right) } & =\frac{\left( 1+m\right) ^{n}-\left( 1-m\right) ^{n}}{%
\left( 1+m\right) ^{n}+\left( 1-m\right) ^{n}}  \label{Mn} \\
\lambda_{\mu}^{\left( n\right) } & =\left( m\right) ^{-1}\left( m^{\left(
n\right) }\right) \lambda_{\mu}  \label{lamn} \\
\mathcal{N}^{\left( n\right) } & =\frac{\mathcal{N}^{n}\exp\left[ \frac{n}{4}%
\bar{\lambda}\sigma m^{-1}\lambda-\frac{1}{4}\bar{\lambda}^{\left( n\right)
}\sigma\left( m^{\left( n\right) }\right) ^{-1}\lambda^{\left( n\right) }%
\right] }{\det\left( \frac{\left( 1+m\right) ^{n}+\left( 1-m\right) ^{n}}{2}%
\right) ^{d/2}}.   \label{powernorm}
\end{align}
The trace is computed from Eq.(\ref{trA})%
\begin{equation}
Tr\left( \left( \mathcal{N}e^{-\bar{\xi}M\xi-\bar{\xi}\lambda}\right)
_{\ast}^{n}\right) =\frac{\left( \mathcal{N}\exp\left[ \frac{1}{4}\bar{%
\lambda}M^{-1}\lambda\right] \right) ^{n}}{\det\left( \left(
1+M\sigma\right) ^{n}-\left( 1-M\sigma\right) ^{n}\right) ^{d/2}}.
\label{trpower}
\end{equation}
As applications of these results we specialize to $\lambda=0$ to compute the
wedge and sliver fields below.

\subsubsection{Wedge states for any $\protect\kappa_{e},\protect\kappa_{o},N$%
}

>From our calculation above, it is now straightforward to give the
representation of the wedge states \cite{r-RSZ} in the Moyal formalism.
Wedge states are defined by two equivalent definitions. One of them is the
surface state defined by the conformal transformation
\begin{equation}
f_{n}(z)=\left( \frac{1+iz}{1-iz}\right) ^{2/\left( n+1\right) }\,\,,
\end{equation}
which illuminates its geometrical nature in conformal field theory. The
other definition is the powers of the perturbative vacuum states, $|0\rangle
\ast\cdots\ast|0\rangle\,.$ In this definition, the algebraic aspect of
wedge states is more clearly illuminated. In the MSFT formalism the wedge
field is given by
\begin{equation}
W_{n}\left( x_{e},p_{e}\right) =\left( A_{0}\right)
_{\ast}^{n}=A_{0}\ast\cdots\ast A_{0}\,\,
\end{equation}
where the vacuum field $A_{0}$ is given in Eq.(\ref{AN0}) with $\lambda=0.$
Using associativity, it is evident that these satisfy the algebra $W_{n}\ast
W_{m}=W_{n+m}.$ Eqs.(\ref{Mn}-\ref{trpower}) with $\lambda=0$ give the wedge
fields explicitly%
\begin{align}
W_{n}\left( x_{e},p_{e}\right) & =\frac{\left( \mathcal{N}_{0}\right)
^{n}\exp\left( -\bar{\xi}\frac{\left( 1+m_{0}\right) ^{n}-\left(
1-m_{0}\right) ^{n}}{\left( 1+m_{0}\right) ^{n}+\left( 1-m_{0}\right) ^{n}}%
\sigma^{-1}\xi\right) }{\det\left( \frac{\left( 1+m_{0}\right) ^{n}+\left(
1-m_{0}\right) ^{n}}{2}\right) ^{d/2}}  \label{Wn} \\
Tr\left( W_{n}\right) & =\frac{\left( \mathcal{N}_{0}\right) ^{n}}{%
\det\left( \left( 1+m_{0}\right) ^{n}-\left( 1-m_{0}\right) ^{n}\right)
^{d/2}},   \label{TWn}
\end{align}
where
\begin{equation}
m_{0}\equiv M_{0}\sigma=\left(
\begin{array}{cc}
0 & \frac{i\theta}{2l_{s}^{2}}\kappa_{e} \\
\frac{-2il_{s}^{2}}{\theta}Z & 0%
\end{array}
\right) ,\;Z=T\kappa_{o}^{-1}\bar{T},\;\mathcal{N}_{0}=\left( \frac {%
\det\left( 16\kappa_{e}\right) }{\det\kappa_{o}}\right) ^{d/4}
\end{equation}
follows from Eq.(\ref{AN0}). In computing the powers of $m_{0}$ we encounter
the expression $\Gamma=Z\kappa_{e}=T\kappa_{o}^{-1}\bar{T}\kappa_{e}$ in the
form
\begin{equation}
m_{0}^{2}\equiv\left(
\begin{array}{cc}
\bar{\Gamma} & 0 \\
0 & \Gamma%
\end{array}
\right) =\left(
\begin{array}{cc}
\kappa_{e}T\kappa_{o}^{-1}\bar{T} & 0 \\
0 & T\kappa_{o}^{-1}\bar{T}\kappa_{e}%
\end{array}
\right) =\left(
\begin{array}{cc}
\kappa_{e}^{-1}\bar{R}\kappa_{o}\bar{T} & 0 \\
0 & T\kappa_{o}R\kappa_{e}^{-1}%
\end{array}
\right) .   \label{msms}
\end{equation}
The properties of $\Gamma$ were given in Eqs.(\ref{AN0}-\ref{gt}).

\subsubsection{Sliver State for any $\protect\kappa_{e},\protect\kappa_{o},N$%
}

{}The sliver field $\Xi\left( x_{e},p_{e}\right) $ is defined as the limit
of an infinite number of star products of the perturbative vacuum field $%
A_{0},$ so it is proportional to $W_{\infty},$ which is in the monoid
\begin{equation}
\Xi\left( x_{e},p_{e}\right) =\mathcal{N}_{s}e^{-\xi M_{s}\xi}\sim
\lim_{n\rightarrow\infty}\left( A_{0}\right) _{\ast}^{n}=\lim_{n\rightarrow
\infty}\left( \mathcal{N}_{0}e^{-\xi M_{0}\xi}\right) _{\ast}^{n}..
\label{sliverpower}
\end{equation}
The overall constant $\mathcal{N}_{s}$ depends on the relative normalization
of $A_{0}$ and $\Xi.$ It is possible to compute this limit by using the
exact results of the previous section as follows. To take the $%
n\rightarrow\infty$ limit, we need to rewrite the wedge state $W_{n}$ in a
form that has a well-defined limit. First rewrite $m^{\left( n\right) }$ in
the form
\begin{equation}
m^{\left( n\right) }=m_{0}f_{n}\left( t\right) ,\quad f_{n}\left( t\right)
=t^{-1}\frac{\left( 1+t\right) ^{n}-\left( 1-t\right) ^{n}}{\left(
1+t\right) ^{n}+\left( 1-t\right) ^{n}},\quad t=\sqrt{m_{0}^{2}}.
\label{Mwedge}
\end{equation}
We note that for finite $n$, $f_{n}(t)$ is really a function of $%
t^{2}=m_{0}^{2}$ rather than the square root $t=\sqrt{m_{0}^{2}}.$ In this
sense, there is no ambiguity coming from taking the square root of the
matrix. Introduction of the extra matrix $t$ appears redundant. However, we
have to note that $m_{0}=M_{0}\sigma$ is an \emph{off-diagonal} matrix and
difficult to handle when $n\rightarrow\infty$, by contrast $t^{2}=m_{0}^{2}$
is block diagonal and written in terms of $\Gamma$ as in Eq.(\ref{msms}). We
can use the fact which we observed in Eq.(\ref{g}), namely that we can
diagonalize $\Gamma$ and that it has positive definite eigenvalues.

If we make a similarity transformation to a basis in which $t^{2}=m_{0}^{2}$
is diagonal, for each eigenvalue of $t^{2}$ the square root can be either
positive or negative, and the function $f_{n}\left( t\right) $ would be
evaluated at that eigenvalue. Now, taking the $n\rightarrow\infty$ limit of $%
f_{n}\left( t\right) $ for each eigenvalue, we see that, since the
square-root of the eigenvalue $\pm\sqrt{t^{2}}$ is real, the result is
\begin{equation}
\lim_{n\rightarrow\infty}f_{n}\left( t\right) =\left\vert t\right\vert ^{-1}
\end{equation}
where $\left\vert t\right\vert $ is the real positive square root. If the
square-root of the eigenvalue $\pm\sqrt{t^{2}}$ were imaginary, the limit
would have oscillated wildly and there would have been no well defined
limit. Therefore, the limit of the whole matrix $\lim_{n\rightarrow\infty}%
\left( m^{\left( n\right) }\right) $ is well defined thanks to the fact that
$m_{0}^{2}$ is a positive definite matrix which is the case as seen in our
analysis. Having established this fact, we can now write that the limit for
the entire matrix, after transforming back to the general non-diagonal
basis, is $f_{\infty}\left( t\right) =\left\vert t\right\vert ^{-1}=\left(
m_{0}^{2}\right) ^{-1/2},$ by which we mean that we keep only the positive
square root of the eigenvalues.

With this analysis, we have now established that the sliver field in Eq.(\ref%
{sliverpower}) is uniquely given by%
\begin{equation}
m_{s}=M_{s}\sigma=m_{0}\left( m_{0}^{2}\right) ^{-1/2}=\left(
\begin{array}{cc}
0 & \frac{i\theta}{2l_{s}^{2}}\kappa_{e} \\
\frac{-2il_{s}^{2}}{\theta}Z & 0%
\end{array}
\right) \left(
\begin{array}{cc}
\sqrt{\bar{R}\kappa_{o}^{-1}\bar{T}\kappa_{e}} & 0 \\
0 & \sqrt{\kappa_{e}T\kappa_{o}^{-1}R}%
\end{array}
\right)   \label{msliver}
\end{equation}
After multiplying with $\sigma,$ we extract the block diagonal $M_{s}$
\begin{equation}
M_{s}=\left(
\begin{array}{cc}
a & 0 \\
0 & \frac{1}{a\theta^{2}}%
\end{array}
\right) ,\quad m_{s}=i\left(
\begin{array}{cc}
0 & a\theta \\
\frac{-1}{a\theta} & 0%
\end{array}
\right) ,\;\;a=\frac{1}{2l_{s}^{2}}\kappa_{e}\sqrt{\kappa_{e}T%
\kappa_{o}^{-1}R},\;\;m_{s}^{2}=1   \label{sM}
\end{equation}
Note that $\kappa_{e}\sqrt{\kappa_{e}^{-1}\bar{R}\kappa_{o}R}\neq\sqrt {%
\kappa_{e}\bar{R}\kappa_{o}R}.$ Note also that $a$ is a symmetric matrix,
and can be rewritten in several forms by using the first relation in Eq.(\ref%
{tr})%
\begin{equation}
a=\frac{1}{2l_{s}^{2}}\kappa_{e}\sqrt{\kappa_{e}T\kappa_{o}^{-1}R}=\frac {1}{%
2l_{s}^{2}}\sqrt{\bar{R}\kappa_{o}^{-1}\bar{T}\kappa_{e}}\kappa_{e}=\frac{1}{%
2l_{s}^{2}}\kappa_{e}^{1/2}\left( \sqrt{\kappa_{e}^{1/2}T\,\kappa_{o}^{-1}%
\bar{T}\,\kappa_{e}^{1/2}}\right) ^{-1}\kappa_{e}^{1/2}.   \label{a1}
\end{equation}
Furthermore, using Eq.(\ref{ktk}) $a$ can also be rewritten in terms of the
eigenvalues $\tau_{k}$%
\begin{equation}
a=\frac{1}{2l_{s}^{2}}\kappa_{e}^{1/2}V^{e}\tau^{-1}\bar{V}^{e}\kappa
_{e}^{1/2},\;\;a^{-1}=2l_{s}^{2}\kappa_{e}^{-1/2}V^{e}\tau\bar{V}%
^{e}\kappa_{e}^{-1/2}   \label{a2}
\end{equation}
Note that $\left( a^{-1}\right) _{ee^{\prime}}$ is well defined at finite $N$
for generic $\kappa_{e},\kappa_{o}.$ Furthermore, at infinite $N$ the
integral $\int_{0}^{\infty}dkV_{e}\left( k\right) \left( \tanh\left( \pi
k/4\right) \right) ^{-1}V_{e^{\prime}}\left( k\right) $ is convergent
despite the zero eigenvalue $\tau=0$ at $k=0,$ because $\lim_{k\rightarrow
0}V_{e}\left( k\right) =O\left( k\right) ,$ therefore $\left( a\right)
_{ee^{\prime}}$ is still well defined at infinite $N$. So, the sliver field
is explicitly given by%
\begin{equation}
\Xi\left( x_{2n},p_{2n}\right) =\left( \prod_{e>0}2^{d}\right) \exp\left(
-x_{e}ax_{e}-p_{e}\frac{1}{a\theta^{2}}p_{e}\right) ,   \label{sliver}
\end{equation}
We have fixed the relative coefficient in Eq.(\ref{sliverpower}) so that the
normalization factor $\mathcal{N}_{s}=\prod_{e>0}2^{d}$ is chosen to satisfy
the projector equation%
\begin{equation}
\Xi\ast\Xi=\Xi,
\end{equation}
as verified through Eqs.(\ref{powers}-\ref{powernorm}) for $\lambda=0$, $%
\left( M_{s}\sigma\right) ^{2}=1,$ and $n=2.$ With this normalization we
compute the trace from Eq.(\ref{trA}) and find%
\begin{equation}
Tr\left( \Xi\right) =1.
\end{equation}
So, the rank of this projector is $1.$ This is a special form of a projector
as can be seen by comparing to Eqs.(\ref{D}-\ref{AD}) in the next section.

\subsection{Projectors}

\subsubsection{Projectors in Monoid}

In noncommutative field theory an important r\^{o}le is played by so-called
noncommutative solitons which satisfy the unipotency relation $f\ast f=f$.
Such solutions are associated with $D$ branes. Using the monoid closure of
Eqs.(\ref{a12}-\ref{n12}) one can find such soliton solutions by the
requirement,
\begin{equation}
M=M_{1}=M_{2}=M_{12},\quad\lambda=\lambda_{1}=\lambda_{2},\quad\mathcal{N}=%
\mathcal{N}_{1}=\mathcal{N}_{2}=\mathcal{N}_{12}
\end{equation}
The first equation reduces to, $m(m^{2}-1)=0$ which implies $M=0$ or $\left(
M\sigma\right) ^{2}=1$. For $M=0$, one must also demand $\lambda=0$. This is
nothing but the identity element. For the second choice, there is no
constraint on $\lambda$ but one needs to impose $\mathcal{N}%
=\det_{e}(2^{d})\exp(-\frac{1}{4}\lambda M^{-1}\lambda)$. The general matrix
$M$ that satisfies the condition $\left( M\sigma\right) ^{2}=1$ will be
denoted $D$. It can be parameterized in terms of blocks as follows%
\begin{equation}
D=\left(
\begin{array}{cc}
a & ab \\
ba & ~~\frac{1}{a\theta^{2}}+bab%
\end{array}
\right) =\left(
\begin{array}{cc}
1 & 0 \\
b & 1%
\end{array}
\right) \left(
\begin{array}{cc}
a & 0 \\
0 & ~\frac{1}{a\theta^{2}}%
\end{array}
\right) \left(
\begin{array}{cc}
1 & b \\
0 & 1%
\end{array}
\right) ,\quad   \label{D}
\end{equation}
with $a,b$ arbitrary $N\times N$ \textit{symmetric} matrices$.$ The norm $%
\mathcal{N}$ is also uniquely determined. Thus a projector, which is a
candidate for a nonperturbative vacuum associated with D-branes is
characterized by a matrix of the form $D$ and an arbitrary $\lambda^{\mu},$
and takes the form
\begin{align}
A_{D,\lambda}\left( \xi\right) & =\left( \prod_{e>0}2^{d}\right) \exp(-\frac{%
1}{4}\bar{\lambda}\sigma D\sigma\lambda)\exp\left( -\bar{\xi}D\xi-\bar{\xi}%
\lambda\right)  \label{AD} \\
A_{D,\lambda} & =A_{D,\lambda}\ast A_{D,\lambda},\;\;Tr\left( A_{D,\lambda
}\right) =1.  \notag
\end{align}
where we have used $D^{-1}=\sigma D\sigma.$ The trace of $%
A_{D,\lambda}\left( \xi\right) $, which corresponds to its rank, is exactly
1 for any $\lambda,a,b.$

We see that the sliver field is a special case with $\lambda=0,$ $b=0,$ and
a particular $a$ given in Eqs.(\ref{a1},\ref{a2}). Another simpler and
natural projector is when $a=\frac{1}{2l_{s}^{2}}\kappa_{e}$ with $\lambda=0,
$ $b=0.$ It takes the explicit form%
\begin{equation}
A_{butterfly}\left( x_{e},p_{e}\right) =\left( \prod_{e>0}2^{d}\right)
\exp\left( -\frac{1}{2l_{s}^{2}}x_{e}\kappa_{e}x_{e}-\frac{2l_{s}^{2}}{%
\theta^{2}}p_{e}\frac{1}{\kappa_{e}}p_{e}\right)   \label{butterfly}
\end{equation}
This is the state that we referred to as the \textquotedblleft false
vacuum\textquotedblright\ in our discussion following Eq.(\ref{spectrum}).
In fact, it corresponds to the product of the vacuua for the left and right
oscillators of the split string formalism, and it was named the
\textquotedblleft butterfly\textquotedblright\ in \cite{rsz3}.

\subsubsection{Closure of products of projectors in monoid}

Consider two projectors $A_{D_{1},\lambda_{1}}$, $A_{D_{2},\lambda_{2}}$ of
the form (\ref{AD}). Their product is found to be proportional to a
projector
\begin{equation}
A_{D_{1},\lambda_{1}}*
A_{D_{2},\lambda_{2}}=C_{12}A_{D_{12},\lambda_{12}},\quad
A_{D_{12},\lambda_{12}}* A_{D_{12},\lambda_{12}}=A_{D_{12},\lambda_{12}},
\end{equation}
where $D_{12},\lambda_{12}$ is given in Eqs.(\ref{m12},\ref{lambda12}$)$ the
overall norm of $A_{D_{12},\lambda_{12}}$ is fixed as in Eq.(\ref{AD}) and
\begin{equation}
C_{12}=\left( \det\left( \frac{1+D_{1}\sigma D_{2}\sigma}{2}\right) \right)
^{-d/2}e^{\frac{1}{4}\left( \lambda_{1}+\lambda_{2}\right) \left(
D_{1}+D_{2}\right) ^{-1}\left( \lambda_{1}+\lambda_{2}\right) -\frac{1}{4}%
\lambda_{1}\left( D_{1}\right) ^{-1}\lambda_{1}-\frac{1}{4}\lambda
_{2}\left( D_{2}\right) ^{-1}\lambda_{2}}.
\end{equation}
To show that $A_{D_{12},\lambda_{12}}$ is a projector we must prove that $%
\left( D_{12}\sigma\right) ^{2}=1$ when $D_{12}$ is given by Eq.(\ref{m12})
and $\left( D_{1}\sigma\right) ^{2}=\left( D_{2}\sigma\right) ^{2}=1.$ We
use an alternate form of Eq.(\ref{m12}) given in the appendix in Eq.(\ref%
{M121})
\begin{equation}
D_{12}\sigma=D_{1}\sigma+\left( 1-D_{1}\sigma\right) \left(
D_{1}\sigma+\left( D_{2}\sigma\right) ^{-1}\right) ^{-1}\left(
1+D_{1}\sigma\right) .
\end{equation}
The square of this expression satisfies $\left( D_{12}\sigma\right) ^{2}=1$
due to $\left( D_{1}\sigma\right) ^{2}=1,$ since the second term squares to
zero and the cross terms cancel each other.

When $D_{1}=D_{2}=D$ but the $\lambda^{\prime}$s are different we get $%
D_{12}=D$ and
\begin{equation}
A_{D,\lambda_{1}}\ast A_{D,\lambda_{2}}=C_{12}\,A_{D,\lambda_{12}},\quad
C_{12}=e^{-\frac{1}{8}\left( \lambda_{1}-\lambda_{2}\right) \left( D\right)
^{-1}\left( \lambda_{1}-\lambda_{2}\right) },\quad\lambda _{12}^{\mu}=\frac{1%
}{2}\left( \lambda_{1}^{\mu}+\lambda_{2}^{\mu}\right) +\frac{m}{2}\left(
\lambda_{1}^{\mu}-\lambda_{2}^{\mu}\right) .
\end{equation}
Furthermore, if $\lambda_{1}=\lambda_{2}$ we get $\lambda_{12}=\lambda$ and $%
C_{12}=1,$ as expected from Eq.(\ref{AD}).

\subsubsection{More general projectors}

Not all projectors are of the monoid form. More general projectors may be
constructed by using generalized Wigner distribution functions \cite{wigner}%
. These have the form%
\begin{equation}
A_{rs}\left( x_{2n},p_{2n}\right) =\int\left( \prod_{n=1}^{\infty}\left(
dy_{2n}\right) \,e^{\frac{i}{\theta}y_{2n}\cdot p_{2n}}\right) \,\,\psi
_{r}\left( x_{2n}+\frac{y_{2n}}{2}\right) \,\,\,\psi_{s}^{\ast}\left( x_{2n}-%
\frac{y_{2n}}{2}\right)   \label{Awigner}
\end{equation}
where $r,s$ denote any set of orthogonal functions
\begin{equation}
\int\prod_{n=1}^{\infty}\left( dy_{2n}\right) \,\psi_{r}\left( y_{2n}\right)
\,\,\,\psi_{s}^{\ast}\left( y_{2n}\right) =\delta_{rs}
\end{equation}
In the literature on deformation quantization one finds many ways of
obtaining a complete set of Wigner functions by using the complete set of
normalizable wavefunctions for any quantum mechanics problem (a particle in
a potential). Of course, the Wigner functions found in the literature are
functions appropriate for a particle, but it is straightforward to
generalize them to our case with many string modes. For example, by
imitating the case of the harmonic oscillator, which behaves like string
modes, the Wigner functions could be taken in the form of polynomials that
multiply the projector in Eq.(\ref{AD}).

As is well known, under Moyal star products, which is equivalent to the
string star product in our case, the Wigner functions satisfy%
\begin{equation}
A_{rs}* A_{kl}\left( x_{2n},p_{2n}\right) =\delta_{sk}\,A_{rl}\left(
x_{2n},p_{2n}\right) .
\end{equation}
Therefore all diagonal Wigner functions $A_{rr}\left( x_{2n},p_{2n}\right) $
are projectors. The trace of the Wigner function is given by (no sum on $r$)%
\begin{equation}
Tr\left( A_{rr}\right) =\int\left( \prod_{n=1}^{\infty}\frac{dx_{2n}dp_{2n}}{%
2\pi\theta}\right) \,A_{rr}\left( x_{2n},p_{2n}\right) =\int
\prod_{n=1}^{\infty}\left( dx_{2n}\right) \,\psi_{r}\left( x_{2n}\right)
\,\,\,\psi_{r}^{\ast}\left( y_{2n}\right) =1
\end{equation}
so the rank of each projector $A_{rr}$ is 1. Presumably the projectors that
are in the monoid (as in the previous subsection) can be rewritten as
special Wigner functions of the form $A_{rr}.$

Multi D-brane states can now be easily constructed by taking sums of
othogonal projectors. Thus a state with $\mathcal{N}$ D-branes is given by%
\begin{equation}
A^{\left( N\right) }\left( x_{2n},p_{2n}\right) =\sum_{r=1}^{N}A_{rr}\left(
x_{2n},p_{2n}\right) ,\quad TrA^{\left( N\right) }=N.
\end{equation}
Then one may choose a set of $\psi_{r}\left( x_{2n}\right) $ that form the
basis for $U\left( N\right) $ transformations which correspond to Chan-Paton
type symmetries associated with $D$-branes at the ends of strings.

\subsection{Products and traces for same $m$ and different $\protect\lambda%
_{i}$}

As argued following Eq.(\ref{Asum}) for computations involving fields built
on a given vacuum, such as the perturbative vacuum, it is sufficient to
compute the star products of elements of the monoid with the same $M,$ but
different $\lambda$'s and $\mathcal{N}$'s. When all $\lambda$'s are
identical the results should coincide with Eqs.(\ref{powers}-\ref{trpower}).
Therefore these products are generalizations of the wedge fields $%
W_{n}\left( x_{e},p_{e}\right) .$ The closure of the monoid gives the form
of the answer%
\begin{align}
A_{\mathcal{N}_{12\cdots n},M^{\left( n\right) },\lambda_{12\cdots n}} &
\equiv A_{\mathcal{N}_{1},M,\lambda_{1}}\ast A_{\mathcal{N}%
_{2},M,\lambda_{2}}\ast\cdots\ast A_{\mathcal{N}_{n},M,\lambda_{n}}
\label{A12n} \\
& =\mathcal{N}_{12\cdots n}\exp\left( -\bar{\xi}M^{\left( n\right) }\xi-\bar{%
\xi}\lambda_{12\cdots n}\right)
\end{align}
>From the general formula for $M_{12}$ in Eq.(\ref{m12}) we see that $M_{12}$
is independent of the $\lambda$'s and $\mathcal{N}$'s. Therefore the product
of $n$ factors produces produces the same result as if all $\lambda$'s and $%
\mathcal{N}$'s are the same. Therefore we have $M^{\left( n\right)
}\sigma=m^{\left( n\right) }$ where $m^{\left( n\right) }$ was already
computed in Eq.(\ref{Mn}).

To obtain the dependence of $\mathcal{N}_{12\cdots n}$ and $%
\lambda_{12\cdots n}$ on the $\lambda$'s, let us first consider the product
of two elements. Using Eqs.(\ref{lambda12},\ref{n12}) we find
\begin{equation*}
A_{\mathcal{N}_{12},M^{\left( n\right) },\lambda_{12}}\left( \xi\right) =%
\frac{\mathcal{N}_{1}\mathcal{N}_{2}\exp\left( \frac{1}{4}K_{12}\right) }{%
\det\left( 1+m^{2}\right) ^{d/2}}\exp\left( -\bar{\xi}\frac{2m}{1+m^{2}}%
\sigma^{-1}\xi-\bar{\xi}\lambda_{12}\right)
\end{equation*}
with%
\begin{align}
\lambda_{12} & =\frac{1+m}{1+m^{2}}\lambda_{1}+\frac{1-m}{1+m^{2}}%
\lambda_{2}, \\
K_{12} & =\bar{\lambda}_{1}\sigma\frac{m}{1+m^{2}}\lambda_{1}+\bar{\lambda }%
_{2}\sigma\frac{m}{1+m^{2}}\lambda_{2}+\bar{\lambda}_{1}\sigma\frac {1}{%
1+m^{2}}\lambda_{2}-\bar{\lambda}_{2}\sigma\frac{1}{1+m^{2}}\lambda_{1}
\end{align}
To compute the case for $n=3$ we can use $A_{123}=A_{12}* A_{3},$ insert the
above result for $A_{12},$ and apply again the general formulas in Eqs.(\ref%
{m12},\ref{n12}) for commuting matrices $m_{12}=m^{\left( 2\right) }$ and $%
m_{3}=m.$ This process is repeated to build the general $\mathcal{N}%
_{12\cdots n},$ $M^{\left( n\right) },\;\lambda_{12\cdots n}$ that appear in
Eq.(\ref{A12n}). In these computations Eqs.(\ref{m12},\ref{n12}) simplify
because the matrices $m_{1},m_{2}$ commute with each other since they are
all functions of the same $m.$ This is explained fully in the next section.
The result for the $n$'th product gives

\begin{align}
M^{\left( n\right) }\sigma & =m^{\left( n\right) }=\frac{J_{n}^{-}}{J_{n}^{+}%
}\,,\quad J_{n}^{\pm}\equiv\frac{(1+m)^{n}\pm(1-m)^{n}}{2} \\
\lambda_{12\cdots n} &
=(J_{n}^{+})^{-1}\sum_{r=1}^{n}(1-m)^{r-1}(1+m)^{n-r}\lambda_{r} \\
\mathcal{N}_{12\cdots n} & =\mathcal{N}_{1}\mathcal{N}_{2}\mathcal{\cdots N}%
_{n}(\det(J_{n}^{+}))^{-d/2}\exp\left( \frac{1}{4}K_{n}\left( \lambda\right)
\right) \\
K_{n}\left( \lambda\right) & =\sum_{r=1}^{n}\bar{\lambda}_{r}\sigma \frac{%
J_{n-1}^{-}}{J_{n}^{+}}\lambda_{r}+2\sum_{r<s}^{n}\bar{\lambda}_{r}\sigma%
\frac{(1-m)^{s-r-1}(1+m)^{n+r-s-1}}{J_{n}^{+}}\lambda_{s}\,\,.
\end{align}

This result may be used in conjunction with Eq.(\ref{Asum}) to compute star
products of any number of arbitrary string fields built around a vacuum. For
example, the cube of a general string field is given by%
\begin{equation}
\left( A* A* A\right) \left( \bar{x},\xi\right) =e^{3i\bar{x}_{27}}\int
d\lambda_{1}d\lambda_{2}d\lambda_{3}\psi\left( \bar{x},\lambda_{1}\right)
\psi\left( \bar{x},\lambda_{2}\right) \psi\left( \bar{x},\lambda _{3}\right)
A_{\mathcal{N}_{123},M^{\left( 3\right) },\lambda_{123}}\left( \xi\right)
\end{equation}
where the midpoint insertions have been made explicit.

The trace of Eq.(\ref{A12n}), which gives the $n$-point vertices, is
straightforward to compute%
\begin{equation}
Tr\left( A_{12\cdots n}\right) =\frac{\mathcal{N}_{1}\mathcal{N}_{2}\cdots%
\mathcal{N}_{n}~\exp\left( \frac{1}{4}\sum_{r,s=1}^{n}\bar{\lambda}_{r}\sigma%
\mathcal{O}_{\left( s-r\right) \func{mod}n}\lambda _{s}\right) }{\det\left(
(1+m)^{n}-(1-m)^{n}\right) ^{d/2}}   \label{tran}
\end{equation}
where
\begin{align}
\mathcal{O}_{0}\left( m\right) & =\frac{\left( 1+m\right) ^{n-1}+\left(
1-m\right) ^{n-1}}{\left( 1+m\right) ^{n}-\left( 1-m\right) ^{n}},
\label{On} \\
\mathcal{O}_{i}\left( m\right) & =2\frac{(1+m)^{n-i-1}(1-m)^{i-1}}{\left(
1+m\right) ^{n}-\left( 1-m\right) ^{n}}\,,\quad(1\leq i\leq n-1).
\label{On1}
\end{align}
In our notation $\mathcal{O}_{\left( -1\right) \func{mod}n}=\mathcal{O}_{n-1}
$, etc. It will be shown in section-V that our computation of the $n$-string
vertex $Tr\left( A_{12\cdots n}\right) $ given above provides a simple
analytic expression of the Neumann coefficients $\left( V_{n}^{rs}\right)
_{kl}$ that are needed in the definition of the $n$-point string vertex in
the oscillator approach.

It is useful to note the following simplifications. For normalized fields
the normalization factors $\mathcal{N}_{i}$ are fixed as follows
\begin{equation}
Tr\left( A_{i}* A_{i}\right) =1\;\rightarrow\;\mathcal{N}_{i}=\left(
\det4m\right) ^{d/4}\exp\left( -\frac{1}{4}\bar{\lambda}_{i}\sigma
m^{-1}\lambda_{i}\right)
\end{equation}
Then the $n$-point vertices depend only on the differences $\left(
\lambda_{i}-\lambda_{j}\right) .$ We give here the explicit forms for the $2$
and 3 point vertices with these normalizations%
\begin{align}
Tr\left( A_{\lambda_{1}}* A_{\lambda_{2}}\right) & =\exp\left( -\frac {1}{8}%
\left( \bar{\lambda}_{1}-\bar{\lambda}_{2}\right) \sigma m^{-1}\left(
\lambda_{1}-\lambda_{2}\right) \right)  \label{transl12} \\
Tr\left( A_{\lambda_{1}}* A_{\lambda_{2}}* A_{\lambda_{3}}\right) & =\frac{%
\det\left( 16m\right) ^{d/4}}{\det\left( 3+m^{2}\right) ^{d/2}}\exp\left(
\begin{array}{c}
-\frac{1}{8}\sum_{i,j=1}^{3}\left( \bar{\lambda}_{i}-\bar{\lambda}%
_{j}\right) \sigma\left( 3m+m^{3}\right) ^{-1}\left(
\lambda_{i}-\lambda_{j}\right) \\
+\frac{1}{4}\sum_{i=1}^{3}\left( \bar{\lambda}_{i}-\bar{\lambda}%
_{i+1}\right) \sigma\left( 3+m^{2}\right) ^{-1}\left( \bar{\lambda}_{i}-\bar{%
\lambda}_{i+2}\right)%
\end{array}
\right)   \label{transl123}
\end{align}
where the indices on the $\lambda$'s should be understood $\func{mod}3.$
Note that if these fields are also projectors satisfying $m^{2}=1$ the
expressions simplify further.

\subsection{Products and traces for commuting $m_{i}$ and arbitrary $\protect%
\lambda_{i}$}

The multiplication formula for the monoid looks rather complicated and it is
difficult to calculate the $n$-th product for arbitrary $m_{i}=M_{i}\sigma$.
However, for certain applications in string field theory, as we saw in the
previous section, one may restrict the form of the monoid to a submonoid
with commuting matrices
\begin{equation}
\lbrack M_{1}\sigma,M_{2}\sigma]=0\,\,.   \label{commute}
\end{equation}
Such cases would arise when we consider products or correlators between
states built on different vacua, such as perturbative states around the
gaussian with $M_{0},$ wedge states or sliver states built around the
gaussians with $M_{0}^{\left( n\right) },M_{s},$ etc., all of which are
functions of the same $M_{0}$ and therefore satisfy the conditions (\ref%
{commute}).

Thus, consider the product of $n$ elements $A_{\mathcal{N}%
_{i},M_{i},\lambda_{i}}$ ($i=1,\cdots,n$) for commuting $m_{i}\equiv
M_{i}\sigma$ which generalize those in the previous section (which had the
same $m$).
\begin{equation}
A_{\mathcal{N}_{12\cdots n},M_{12\cdots n},\lambda_{12\cdots N}}\equiv A_{%
\mathcal{N}_{1},M_{1},\lambda_{1}}\ast A_{\mathcal{N}_{2},M_{2},\lambda
_{2}}\ast\cdots\ast A_{\mathcal{N}_{n},M_{n},\lambda_{n}},
\end{equation}
Using the closure property we know that the result in an element of the
monoid labelled by $\mathcal{N}_{12\cdots n}$, $M_{12\cdots n}$, $%
\lambda_{12\cdots N}$.. To derive these expressions we use associativity to
write $A_{12\cdots \left( n+1\right) }=A_{12\cdots n}\ast A_{n+1}.$ Applying
Eqs.(\ref{m12},\ref{lambda12},\ref{n12}) for commuting $m$'s we set up the
recursion relations
\begin{align}
m_{12\cdots\left( n+1\right) } & =\frac{m_{12\cdots n}+m_{n+1}}{%
1+m_{12\cdots n}m_{n+1}} \\
\lambda_{12\cdots\left( n+1\right) } & =\left[ 1+m_{12\cdots n}m_{n+1}\right]
^{-1}\left[ (1-m_{12\cdots n})\lambda_{n+1}+(1+m_{n+1})\lambda_{12\cdots n}%
\right] \\
K_{12\cdots\left( n+1\right) } & =K_{12\cdots n}+(\bar{\lambda}_{12\cdots n}+%
\bar{\lambda}_{n+1})\sigma(m_{12\cdots n}+m_{n+1})^{-1}(\lambda_{12\cdots
n}+\lambda_{n+1}) \\
& -\bar{\lambda}_{12\cdots\left( n+1\right) }\sigma m_{12\cdots\left(
n+1\right) }^{-1}\lambda_{12\cdots\left( n+1\right) }.
\end{align}
For the overall normalization constant the recursion formula is,
\begin{equation}
\mathcal{N}_{12\cdots\left( n+1\right) }=\frac{\mathcal{N}_{12\cdots n}%
\mathcal{N}_{n+1}}{\det\left( 1+m_{12\cdots n}m_{n+1}\right) ^{d/2}}e^{\frac{%
1}{4}\left( K_{12\cdots\left( n+1\right) }-K_{12\cdots n}\right) }
\end{equation}
We will prove that the solution of these recursion relations is
\begin{align}
m_{12\cdots n} & \equiv M_{12\cdots n}\sigma=\frac{J_{12\cdots n}^{-}}{%
J_{12\cdots n}^{+}}\ ,\quad J_{12\cdots n}^{\pm}\equiv\frac{1}{2}\left(
\prod_{k=1}^{n}(1+m_{k})\pm\prod_{k=1}^{n}(1-m_{k})\right)  \label{solm} \\
\lambda_{12\cdots n} & =\sum_{i=1}^{n}\frac{\prod_{k=1}^{i-1}\left(
1-m_{k}\right) \prod_{l=i+1}^{n}\left( 1+m_{l}\right) }{J_{12\cdots n}^{+}}%
\lambda_{i}  \label{soll} \\
\mathcal{N}_{12\cdots n} & =\frac{\mathcal{N}_{1}\mathcal{N}_{2}\cdots%
\mathcal{N}_{n}}{\det\left( J_{12\cdots n}^{+}\right) ^{d/2}}\exp\left(
\frac{1}{4}K_{12\cdots n}\right)
\end{align}
where%
\begin{align}
K_{12\cdots n} & \equiv\sum_{i=i}^{n}\bar{\lambda}_{i}\sigma\frac {%
J_{12\cdots n}^{-\left( i\right) }}{J_{12\cdots n}^{+}}\lambda_{i}+\sum_{i%
\neq j}^{n}\bar{\lambda}_{i}\sigma\frac{\prod_{k=1}^{i-1}\left(
1+m_{k}\right) \prod_{r=i+1}^{j-1}\left( 1-m_{r}\right)
\prod_{l=j+1}^{n}\left( 1+m_{l}\right) }{J_{12\cdots n}^{+}\,\,sign\left(
i-j\right) \,}\lambda_{j}  \label{solK} \\
J_{12\cdots n}^{-\left( i\right) } & \equiv\left( J_{12\cdots n}^{-}\right)
|_{m_{i}=0}   \label{solJi}
\end{align}
In the last line only one of the $m$'s is set to zero ($m_{i}=0)$, which
means that $J_{12\cdots n}^{-\left( i\right) }$ is defined by omitting the
factors $\left( 1\pm m_{i}\right) $ in the $J_{12\cdots n}^{-}$ of Eq.(\ref%
{solm}).

The trace of $A_{\mathcal{N}_{12\cdots n},M_{12\cdots n},\lambda_{12\cdots
N}}$ computed according to Eq.(\ref{trA}) takes the following form,
\begin{equation}
Tr\left( A_{12\cdots n}\right) =\frac{\mathcal{N}_{1}\mathcal{N}_{2}\cdots%
\mathcal{N}_{n}}{\det\left( 2J_{12\cdots n}^{-}\right) ^{d/2}}\exp\left(
\frac{1}{4}Q_{12\cdots n}\right)
\end{equation}
where%
\begin{align}
Q_{12\cdots n} & \equiv K_{12\cdots n}+\sum_{i=i}^{n}\bar{\lambda}%
_{i}\sigma\left( m_{12\cdots n}^{-1}\right) \lambda_{i}  \label{traceQ} \\
& =\sum_{i=i}^{n}\bar{\lambda}_{i}\sigma\frac{J_{12\cdots n}^{+\left(
i\right) }}{J_{12\cdots n}^{-}}\lambda_{i}+\sum_{i\neq j}^{n}\bar{\lambda }%
_{i}\sigma\frac{\prod_{k=1}^{i-1}\left( 1+m_{k}\right)
\prod_{r=i+1}^{j-1}\left( 1-m_{r}\right) \prod_{l=j+1}^{n}\left(
1+m_{l}\right) }{J_{12\cdots n}^{-}\,\,\,}\lambda_{j}
\end{align}

\subsection{Angle variables and $K_{1}$}

To check the recursion relations is straightforward but rather tedious. Some
aspects of the recursion formulae can be more illuminating if we make a
change of variables. The recursion formula for $m$ becomes much simpler if
we introduce the ``angle'' variables (which are commuting matrices)
\begin{equation}
\Theta_{\ell}=\tan^{-1}(-im_{\ell})=\frac{1}{2i}\log\frac{1+m_{\ell}}{%
1-m_{\ell}},\quad\Theta_{12\cdots n}\equiv\tan^{-1}(-im_{12\cdots n}).
\end{equation}
With these variables, the above relations can be simply written as $%
\Theta_{12\cdots n+1}=\Theta_{12\cdots n}+\Theta_{n+1}$. Since this is a
linear relation, one can immediately solve it as
\begin{equation}  \label{soltheta}
\Theta_{12\cdots n}=\sum_{l=1}^{n}\Theta_{l}.
\end{equation}
By using the elementary relations between $\Theta$ and $m$,
\begin{equation*}
\cos\Theta_{\ell}=\frac{1}{\sqrt{1-m_{\ell}^{2}}},\quad\sin\Theta_{\ell}=%
\frac{-im_{\ell}}{\sqrt{1-m_{\ell}^{2}}},\quad e^{\pm i\Theta_{\ell}}=\frac{%
1\pm m_{\ell}}{\sqrt{1-m_{\ell}^{2}}}
\end{equation*}
we find%
\begin{equation}
J_{12\cdots n}^{+}=\frac{\cos(\sum_{\ell=1}^{n}\Theta_{\ell})}{\prod_{\ell
=1}^{n}\cos\Theta_{\ell}},\;\;J_{12\cdots n}^{-}=i\frac{\sin(\sum_{%
\ell=1}^{n}\Theta_{\ell})}{\prod_{\ell=1}^{n}\cos\Theta_{\ell}}
\end{equation}
This immediately gives (\ref{solm}) by rewriting (\ref{soltheta}) in terms
of the variable $m$.

To derive $\lambda_{12\cdots n}$, we rewrite the recursion relation (\ref%
{lambda12}) in terms of a new variable $\tilde{\lambda}_{\ell}\equiv \frac{%
\lambda_{\ell}}{\cos\left( \Theta_{\ell}\right) }$, which simplifies the
relation
\begin{equation}
\tilde{\lambda}_{12}=e^{-i\Theta_{1}}\tilde{\lambda}_{2}+e^{i\Theta_{2}}%
\tilde{\lambda}_{1}.
\end{equation}
Then, one can derive $\lambda_{12\cdots n}$ from the simpler recursion
\begin{equation}
\tilde{\lambda}_{12\cdots n+1}=e^{i\Theta_{n+1}}\tilde{\lambda}_{12\cdots
n}+e^{-i\Theta_{12\cdots n}}\tilde{\lambda}_{n+1}
\end{equation}
The result is%
\begin{equation}
\tilde{\lambda}_{12\cdots n}=\sum_{\ell=1}^{n}\exp\left( -i\sum_{k<\ell
}\Theta_{k}+i\sum_{k>\ell}\Theta_{k}\right) \tilde{\lambda}_{\ell}.
\end{equation}
Coming back to the original variables $\lambda_{i}$, it is easy to see that
we arrive at (\ref{soll}). The derivation of $K_{12\cdots n}$ and $%
Q_{12\cdots n}$ is more complicated but can be done with similar arguments.
We give the angular variable version of these formulae:

\begin{align}
K_{12\cdots n} & =i\sum_{i=1}^{n}\bar{\tilde{\lambda}}_{i}\sigma(\tan
\Theta_{12\cdots n}-\tan\Theta_{i})\tilde{\lambda}_{i}-2\sum_{i<j}\bar {%
\tilde{\lambda}}_{i}\sigma\frac{e^{-i\sum_{k=i+1}^{j-1}\Theta_{k}+i%
\sum_{k=j+1}^{i-1}\Theta_{k}}}{\cos(\Theta_{12\cdots n})}\tilde{\lambda}_{j}
\\
iQ_{12\cdots n} & =\sum_{i=1}^{n}\bar{\tilde{\lambda}}_{i}\sigma(\cot
\Theta_{12\cdots n}+\tan\Theta_{i})\tilde{\lambda}_{i}+2\sum_{i<j}\bar {%
\tilde{\lambda}}_{i}\sigma\frac{e^{i\sum_{k=j+1}^{i-1}\Theta_{k}-i%
\sum_{k=i+1}^{j-1}\Theta_{k}}}{\sin(\Theta_{12\cdots n})}\tilde{\lambda}_{j}
\end{align}
To derive these formulae, we use the relation $\bar{m}\sigma=\left(
M\sigma\right) ^{T}\sigma=-\sigma M\sigma=-\sigma m$ for a symmetric $M$ and
antisymmetric $\sigma,$ and its extension to functions of $m$ as follows
\begin{equation}
(f(m)\lambda)^{T}\sigma=\bar{\lambda}\sigma f(-m),\quad\bar{\lambda}%
_{1}\sigma f(m)\lambda_{2}=-\bar{\lambda}_{2}\sigma f(-m)\lambda_{1}.
\end{equation}

Actually the angle variable $\Theta$ turns out to be more than a
computational device which simplifies the recursion formula. In section 6,
we will give an explicit formula of the three string Neumann coefficients in
terms of $m_{0}$. Through that relation, in the notation of \cite{-rRSZ2},
the spectrum of $m_{0}$ is identified as $\tanh\left( \frac{\pi\kappa}{4}%
\right) $ where $\kappa$ is the spectrum of $K_{1}\equiv L_{1}+L_{-1}$. If
we write $\Theta_{0}=\tan^{-1}(-im_{0})$, the spectrum of $\Theta_{0}$ is
identified with $\frac{-\pi i}{4}\kappa$. It implies that $\Theta_{0}$
equals to $\frac{-\pi i}{4} K_{1}$ up to a similarity transformation. We
note that $K_{1}$ is the basic matrix from where the essential properties of
the Neumann coefficients are derived in the infinite $N$ limit as well as in
the level truncation regularization.

\subsection{Products and traces with general $m_{i}$ and $\protect\lambda_{i}
$}

In certain computations in string field theory we anticipate also gaussians
with noncommuting $M_{i}\sigma.$ For example, this may occur when we would
like to compute products or correlators for string states in the presence of
different D-branes, such as those described by gaussians of the form (\ref{D}%
). In this section we analyze properties of such products.

\subsubsection{two points}

The product for two general generating functions is given in (\ref{a12}).
The 2-point vertex is given by its trace%
\begin{equation}
T_{12}\equiv Tr\left( A_{M_{1},\lambda_{1},\mathcal{N}_{1}}*
A_{M_{2},\lambda_{2},\mathcal{N}_{2}}\right) =\frac{\mathcal{N}_{12}\,e^{%
\frac{1}{4}\lambda_{12}M_{12}^{-1}\lambda_{12}}}{\left( \det2\sigma
M_{12}\right) ^{d/2}}
\end{equation}
This expression simplifies since the star can be dropped in evaluating the
integral. Then we obtain the relation which was shown in the Appendix
\begin{equation}
T_{12}=\frac{\mathcal{N}_{12}\,e^{\frac{1}{4}\bar{\lambda}%
_{12}M_{12}^{-1}\lambda_{12}}}{\left( \det2M_{12}\sigma\right) ^{d/2}}=%
\frac {\mathcal{N}_{1}\mathcal{N}_{2}\exp\left( \frac{1}{4}\left( \bar{%
\lambda }_{1}+\bar{\lambda}_{2}\right) \left( M_{1}+M_{2}\right) ^{-1}\left(
\lambda_{1}+\lambda_{2}\right) \right) }{\left( \det\left( 2\left(
M_{1}+M_{2}\right) \sigma\right) \right) ^{d/2}}   \label{t12}
\end{equation}
We now specialize to $M_{1}=M_{2}=M$ as in a previous section, but still
keep $\lambda_{1},\lambda_{2}$ arbitrary. The matrix $M$ represents some
vacuum state. This could be the perturbative vacuum $M_{0}$ given in Eq.(\ref%
{M0}) or a nonperturbative D-brane vacuum represented by a matrix $D$ as in (%
\ref{D}), for example the sliver vacuum $M_{s}$ as in Eq.(\ref{sM}). We also
use the $\mathcal{N}_{1},\mathcal{N}_{2}$ consistent with normalized states
for arbitrary $\lambda$'s, $Tr\left( A_{1}\right) ^{2}=1=Tr\left(
A_{2}\right) ^{2}$ as follows%
\begin{equation}
A_{1}=\left( \det4M\sigma\right) ^{d/4}e^{-\frac{1}{4}\lambda_{1}M^{-1}%
\lambda_{1}}\,e^{-\xi M\xi-\xi\lambda_{1}},\quad A_{2}=\left(
\det4M\sigma\right) ^{d/4}e^{-\frac{1}{4}\lambda_{2}M^{-1}\lambda_{2}}\,e^{-%
\xi M\xi-\xi\lambda_{2}}   \label{Anormalized}
\end{equation}
For these, the two point vertex becomes%
\begin{equation}
T_{12}^{\left( M\right) }=\exp\left( -\frac{1}{8}\left( \lambda
_{1}-\lambda_{2}\right) M^{-1}\left( \lambda_{1}-\lambda_{2}\right) \right)
.   \label{T12M}
\end{equation}

The center of mass mode may also be included. For example, for tachyon
waves, it takes the form%
\begin{equation}
e^{ik_{1}\cdot x_{0}}\mathcal{N}_{1}e^{-\xi
M\xi-\xi\lambda_{1}}=e^{ik_{1}\cdot\bar{x}}\mathcal{N}_{1}e^{-\xi
M\xi-\xi\lambda_{1}^{\prime}},\;\lambda_{1\mu}^{\prime}=\lambda_{1\mu}-ik_{1%
\mu}\left(
\begin{array}{c}
w \\
0%
\end{array}
\right)
\end{equation}
Therefore, $\lambda$ gets replaced by $\lambda^{\prime}$ in the previous
discussion which otherwise remains unchanged. In addition to the trace there
is also the integral $\int\frac{d^{d}\bar{x}}{\left( 2\pi\right) ^{d}}.$
This additional integration produces the Dirac delta function as an overall
factor%
\begin{equation}
Tr\left( \left( e^{ik_{1}\cdot x_{0}}A_{1}\right) \ast\left( e^{ik_{2}\cdot
x_{0}}A_{2}\right) \right) =\delta^{d}\left( k_{1}+k_{2}\right)
\,T_{12}^{\left( M\right) }
\end{equation}
where $T_{12}^{\left( M\right) }$ is given above.

\subsubsection{three points}

To compute the three point vertex, use associativity and the cyclicity of
the trace to find three different expressions%
\begin{align}
& T_{123}=Tr\left( A_{M_{1},\lambda_{1},\mathcal{N}_{1}}*
A_{M_{2},\lambda_{2},\mathcal{N}_{2}}* A_{M_{3},\lambda_{3},\mathcal{N}%
_{3}}\right) \\
& =Tr\left( \,A_{M_{12},\lambda_{12},\mathcal{N}_{12}}* A_{M_{3},\lambda
_{3},\mathcal{N}_{3}}\right) \\
& =Tr\left( \,A_{M_{23},\lambda_{23},\mathcal{N}_{23}}* A_{M_{1},\lambda
_{1},\mathcal{N}_{1}}\right) \\
& =Tr\left( \,A_{M_{31},\lambda_{31},\mathcal{N}_{31}}* A_{M_{2},\lambda
_{2},\mathcal{N}_{2}}\right)
\end{align}
and then use the result for the two point function, to write the three
expressions%
\begin{align}
& T_{123}=Tr\left( A_{M_{1},\lambda_{1},\mathcal{N}_{1}}*
A_{M_{2},\lambda_{2},\mathcal{N}_{2}}* A_{M_{3},\lambda_{3},\mathcal{N}%
_{3}}\right) \\
& =\frac{\mathcal{N}_{1}\mathcal{N}_{23}\exp\left( \frac{1}{4}\left(
\lambda_{1}+\lambda_{23}\right) \left( M_{1}+M_{23}\right) ^{-1}\left(
\lambda_{1}+\lambda_{23}\right) \right) }{\left( \det\left( 2\sigma\left(
M_{1}+M_{23}\right) \right) \right) ^{d/2}} \\
& =\frac{\mathcal{N}_{2}\mathcal{N}_{31}\exp\left( \frac{1}{4}\left(
\lambda_{2}+\lambda_{31}\right) \left( M_{2}+M_{31}\right) ^{-1}\left(
\lambda_{2}+\lambda_{31}\right) \right) }{\left( \det\left( 2\sigma\left(
M_{2}+M_{31}\right) \right) \right) ^{d/2}} \\
& =\frac{\mathcal{N}_{3}\mathcal{N}_{12}\exp\left( \frac{1}{4}\left(
\lambda_{3}+\lambda_{12}\right) \left( M_{3}+M_{12}\right) ^{-1}\left(
\lambda_{3}+\lambda_{12}\right) \right) }{\left( \det\left( 2\sigma\left(
M_{3}+M_{12}\right) \right) \right) ^{d/2}}
\end{align}
Each form makes explicit the dependence on the parameters of strings 1,2,3
respectively.

The product for three generating functions may also be evaluated as follows%
\begin{align}
A_{M_{1},\lambda_{1},\mathcal{N}_{1}}* A_{M_{2},\lambda_{2},\mathcal{N}%
_{2}}* A_{M_{3},\lambda_{3},\mathcal{N}_{3}} & =A_{M_{12},\lambda_{12},%
\mathcal{N}_{12}}* A_{M_{3},\lambda_{3},\mathcal{N}_{3}} \\
& =A_{M_{1},\lambda_{1},\mathcal{N}_{1}}* A_{M_{23},\lambda_{23},\mathcal{N}%
_{23}} \\
& =A_{M_{123},\lambda_{123},\mathcal{N}_{123}}
\end{align}
with two different, but equivalent (dual), expressions for each quantity $%
M_{123},\lambda_{123},\mathcal{N}_{123}$
\begin{align}
M_{123} & =\left( M_{12}+M_{3}\sigma M_{12}\right) \left( 1+\sigma
M_{3}\sigma M_{12}\right) ^{-1}+\left( M_{3}-M_{12}\sigma M_{3}\right)
\left( 1+\sigma M_{12}\sigma M_{3}\right) ^{-1} \\
& =\left( M_{1}+M_{23}\sigma M_{1}\right) \left( 1+\sigma M_{23}\sigma
M_{1}\right) ^{-1}+\left( M_{23}-M_{1}\sigma M_{23}\right) \left( 1+\sigma
M_{1}\sigma M_{23}\right) ^{-1} \\
\lambda_{123}^{\mu} & =\left( 1+M_{3}\sigma\right) \left( 1+M_{12}\sigma
M_{3}\sigma\right) ^{-1}\lambda_{12}^{\mu}+\left( 1-M_{12}\sigma\right)
\left( 1+M_{3}\sigma M_{12}\sigma\right) ^{-1}\lambda_{3}^{\mu} \\
& =\left( 1+M_{23}\sigma\right) \left( 1+M_{1}\sigma M_{23}\sigma\right)
^{-1}\lambda_{1}^{\mu}+\left( 1-M_{1}\sigma\right) \left( 1+M_{23}\sigma
M_{1}\sigma\right) ^{-1}\lambda_{23}^{\mu}
\end{align}
and%
\begin{align}
\mathcal{N}_{123} & =\mathcal{N}_{12}\mathcal{N}_{3}\left( \frac {\det\left(
2M_{123}\sigma\right) }{\det\left( 2\left( M_{12}+M_{3}\right) \sigma\right)
}\right) ^{d/2}\,e^{\frac{1}{4}\left( \lambda_{12}+\lambda_{3}\right) \left(
M_{12}+M_{3}\right) ^{-1}\left( \lambda_{12}+\lambda_{3}\right) -\frac{1}{4}%
\left( \lambda_{123}\right) ^{T}\left( M_{123}\right) ^{-1}\lambda_{123}} \\
& =\mathcal{N}_{1}\mathcal{N}_{23}\left( \frac{\det\left(
2M_{123}\sigma\right) }{\det\left( 2\left( M_{1}+M_{23}\right) \sigma\right)
}\right) ^{d/2}\,e^{\frac{1}{4}\left( \lambda_{1}+\lambda_{23}\right) \left(
M_{1}+M_{23}\right) ^{-1}\left( \lambda_{1}+\lambda_{23}\right) -\frac{1}{4}%
\left( \lambda_{123}\right) ^{T}\left( M_{123}\right) ^{-1}\lambda_{123}}
\end{align}
Furthermore, we must have the three point vertex%
\begin{equation}
T_{123}=\frac{\mathcal{N}_{123}\,e^{\frac{1}{4}\lambda_{123}M_{123}^{-1}%
\lambda_{123}}}{\left( \det\left( 2M_{123}\sigma\right) \right) ^{d/2}}
\end{equation}
which must be equal to the\thinspace expressions for\thinspace the\thinspace
3-point \thinspace vertex given above. This gives\thinspace a \thinspace lot
of \thinspace identities, in particular many relations are obtained by
comparing the\thinspace quadratics\thinspace in various $\lambda$'s in the
exponent.

The expressions simplify, by inserting $M_{1}=M_{2}=M_{3}=M,$ but keeping
the $\lambda$'s different as in previous sections. When $M=M_{0}$ the
expressions are appropriate for computing the perturbative 3-point function.
When $M=M_{s}$ (the sliver field) it will be appropriate for computing
non-perturbative 3-point vertex, etc.

\subsubsection{$n$-points}

We can compute the 4-point vertex by using associativity and the cyclic
property of the trace, to obtain the following forms%
\begin{align}
& Tr\left( A_{M_{1},\lambda_{1},\mathcal{N}_{1}}* A_{M_{2},\lambda _{2},%
\mathcal{N}_{2}}* A_{M_{3},\lambda_{3},\mathcal{N}_{3}}*
A_{M_{4},\lambda_{4},\mathcal{N}_{4}}\right) \\
& =\frac{\mathcal{N}_{12}\mathcal{N}_{34}\exp\left( \frac{1}{4}\left(
\lambda_{12}+\lambda_{34}\right) \left( M_{12}+M_{34}\right) ^{-1}\left(
\lambda_{12}+\lambda_{34}\right) \right) }{\left( \det\left( 2\left(
M_{12}+M_{34}\right) \sigma\right) \right) ^{d/2}} \\
& =\frac{\mathcal{N}_{23}\mathcal{N}_{41}\exp\left( \frac{1}{4}\left(
\lambda_{23}+\lambda_{41}\right) \left( M_{23}+M_{41}\right) ^{-1}\left(
\lambda_{23}+\lambda_{41}\right) \right) }{\left( \det\left( 2\left(
M_{23}+M_{41}\right) \sigma\right) \right) ^{d/2}} \\
& =\frac{\mathcal{N}_{123}\mathcal{N}_{4}\exp\left( \frac{1}{4}\left(
\lambda_{123}+\lambda_{4}\right) \left( M_{123}+M_{4}\right) ^{-1}\left(
\lambda_{123}+\lambda_{4}\right) \right) }{\left( \det\left( 2\left(
M_{123}+M_{4}\right) \sigma\right) \right) ^{d/2}} \\
& =cyclic\quad permutations
\end{align}
These can be further computed by inserting the formulas for $\mathcal{N}%
_{ij},\lambda_{ij},M_{ij}$ given in the previous section. These two forms of
the 4-point vertex is compatible with duality (the original duality of the
Veneziano amplitude).

Similarly, the 5-point vertex is computed in several forms by using the star
product and the result for the 3-point vertex%
\begin{align}
& Tr\left( A_{M_{1},\lambda_{1},\mathcal{N}_{1}}* A_{M_{2},\lambda _{2},%
\mathcal{N}_{2}}* A_{M_{3},\lambda_{3},\mathcal{N}_{3}}*
A_{M_{4},\lambda_{4},\mathcal{N}_{4}}* A_{M_{5},\lambda_{5},\mathcal{N}%
_{5}}\right) \\
& =\frac{\mathcal{N}_{123}\mathcal{N}_{45}\exp\left( \frac{1}{4}\left(
\lambda_{123}+\lambda_{45}\right) \left( M_{123}+M_{45}\right) ^{-1}\left(
\lambda_{123}+\lambda_{45}\right) \right) }{\left( \det\left( 2\left(
M_{123}+M_{45}\right) \sigma\right) \right) ^{d/2}} \\
& =\frac{\mathcal{N}_{1234}\mathcal{N}_{5}\exp\left( \frac{1}{4}\left(
\lambda_{1234}+\lambda_{5}\right) \left( M_{1234}+M_{5}\right) ^{-1}\left(
\lambda_{1234}+\lambda_{5}\right) \right) }{\left( \det\left( 2\left(
M_{1234}+M_{5}\right) \sigma\right) \right) ^{d/2}} \\
& =more\,\,\,forms\,\,by\,\,cyclic\,\,permutations\,\,of\,\,1,2,3,4,5
\end{align}
The process is similar for $n$ points. These forms are compatible with
duality of the $n$-point function.

As before these expressions reduce to the computations in the previous
sections when we take $M_{1}=\cdots=M_{n}=M,$ but still keep the $\lambda$'s
different.

\section{Neumann coefficients}

As we have seen, the MSFT formulation gives a simple mathematical framework
to calculate all string vertices of the open string field theory. In this
section we apply our results to derive a new and simpler expression for the
Neumann coefficients for all string vertices in the oscillator formalism. We
will do this for any oscillator frequencies $\kappa_{e},\kappa_{o}$ and any
number of oscillators. In this process we also establish a more explicit
connection between our results and the oscillator approach. One of the
purposes of this computation is to derive an explicit regularized formula
for physical quantities, such as the brane tension, tachyon mass and so on,
in terms of our regularization scheme.

\subsection{Computation of Neumann coefficients in MSFT}

In the operator approach to the open string field theory \cite{GJ}, the $n$%
-vertex is written in terms of the open string oscillator in the form,
\begin{equation}
\langle V_{n}|=\langle p|\exp\sum_{r,s=1}^{n}\left( {\frac{1}{2}}%
\sum_{k,l\geq1}\frac{\alpha_{k}^{(r)}}{\sqrt{\kappa_{k}}}\left( V_{n}^{\left[
rs\right] }\right) _{kl}\frac{\alpha_{l}^{(s)}}{\sqrt {\kappa_{l}}}%
+\sum_{k\geq1}\frac{\alpha_{k}^{(r)}}{\sqrt{\kappa_{k}}}\left( V_{n}^{\left[
rs\right] }\right) _{k0}p^{(s)}+\frac{1}{2}p^{(r)}\left( V_{n}^{\left[ rs%
\right] }\right) _{00}p^{(s)}\right) ,
\end{equation}
times a momentum conservation delta function $\left( 2\pi\right)
^{d}\delta\left( \sum_{r}p_{\mu}^{\left( r\right) }\right) ,$ where we have
taken a finite number of modes $N$ and inserted arbitrary frequencies $%
\kappa_{k}.$ The square root factors in the exponential are added because
our normalization of the oscillators is different from \cite{GJ}. We will
compute the coefficients $\left( V_{n}^{[rs]}\right) _{kl}$ for any $\kappa
_{e},\kappa_{o},N$ by using the methods of the Moyal star product. Our
results provide a simple expression for these Neumann coefficients. We
mention that other closed expression for Neumann coefficients have been
given in the recent literature \cite{furuuchiokuyama}.

The $n$-point off-shell amplitudes are written in terms of $\langle V_{n}|$
as
\begin{equation}
\langle V_{n}||\Psi_{1}\rangle_{1}\otimes\cdots\otimes|\Psi_{n}\rangle
_{n}\equiv\int\Psi_{1}* \cdots* \Psi_{n}\,\,,
\end{equation}
for the $n$ elements $|\Psi\rangle$ in the Hilbert space. In \cite{GJ}, $%
\langle V_{n}|$ is uniquely determined from the overlap conditions (here we
write them in terms of the split string variables) for the star product,
\begin{equation}
(r_{2n}\Psi_{1})\star\Psi_{2}=\Psi_{1}\star(l_{2n}\Psi_{2})\,\,,\quad\left(
\frac{\partial}{\partial r_{2n}}\Psi_{1}\right) \star\Psi_{2}
=-\Psi_{1}\star\left( \frac{\partial}{\partial l_{2n}}\Psi_{2}\right) \,\,.
\end{equation}
In the Moyal formalism, it can be verified that these condition reduce to
the associativity of the Moyal star product,
\begin{equation}
\left( A_{1}\ast x_{2n}\right) \ast A_{2}=A_{1}\ast\left( x_{2n}\ast
A_{2}\right) \,\,,\qquad\left( A_{1}\ast p_{2n}\right) \ast
A_{2}=A_{1}\ast\left( p_{2n}\ast A_{2}\right) \,\,.
\end{equation}
In this sense, the equivalence between the two formalism is by definition
ensured, and up to an overall constant we should have the relation
\begin{equation}
\langle V_{n}||\Psi_{1}\rangle_{1}\otimes\cdots\otimes|\Psi_{N}\rangle
_{N}\;\sim\,\;\int d\bar{x}\,Tr\left( A_{1}\ast\cdots\ast A_{N}\right) .
\label{equiv}
\end{equation}
We will use this correspondence to compute the Neumann coefficients $\left(
V_{n}\right) _{kl}^{[rs]},$ $\left( V_{n}^{\left[ rs\right] }\right) _{k0},$
$\left( V_{n}^{\left[ rs\right] }\right) _{00}$ for any $\kappa_{e},%
\kappa_{o},N$ in terms of the expressions we obtained in the previous
sections.

In the following computation, we take $\Psi_{i}$ as the coherent states
\begin{equation}
|\Psi_{i}\rangle=\exp\left( \sum_{n=1}^{\infty}\mu_{n}^{(i)}\left(
a_{n}^{(i)}\right) ^{\dagger}\right) |p^{(i)}\rangle=\exp\left(
ip^{(i)}x_{0}+\sum_{n=1}^{\infty}\kappa_{n}^{-1/2}\mu_{n}^{(i)}\alpha
_{-n}^{(i)}\right) |0\rangle.
\end{equation}
$\langle V_{n}||\Psi_{1}\rangle_{1}\otimes\cdots\otimes|\Psi_{n}\rangle_{n}$
is easily computed using the property $\alpha_{n}^{\left( i\right)
}|\Psi_{i}\rangle=\sqrt{\kappa_{n}}\mu_{n}^{(i)}|\Psi_{i}\rangle$ of
coherent states
\begin{equation}
\exp\sum_{r,s=1}^{n}\left( {\frac{1}{2}}\sum_{k.l\geq1}\mu_{k}^{(r)}\left(
V_{n}^{\left[ rs\right] }\right) _{kl}\mu_{l}^{(s)}+\sum_{k\geq1}\mu
_{k}^{(r)}\left( V_{n}^{\left[ rs\right] }\right) _{k0}p^{(s)}+\frac{1}{2}%
p^{(r)}\left( V_{n}^{\left[ rs\right] }\right) _{00}p^{(s)}\right)
\label{expV}
\end{equation}
This gives enough information since the factors of $\mu_{k}^{(r)}$ or $%
p^{(r)}$ in the exponent identify the Neumann coefficients $%
V_{kl}^{[rs]},V_{k0}^{[rs]},V_{00}^{[rs]}$ .

To perform the equivalent Moyal computation, we use the $p$-basis given in (%
\ref{p}) to obtain the field $A_{r}\left( \bar{x},x_{e},p_{e}\right) =\langle%
\bar{x},x_{e},p_{e}|\Psi_{r}\rangle$ for the coherent state
\begin{equation}
A_{r}\left( \bar{x},x_{e},p_{e}\right) =\left( \frac{\det4\kappa_{o}}{%
\det\kappa_{e}}\right) ^{d/4}e^{ip^{\left( r\right) }\bar{x}}~e^{\frac {1}{2}%
\left( \bar{\mu}_{e}^{\left( r\right) }\mu_{e}^{\left( r\right) }-\bar{\mu}%
_{o}^{\left( r\right) }\mu_{o}^{\left( r\right) }\right) }\exp\left( -\bar{%
\xi}M_{0}\xi-\bar{\xi}\lambda^{\left( r\right) }\right)
\end{equation}
where we have used $\langle\bar{x}|p^{\left( r\right) }\rangle=\exp\left( i%
\bar{x}\cdot p^{\left( r\right) }\right) $. Then, as in Eq.(\ref{lambda}) $%
\lambda^{\left( r\right) }$ is given by
\begin{equation}
\lambda^{(r)}\left( \mu,p\right) =\left(
\begin{array}{c}
-\frac{\sqrt{2}i}{l_{s}}\sqrt{\kappa_{e}}\mu_{e}^{\left( r\right)
}-iw_{e}p^{\left( r\right) } \\
-\frac{2\sqrt{2}l_{s}}{\theta}\sum_{o>0}T_{e,o}\kappa_{o}^{-1/2}\mu
_{o}^{\left( r\right) }%
\end{array}
\right) =2K\left( \mu^{\left( r\right) }+Wp^{\left( r\right) }\right) ,\;\;
\label{lamr}
\end{equation}
where in the right hand side we have defined $K,W$%
\begin{equation}
K\equiv\left(
\begin{array}{cc}
\frac{-i}{l_{s}}\sqrt{\frac{\kappa_{e}}{2}} & 0 \\
0 & \frac{-l_{s}}{\theta}T\sqrt{\frac{2}{\kappa_{o}}}%
\end{array}
\right) \,\,,\;W\equiv\left(
\begin{array}{c}
\frac{l_{s}}{\sqrt{2\kappa_{e}}}w \\
0%
\end{array}
\right)   \label{K}
\end{equation}

This $A_{r}(x,p)$ has the standard form of the monoid elements. Therefore,
the right hand side of Eq.(\ref{equiv}) is easily computed through Eq.(\ref%
{tran})
\begin{equation}
Tr\left( A_{1}\ast\cdots\ast A_{n}\right) =\frac{\exp\left[
\sum_{r=1}^{n}\left( ip^{\left( r\right) }\bar{x}+\frac{1}{2}\bar{\mu}%
_{e}^{\left( r\right) }\mu_{e}^{\left( r\right) }-\frac{1}{2}\bar{\mu}%
_{o}^{\left( r\right) }\mu_{o}^{\left( r\right) }\right) +\frac{1}{4}%
Q_{n}\left( \lambda\left( \mu,p\right) \right) \right] }{\det\left(
(1+m_{0})^{n}-(1-m_{0})^{n}\right) ^{d/2}}   \label{expQ}
\end{equation}
where $Q_{n}$ was computed in Eq.(\ref{tran})
\begin{equation}
Q_{n}\left( \lambda\right) =\sum_{r,s=1}^{n}\bar{\lambda}_{r}\sigma \mathcal{%
O}_{\left( s-r\right) \func{mod}n}\left( m_{0}\right) \lambda_{s}
\label{qn}
\end{equation}
but now $\lambda_{r}\left( \mu,p\right) $ is replaced by Eq.(\ref{lamr}).
The $\mathcal{O}_{\left( s-r\right) \func{mod}n}$ were given explicitly in
Eq.(\ref{On}). All together $\int d\bar{x}Tr\left( A_{1}\ast\cdots\ast
A_{n}\right) $ has an overall momentum conservation delta function $\left(
2\pi\right) ^{d}\delta\left( \sum_{r}p_{\mu}^{\left( r\right) }\right) ,$
times a factor of the form Eq.(\ref{expV}). Matching the exponents in (\ref%
{equiv},\ref{expV},\ref{expQ}), we see from the structure of Eq.(\ref{qn})
that the Neumann coefficients $V_{n}^{\left[ r,s\right] }$ must depend only
on the difference $\left( s-r\right) \func{mod}n.$ Therefore we define the
matrix $\mathcal{M}_{kl}^{i}$, vector $\mathcal{V}_{k}^{i}$ and scalar $%
\mathcal{C}^{i}$ for $i\func{mod}n$ as follows%
\begin{equation}
\left( \left( V_{n}^{\left[ r,s\right] }\right) _{kl},\left( V_{n}^{\left[
r,s\right] }\right) _{k0},\left( V_{n}^{\left[ r,s\right] }\right)
_{00}\right) \equiv-\left( \left( C\mathcal{M}_{\left( s-r\right) }\right)
_{kl},\left( \mathcal{V}_{\left( s-r\right) }\right) _{k},\mathcal{C}%
_{\left( s-r\right) }\right) ,   \label{fullNeum}
\end{equation}
where $C_{kl}=\left( -1\right) ^{k}\delta_{kl}$, and from our explicitly
computation in Eq.(\ref{expQ}) we obtain
\begin{align}
\mathcal{M}_{i} & =2\tilde{m}_{0}\mathcal{O}_{i}(\tilde{m}_{0})-\delta
_{i,0},\;  \label{neum} \\
\mathcal{V}_{i} & =\left( -2\tilde{m}_{0}\mathcal{O}_{i}(-\tilde{m}_{0})-%
\frac{2}{n}\right) W,  \label{neum2} \\
\mathcal{C}_{i} & =\bar{W}\left( 2\tilde{m}_{0}\mathcal{O}_{i}(\tilde {m}%
_{0})-\frac{2}{n}\right) W.   \label{neum3}
\end{align}
with the functions $\mathcal{O}_{i}(m)$ defined in (\ref{On},\ref{On1}).
Actually a naive comparison of the zero mode coefficient gives only the
first terms in Eqs.(\ref{neum2},\ref{neum3}). However from the momentum
conservation of the $n$-vertex, we have some arbitrariness in choosing them
up to the translations $\left( \mathcal{V}_{i}\right) _{k}\rightarrow\left(
\mathcal{V}_{i}\right) _{k}+q_{k}$, and $\mathcal{C}_{i}\rightarrow \mathcal{%
C}_{i}$ $+c$, for any constants $q_{k}$ and $c$. We have used this freedom
to ensure,
\begin{equation}
\sum_{i}\mathcal{V}_{i}=0,\qquad\sum_{i}\mathcal{C}_{^{i}}=0\,\,.
\end{equation}
by using the identity $\tilde{m}_{0}\sum_{i}\mathcal{O}_{i}\left( \tilde
{m}%
_{0}\right) =1.$ We emphasize that this compact form depends only on the
matrix $\tilde{m}_{0}$ which is described below.

Note that initially the $\mathcal{O}_{i}\left( m_{0}\right) $ in Eqs.(\ref%
{qn}) are functions of $m_{0}=M_{0}\sigma$ for the vacuum state Eq.(\ref{M0}%
), not $\tilde{m}_{0}.$ To arrive at the above forms we have used $\bar{K}%
\sigma=-CK^{-1}m_{0},$ and performed the similarity transformation $K^{-1}%
\mathcal{O}_{i}\left( m_{0}\right) K$ which resulted in the above
expressions for $\mathcal{O}_{i}\left( \tilde{m}_{0}\right) $ written in
terms of $\tilde{m}_{0}$
\begin{equation}
\tilde{m}_{0}=K^{-1}m_{0}K=K^{-1}\left( M_{0}\sigma\right) K.
\end{equation}
Using our expressions for $K,M_{0}$ in Eqs.(\ref{K},\ref{M0}), $\tilde{m}_{0}
$ and $\tilde{m}_{0}^{2}$ take the following more explicit forms in terms of
$\kappa_{e}^{1/2}T\kappa_{o}^{-1/2}$ or its diagonalized version $V^{e}\tau%
\bar{V}^{o}$ given in Eq.(\ref{ktk}),
\begin{align}
\tilde{m}_{0} & =\left(
\begin{array}{cc}
0 & \kappa_{e}^{1/2}T\kappa_{o}^{-1/2} \\
\kappa_{o}^{-1/2}\bar{T}\kappa_{e}^{1/2} & 0%
\end{array}
\right) =\left(
\begin{array}{cc}
V^{e} & 0 \\
0 & V^{o}%
\end{array}
\right) \left(
\begin{array}{cc}
0 & \tau \\
\tau & 0%
\end{array}
\right) \left(
\begin{array}{cc}
\bar{V}^{e} & 0 \\
0 & \bar{V}^{o}%
\end{array}
\right)  \label{mtildezero} \\
\tilde{m}_{0}^{2} & =\left(
\begin{array}{cc}
\kappa_{e}^{1/2}T\kappa_{o}^{-1}\bar{T}\kappa_{e}^{1/2} & 0 \\
0 & \kappa_{o}^{-1/2}\bar{T}\kappa_{e}T\kappa_{o}^{-1/2}%
\end{array}
\right) =\left(
\begin{array}{cc}
V^{\left( e\right) } & 0 \\
0 & V^{\left( o\right) }%
\end{array}
\right) \left(
\begin{array}{cc}
\tau^{2} & 0 \\
0 & \tau^{2}%
\end{array}
\right) \left(
\begin{array}{cc}
\bar{V}^{\left( e\right) } & 0 \\
0 & \bar{V}^{\left( o\right) }%
\end{array}
\right)   \label{msquare}
\end{align}
Recalling that $T_{eo}$ is determined in Eq.(\ref{TRcut}), $%
\kappa_{e}^{1/2}T_{eo}\kappa_{o}^{-1/2}=\kappa_{e}^{1/2}w_{e}v_{o}%
\kappa_{o}^{3/2}\left( \kappa_{e}^{2}-\kappa_{o}^{2}\right) ^{-1},$ we see
that we have explicitly computed in Eq.(\ref{neum}) the $2N\times2N$
regularized Neumann coefficients for any $\kappa_{e},\kappa_{o},N.$
Furthermore, the diagonal forms of $\tilde{m}_{0},\tilde{m}_{0}^{2}$ in Eqs.(%
\ref{mtildezero},\ref{msms}) give the spectroscopy for Neumann coefficients
for all string vertices. In the large $N$ limit $\kappa_{e}^{1/2}T_{eo}%
\kappa_{o}^{-1/2}$ is given in Eqs.(\ref{Tinf}-\ref{kinf}), or $%
\tau,V^{e},V^{o}$ are given in Eq.(\ref{tau}), and therefore all matrix
elements of $\tilde{m}_{0},$ and hence all Neumann coefficients, are fully
determined.

In \cite{-rRSZ2} the spectroscopy of Neumann matrices $\mathcal{M}_{i\func{%
mod}3}$ for the 3-point vertex was computed. We may compare our spectroscopy
for $n=3$ and infinite $N$ to their results by using the eigenvalues $%
\tau\left( k\right) =\tanh\left( \pi k/4\right) $ explained in Eq.(\ref{tau}%
), and find full agreement. This is seen in the more explicit expressions
for $n=3$ given below in Eqs.(\ref{M00}). This confirmation provides
confidence that our formulas correctly give all Neumann coefficients
consistently for all $n$-point string vertices either in the finite $N$ or
the infinite $N$ theory.

As in \cite{GJ}, we obtain a simpler expression if we perform the following
discrete Fourier transformation for the oscillators,
\begin{equation}
\left( \tilde{\alpha}_{J}\right) _{k}=\frac{1}{\sqrt{n}}\sum_{i=1}^{n}%
\omega^{J(i-1)}\left( \alpha_{i}\right) _{k}\,\,,\quad\omega=e^{2\pi
i/n}\,\,,\quad J\in\left\{ 0,1,2,\cdots,n-1\right\} =\mathbf{Z}_{n}\,\,.
\end{equation}
With respect to this combination, the overlap conditions become diagonal. We
use the similar recombination for the source, $\left( \tilde{\mu}_{J}\right)
_{k}=\frac{1}{\sqrt{n}}\sum_{r}\omega^{J(r-1)}\left( \mu_{r}\right) _{k}$.
In terms of this variables, the Neumann function is transformed to,%
\begin{equation}
\left( V_{n}^{\left[ I,J\right] }\right) =\frac{1}{n}\sum_{I,J\in \mathbf{Z}%
_{n}}\omega^{-I(r-1)-J(s-1)}\left( V_{n}^{[r,s]}\right)
\end{equation}
In the Moyal basis, the discrete Fourier transformation gives,
\begin{equation*}
\sum_{r,k}\bar{\lambda}_{r}\sigma\mathcal{O}_{k}\lambda_{k+r}=\frac{1}{n}%
\sum_{r,k}\sum_{I,J}\omega^{-(I+J)(r-1)-Jk}\overline{\tilde{\lambda}^{(I)}}%
\sigma O_{k}\tilde{\lambda}^{(J)}=\sum_{I\in\mathbf{Z}_{n}}\overline {\tilde{%
\lambda}^{(I)}}\sigma\tilde{\mathcal{O}}_{I}\tilde{\lambda}^{(-I)}
\end{equation*}
where%
\begin{equation}
\tilde{\mathcal{O}}_{I}\left( \tilde{m}_{0}\right) \equiv\sum_{r}\omega ^{Ir}%
\mathcal{O}_{r}\left( \tilde{m}_{0}\right) =\frac{1+\omega^{I}}{(1+\tilde{m}%
_{0})-(1-\tilde{m}_{0})\omega^{I}}=\frac{1}{\tilde{m}_{0}-i\tan(\frac{\pi I}{%
n})}\,\,.
\end{equation}
Then the Neumann coefficients take a much simpler form because $\left(
V_{n}^{\left[ I,J\right] }\right) $ become diagonal (proportional to $%
\delta_{I+J}$). Therefore we define $\tilde{\mathcal{M}}_{kl}^{I},\mathcal{%
\tilde{V}}_{k}^{I},\mathcal{\tilde{C}}^{I}$
\begin{equation}
\left( \left( V_{n}^{[I,J]}\right) _{kl},\left( V_{n}^{[I,J]}\right)
_{k0},\left( V_{n}^{[I,J]}\right) _{00}\right) \equiv-\left( \left( C%
\mathcal{\tilde{M}}_{I}\right) _{kl},\left( \mathcal{\tilde{V}}_{I}\right)
_{k},\mathcal{\tilde{C}}_{I}\right) \delta_{I+J}\,\,
\end{equation}
where we obtain for $I=0,1,\cdots,n-1$%
\begin{align}
\tilde{\mathcal{M}}_{I}\left( \tilde{m}_{0}\right) & =\frac{2\tilde{m}_{0}}{%
\tilde{m}_{0}-i\tan(\frac{\pi I}{n})}-\delta_{I,0},\;\;  \label{neuman} \\
\mathcal{\tilde{V}}_{I}\left( \tilde{m}_{0}\right) & =\left( \frac{2\tilde{m}%
_{0}}{\tilde{m}_{0}+i\tan(\frac{\pi I}{n})}-2\delta _{I,0}\right) W,\;\; \\
\mathcal{\tilde{C}}_{I}\left( \tilde{m}_{0}\right) & \mathfrak{=}\bar {W}%
\left( \frac{2\tilde{m}_{0}}{\tilde{m}_{0}-i\tan(\frac{\pi I}{n})}%
-2\delta_{I,0}\right) W
\end{align}
As already argued above this is an explicit form of the Neumann coefficients
for all string vertices. They depend on a single matrix $\tilde{m}_{0}$
which we have determined in either the finite $N$ or infinite $N$ theory.

The $\tilde{\mathcal{M}}_{I}$ can also be rewritten in terms of the
eigenvalues $\tau_{k}$ and the orthogonal matrices $V_{ek}^{\left( e\right)
},V_{ok}^{\left( o\right) }$ by using the second form of $\tilde{m}_{0}$ in
Eq.(\ref{mtildezero}) and inserting it in Eq.(\ref{neuman})
\begin{equation}
\tilde{\mathcal{M}}_{I}=\left(
\begin{array}{cc}
V^{\left( e\right) } & 0 \\
0 & V^{\left( o\right) }%
\end{array}
\right) \left(
\begin{array}{cc}
\frac{2\tau^{2}}{\tau^{2}+\tan^{2}\left( \frac{\pi I}{N}\right) }-\delta_{I0}
& \frac{i\theta}{l_{s}^{2}}\frac{\tau\tan(\frac{\pi I}{N})}{%
\tau^{2}+\tan^{2}\left( \frac{\pi I}{N}\right) } \\
\frac{l_{s}^{2}}{i\theta}\frac{\tau\tan(\frac{\pi I}{N})}{\tau^{2}+\tan
^{2}\left( \frac{\pi I}{N}\right) } & \frac{2\tau^{2}}{\tau^{2}+\tan
^{2}\left( \frac{\pi I}{N}\right) }-\delta_{I0}%
\end{array}
\right) \left(
\begin{array}{cc}
\bar{V}^{\left( e\right) } & 0 \\
0 & \bar{V}^{\left( o\right) }%
\end{array}
\right)
\end{equation}
and similarly for $\mathcal{\tilde{V}}_{I}\left( \tilde{m}_{0}\right) $ and $%
\mathcal{\tilde{C}}_{I}\left( \tilde{m}_{0}\right) .$ If we further insert
the perturbative frequencies $\kappa_{e}=e,$ $\kappa_{o}=o,$ and an infinite
number of oscillators, we obtain our results in the continuous Moyal basis
given by the eigenvalues $\tau\left( k\right) =\tanh\left( \pi k/4\right) $
and the functions $V_{e}\left( k\right) $, $V_{o}\left( k\right) $ given in
\cite{DLMZ}.

\subsection{Properties of Neumann coefficients}

>From these expressions, we may observe the following properties of the
Neumann matrices. These are standard in the literature in the case of the
large $N$ theory, but in our case they hold for any $\kappa_{e},\kappa_{o},N,
$ which seems remarkable.

\begin{itemize}
\item We note that, for any $n$-vertex, the Neumann matrices $\mathcal{%
\tilde
{M}}^{I}$ or $\mathcal{M}^{i}$ are written in terms of the $%
2N\times2N$ matrix $\tilde{m}_{0}$ in Eq.(\ref{mtildezero}). This
automatically implies that they commute for any $I,J\func{mod}n$ or any $i,j%
\func{mod}n$
\begin{equation}
\left[ \mathcal{\tilde{M}}_{I},\mathcal{\tilde{M}}_{J}\right] =0,\;\left[
\mathcal{\tilde{M}}_{I},\mathcal{M}_{j}\right] =0,\;\left[ \mathcal{M}_{i},%
\mathcal{M}_{j}\right] =0.
\end{equation}

\item The $\mathcal{O}_{k}(\tilde{m}_{0})$ satisfy the following nontrivial
identities,
\begin{equation}
2\tilde{m}_{0}\sum_{t=0}^{n-1}\mathcal{O}_{t}(-\tilde{m}_{0})\mathcal{O}%
_{s-t}(\tilde{m}_{0})+\mathcal{O}_{s}(\tilde{m}_{0})-\mathcal{O}_{s}(-\tilde{%
m}_{0})=0\,\,,
\end{equation}
These translate to the following relations among the Neumann coefficients
\begin{equation}
\sum_{t=1}^{n}\sum_{b=1}^{N}V_{ab}^{[rt]}V_{bc}^{[ts]}=\delta_{r,s}%
\delta_{a,c},\quad\sum_{t=1}^{n}%
\sum_{b=1}^{N}V_{ab}^{[rt]}V_{b0}^{[ts]}=V_{a0}^{[rs]},\quad\sum_{t=1}^{n}%
\sum_{b=1}^{N}V_{0b}^{[rt]}V_{b0}^{[ts]}=2V_{00}^{[rs]}\,\,.
\end{equation}

\item For $I=0$, the matrix $\tilde{\mathcal{M}}_{0}$ becomes particularly
simple for any $n$-vertex, $\tilde{\mathcal{M}}_{0}=1$.

\item For $n=3$, the Neumann coefficients $\mathcal{M}_{i},\mathcal{V}_{i},%
\mathcal{C}_{i}$ of Eq.(\ref{neum}) become (using the notation $\left(
-1,0,1\right) \func{mod}3=\left( 2,0,1\right) $)%
\begin{align}
& \mathcal{M}_{0}=\frac{\tilde{m}_{0}^{2}-1}{\tilde{m}_{0}^{2}+3},\,\;%
\mathcal{M}_{+}=2\frac{1+\tilde{m}_{0}}{\tilde{m}_{0}^{2}+3}\,,\,\;\mathcal{M%
}_{-}=2\frac{1-\tilde{m}_{0}}{\tilde{m}^{2}+3},  \label{M00} \\
& \mathcal{V}_{0}=\frac{4\tilde{m}_{0}^{2}}{3(3+\tilde{m}_{0}^{2})}W,\quad%
\mathcal{V}_{+}=-\frac{2\tilde{m}_{0}(3+\tilde{m}_{0})}{3(3+\tilde {m}%
_{0}^{2})}W,\;\;\mathcal{V}_{-}=\frac{2\tilde{m}_{0}(3-\tilde{m}_{0})}{3(3+%
\tilde{m}_{0}^{2})}W,  \label{V0} \\
& \mathcal{C}_{0}=-2\mathcal{C}_{+}=-2\mathcal{C}_{-}=\bar{W}\frac{4\tilde {m%
}_{0}^{2}}{3(3+\tilde{m}_{0}^{2})}W\equiv\frac{2}{3}V_{00}   \label{C0}
\end{align}
It is also convenient to define the following combinations that appeared in
the literature, which have even or odd powers of $\tilde{m}_{0}$ (there are
called twist even/odd in the literature)
\begin{align}
\mathcal{M}_{even} & =\mathcal{M}_{+}+\mathcal{M}_{-}=\frac{4}{\tilde{m}%
_{0}^{2}+3},\;\;\;\mathcal{M}_{odd}=\mathcal{M}_{+}-\mathcal{M}_{-}=\frac{4%
\tilde{m}_{0}}{\tilde{m}_{0}^{2}+3},\;  \label{modd} \\
\mathcal{V}_{even} & =\mathcal{V}_{+}+\mathcal{V}_{-}=\frac{-4\tilde{m}%
_{0}^{2}}{3\left( \tilde{m}_{0}^{2}+3\right) }W,\;\;\;\mathcal{V}_{odd}=%
\mathcal{V}_{+}-\mathcal{V}_{-}=\frac{-4\tilde{m}_{0}}{\tilde{m}_{0}^{2}+3}W
\label{vodd}
\end{align}
We note that due to momentum conservation $\sum_{r=1}^{3}p^{(r)}=0$ we can
rewrite
\begin{equation}
\frac{1}{2}\sum_{r,s=1}^{3}p^{(r)}\cdot p^{(s)}\left( V_{3}^{\left[ rs\right]
}\right) _{00}=\frac{1}{2}\sum_{r,s=1}^{3}p^{(r)}\cdot p^{(s)}\mathcal{C}%
_{\left( s-r\right) \func{mod}3}=\frac{3}{4}\mathcal{C}_{0}%
\sum_{r=1}^{3}p^{(r)}\cdot p^{(r)}
\end{equation}
>From this parametrization, it is straightforward to verify some relations
that have been noticed before in the literature without having our explicit
formulas for the Neumann coefficients
\begin{align}
\mathcal{M}_{0}+\mathcal{M}_{+}+\mathcal{M}_{-} & =1\,,\quad\mathcal{M}_{+}%
\mathcal{M}_{-}=\mathcal{M}_{0}^{2}-\mathcal{M}_{0}  \label{cM0} \\
\mathcal{M}_{0}^{2}+\mathcal{M}_{+}^{2}+\mathcal{M}_{-}^{2} & =1\,,\quad
\mathcal{M}_{0}\mathcal{M}_{+}+\mathcal{M}_{+}\mathcal{M}_{-}+\mathcal{M}_{-}%
\mathcal{M}_{0}=0 \\
\mathcal{M}_{\pm}^{2}-\mathcal{M}_{\pm} & =\mathcal{M}_{0}\mathcal{M}_{\mp
},\;\;.\mathcal{M}_{odd}^{2}=\left( 1-\mathcal{M}_{0}\right) \left( 1+3%
\mathcal{M}_{0}\right)  \label{cM1} \\
3\left( 1-\mathcal{M}_{0}\right) \mathcal{V}_{0} & =-\mathcal{M}_{odd}%
\mathcal{V}_{odd},\;\;3\mathcal{M}_{odd}\mathcal{V}_{0}=-\left( 1+3\mathcal{M%
}_{0}\right) \mathcal{V}_{odd} \\
2V_{00} & =\frac{9}{4}\mathcal{\bar{V}}_{0}\mathcal{V}_{0}+\frac{3}{4}%
\mathcal{\bar{V}}_{odd}\mathcal{V}_{odd}   \label{cc0}
\end{align}
We emphasize that in our case these results hold for any set of frequencies $%
\kappa_{e},\kappa_{o}$ and any number of oscillators $N$. They were obtained
using the \textit{associative} star product in complete compliance with
gauge invariance.
\end{itemize}

\subsection{Ghost Neumann coefficients}

In the operator formulation of string field theory, the Neumann coefficients
of the ghost field are also key ingredients. Since they can be related to
the Neumann coefficients of the matter sector, one may derive the
regularized expression for them for any $\kappa_{e},\kappa_{o},N$. We write
the ghost part of the three string vertex in the following form,
\begin{equation}
|V_{3}\rangle=\exp\left(
-\sum_{r,s=1}^{3}\sum_{n\geq1,\,m%
\geq0}c_{-n}^{(r)}X_{nm}^{[r,s]}b_{-m}^{(s)}\right) |0\rangle
\end{equation}
The matrix $X$ can be written in terms of the matter Neumann coefficient of
the six string vertex as \cite{r-GT}\cite{r-Kishimoto},
\begin{align}
\frac{1}{\sqrt{\kappa_{k}}}X_{kl}^{[r,s]}\sqrt{\kappa_{l}} &
=(-1)^{r+s}(V_{6}^{[r,s]}-V_{6}^{[r,s+3]})_{kl}, \\
\frac{1}{\sqrt{\kappa_{k}}}X_{k0}^{[r,s]} &
=(-1)^{r+s}(V_{6}^{[r,s]}-V_{6}^{[r,s+3]})_{k}l_{s}^{-1}
\end{align}
Defining%
\begin{equation}
\left( X^{[r,s]}\right) _{kl}=-\left( C\mathcal{M}_{\left( s-r\right) \func{%
mod}3}^{gh}\right) _{kl},\;\;\ \left( X^{[r,s]}\right) _{k0}=-\left(
\mathcal{V}_{\left( s-r\right) \func{mod}3}^{gh}\right) _{k}
\end{equation}
and inserting the explicit formula of the six string vertex, which is given
in Eq.(\ref{neum}) for $n=6,$ we obtain (using the notation $\left(
-1,0,1\right) \func{mod}3=\left( 2,0,1\right) $)
\begin{align}
\mathcal{M}_{0}^{gh} & =\frac{1-\hat{m}_{0}^{2}}{3\hat{m}_{0}^{2}+1},\;\;%
\mathcal{M}_{+}^{gh}=2\hat{m}_{0}\frac{1+\hat{m}_{0}}{3\hat{m}_{0}^{2}+1}%
,\;\;\mathcal{M}_{-}^{gh}=2\hat{m}_{0}\frac{-1+\hat{m}_{0}}{3\hat {m}%
_{0}^{2}+1},  \label{neumghost} \\
\mathcal{V}_{0}^{gh} & =-\frac{4\hat{m}_{0}^{2}}{ 1+3\hat{m}_{0}^{2} }\hat{W}%
,\;\;\mathcal{V}_{+}^{gh}=-\frac{2(1-\hat{m}_{0})\hat{m}_{0}}{\left( 1+3\hat{%
m}_{0}^{2}\right) }\hat{W},\;\;\mathcal{V}_{-}^{gh}=\frac{2\hat {m}_{0}(1+%
\hat{m}_{0})}{1+3\hat{m}_{0}^{2}}\hat{W}
\end{align}
where
\begin{align}
\hat{m}_{0} & =\sqrt{\kappa}\left( \tilde{m}_{0}\right) \frac{1}{\sqrt{\kappa%
}}=\left(
\begin{array}{cc}
\sqrt{\kappa_{e}}V^{e} & 0 \\
0 & \sqrt{\kappa_{o}}V^{o}%
\end{array}
\right) \left(
\begin{array}{cc}
0 & \frac{2l_{s}^{2}}{\theta}\tau \\
\frac{\theta}{2l_{s}^{2}}\tau & 0%
\end{array}
\right) \left(
\begin{array}{cc}
\bar{V}^{e}\frac{1}{\sqrt{\kappa_{e}}} & 0 \\
0 & \bar{V}^{o}\frac{1}{\sqrt{\kappa_{o}}}%
\end{array}
\right)  \label{mhat} \\
\hat{W} & =\sqrt{\kappa}Wl_{s}^{-1}=\left(
\begin{array}{c}
\frac{1}{\sqrt{2}}w \\
0%
\end{array}
\right)
\end{align}
We see that $\mathcal{M}_{i}^{gh}$ has exactly the same form as Eq.(\ref{M00}%
) if we replace $\hat{m}_{0}$ by $\tilde{m}_{0}^{-1}$. This implies that the
matrices $\mathcal{M}_{i}^{gh}$ automatically satisfy the same relations as
Eq.(\ref{cM0}--\ref{cM1}) for any $\kappa_{e},\kappa_{o},N$. On the other
hand, the zero mode part satisfies modified nonlinear relations,
\begin{equation}
(1-\mathcal{M}_{0}^{gh})\mathcal{V}_{-}^{gh}+\mathcal{M}_{+}^{gh}\mathcal{V}%
_{0}^{gh}=0\,,\quad(1-\mathcal{M}_{0}^{gh})\mathcal{V}_{+}^{gh}+\mathcal{M}%
_{-}^{gh}\mathcal{V}_{0}^{gh}=0\,,
\end{equation}
which are again famous in the literature \cite{GJ}\cite{r-Kishimoto}.

\subsection{Regularized sliver matrices}

Another quantity which has appeared often in the literature is the
description of the sliver state in terms of a matrix $C\mathcal{T}$ written
in terms of another matrix $\mathcal{Z}\,$
\begin{equation}
\mathcal{T}=\frac{1}{2\mathcal{Z}}\left( 1+\mathcal{Z}-\sqrt{(1+3\mathcal{Z}%
)(1-\mathcal{Z})}\right) \,\,.
\end{equation}
In our parametrization, for any $\kappa_{e},\kappa_{o}$ the $2N\times2N$
matrix $\mathcal{Z}$ is given by the Neumann coefficient $\mathcal{M}_{0}$
of Eq.(\ref{M00}), $\mathcal{Z}$=$\mathcal{M}_{0}\left( \tilde{m}\right) $.
Then $\mathcal{Z}$,$\mathcal{T}$ ~become
\begin{equation}
\mathcal{Z}\left( \tilde{m}_{0}\right) =\frac{\tilde{m}_{0}^{2}-1}{\tilde {m}%
_{0}^{2}+3},\;\;\mathcal{T}\left( \tilde{m}_{0}\right) =\frac {\sqrt{\tilde{m%
}_{0}^{2}}-1}{\sqrt{\tilde{m}_{0}^{2}}+1}.   \label{cT}
\end{equation}
For finite $N$ these are the regularized matrices. In terms of the
eigenvalues $\tau_{k}$ we have%
\begin{equation}
\mathcal{T}=\left(
\begin{array}{cc}
V^{\left( e\right) } & 0 \\
0 & V^{\left( o\right) }%
\end{array}
\right) \left(
\begin{array}{cc}
\frac{\tau-1}{\tau+1} & 0 \\
0 & \frac{\tau-1}{\tau+1}%
\end{array}
\right) \left(
\begin{array}{cc}
\bar{V}^{\left( e\right) } & 0 \\
0 & \bar{V}^{\left( o\right) }%
\end{array}
\right)
\end{equation}
and%
\begin{equation}
\mathcal{Z=}\left(
\begin{array}{cc}
V^{\left( e\right) } & 0 \\
0 & V^{\left( o\right) }%
\end{array}
\right) \left(
\begin{array}{cc}
\frac{\tau^{2}-1}{\tau^{2}+3} & 0 \\
0 & \frac{\tau^{2}-1}{\tau^{2}+3}%
\end{array}
\right) \left(
\begin{array}{cc}
\bar{V}^{\left( e\right) } & 0 \\
0 & \bar{V}^{\left( o\right) }%
\end{array}
\right)
\end{equation}
Showing clearly the eigenvalue structure of these much discussed matrices.

In the ghost sector, we replace $\tilde{m}_{0}$ by $\hat{m}_{0}$ in
comparing the matter/ghost Neumann coefficients, as seen from Eqs.(\ref{M00},%
\ref{neumghost}). This gives the sliver matrices $\mathcal{Z}\left( \hat
{m}%
_{0}\right) $ and $\mathcal{T}$ $\left( \hat{m}_{0}\right) $ with $\hat{m}%
_{0}$ of Eq.(\ref{mhat}) replacing $\tilde{m}_{0}$ in the expressions of Eq.(%
\ref{cT}). In terms of the eigenvalues, we get%
\begin{equation}
\mathcal{T}^{gh}=\left(
\begin{array}{cc}
\sqrt{\kappa_{e}}V^{e} & 0 \\
0 & \sqrt{\kappa_{o}}V^{o}%
\end{array}
\right) \left(
\begin{array}{cc}
\frac{1-\tau}{\tau+1} & 0 \\
0 & \frac{1-\tau}{\tau+1}%
\end{array}
\right) \left(
\begin{array}{cc}
\bar{V}^{e}\frac{1}{\sqrt{\kappa_{e}}} & 0 \\
0 & \bar{V}^{o}\frac{1}{\sqrt{\kappa_{o}}}%
\end{array}
\right)
\end{equation}%
\begin{equation}
\mathcal{Z}^{gh}\mathcal{=}\left(
\begin{array}{cc}
\sqrt{\kappa_{e}}V^{e} & 0 \\
0 & \sqrt{\kappa_{o}}V^{o}%
\end{array}
\right) \left(
\begin{array}{cc}
\frac{1-\tau^{2}}{1+3\tau^{2}} & 0 \\
0 & \frac{1-\tau^{2}}{1+3\tau^{2}}%
\end{array}
\right) \left(
\begin{array}{cc}
\bar{V}^{e}\frac{1}{\sqrt{\kappa_{e}}} & 0 \\
0 & \bar{V}^{o}\frac{1}{\sqrt{\kappa_{o}}}%
\end{array}
\right)
\end{equation}
Again, all of our expressions are valid for any $\kappa_{e},\kappa_{o},N.$

\section{VSFT and Associativity Anomaly}

In this section we first show that any fluctuation around a D-brane vacuum
becomes pure gauge if we use the associativity of the star product. Some of
our arguments here overlap with section (5.3) of \cite{r-RSZ2} but our
emphasis is on associativity. This implies that there is no physical
excitation in the D-brane vacuum. Since associativity is part of the gauge
invariance, the undesired result has implications on the principles
underlying the definition of gauge invariance in VSFT. Once we understand
the issue which is rigorous for any nonsingular sliver-like projector, we
will point toward a possible solution at \textit{infinite} $N$ through the
introduction of an associativity anomaly which arises from the singular
nature of the sliver state at infinite $N.$ VSFT is not well defined until
the singularity is universally defined and the gauge invariance principle
understood.

A need for the anomaly is anticipated \cite{r-Strominger}\cite{r-BarsMatsuo}
from the fact that the object which VSFT describes at the outset is supposed
to be closed string excitations around the closed string vacuum.
Furthermore, the D-brane itself is the soliton of \textquotedblleft closed
strings\textquotedblright. We will clarify the associativity issues in the
framework of MSFT. The reconciliation of the gauge invariance, associativity
anomaly, and nonperturbative string physics in VSFT remains as a challenge
that we leave to future work.

\subsection{Fluctuations around a D-brane vacuum}

The action of VSFT has the form of Eq.(\ref{action}) with $\mathcal{Q}$
constructed purely from ghosts. Then associativity and gauge invariance
holds exactly for generic $\kappa_{e},\kappa_{o},N$ as discussed following
Eq.(\ref{action}). In a sector in which the field is the product of a ghost
and matter parts $A=A^{\left( g\right) }A^{\left( m\right) },$ the equations
of motion separate%
\begin{equation}
\mathcal{Q}A^{\left( g\right) }=-A^{\left( g\right) }\ast A^{\left( g\right)
},\;\;A^{\left( m\right) }\ast A^{\left( m\right) }=A^{\left( m\right) }.
\end{equation}
We take the D$_{25}$-brane solution to be a projector that is $\bar{x}$
independent, Lorentz invariant in 26 dimensions, and trace one (a single
D-brane). Any projector of the type of Eq.(\ref{AD}) at $\lambda=0$ is such
a solution, but in the literature there is evidence that the sliver $%
A^{\left( m\right) }=\Xi\left( x,p\right) $ given in Eq.(\ref{sliver}) is
the candidate for the $D_{25}$-brane
\begin{equation}
\Xi=\det\left( 2^{d/2}\right) \exp\left( -\bar{\xi}m_{s}\sigma^{-1}\xi%
\right) ,\;m_{s}=i\left(
\begin{array}{cc}
0 & a\theta \\
\frac{-1}{a\theta} & 0%
\end{array}
\right) ,\;m_{s}^{2}=1,   \label{SSS}
\end{equation}
and $a$ is given explicitly in Eqs.(\ref{a1},\ref{a2}) for any $\kappa
_{e},\kappa_{o},N.$
\begin{equation}
a=\frac{1}{2l_{s}^{2}}\kappa_{e}^{1/2}\left( \sqrt{\kappa_{e}^{1/2}T\,%
\kappa_{o}^{-1}\bar{T}\,\kappa_{e}^{1/2}}\right) ^{-1}\kappa_{e}^{1/2}=\frac{%
1}{2l_{s}^{2}}\kappa_{e}^{1/2}V^{e}\tau^{-1}\bar{V}^{e}\kappa _{e}^{1/2}.
\end{equation}
The ghost part $A^{\left( g\right) }$ also has a solution related to the
sliver as discussed in \cite{r-HK}.

The next step is to study fluctuations around the D$_{25}$-brane and
interpret them as open string states. If one seeks fluctuations that have
the same universal ghost factor, as advocated in \cite{r-RSZ}, then
effectively one has to study the action reduced to the following form%
\begin{equation}
S=-KTr_{g}\left( A^{\left( g\right) }\ast A^{\left( g\right) }\right) \int d%
\bar{x}Tr_{m}\left( \frac{1}{2}A^{\left( m\right) }\ast A^{\left( m\right) }-%
\frac{1}{3}A^{\left( m\right) }\ast A^{\left( m\right) }\ast A^{\left(
m\right) }\right) .
\end{equation}
Expanding the matter field around the projector $A^{\left( m\right) }=\Xi+T,$
and using
\begin{equation}
\Xi\ast\Xi=\Xi,   \label{SS}
\end{equation}
gives a quadratic and cubic term in $T$%
\begin{equation}
S=S\left( \Xi\right) -\tilde{K}\int d\bar{x}Tr_{m}\left( T\ast\left( \frac{1%
}{2}-\Xi\right) \ast T-\frac{1}{3}T\ast T\ast T\right)
\end{equation}
where $\tilde{K}=KTr_{g}\left( A^{\left( g\right) }\ast A^{\left( g\right)
}\right) .$ The value of the action of the classical solution $S\left(
\Xi\right) $ is related to the $D_{25}$-brane tension. To determine its
absolute value one notices that the overall coefficient $\tilde{K}$ is
related to the cubic coupling. Therefore one would like to extract the value
of the cubic coupling, say for the fluctuation that corresponds to the
tachyon state (at the top of the effective tachyon potential) thereby fixing
$\tilde{K}$ and consequently computing the absolute value of the $D_{25}$%
-brane tension.

In a perturbative expansion $T=T_{1}+T_{2}+\cdots,$ the quadratic part
determines the spectrum of the fluctuations $T_{1}$, while the cubic part
determines their coupling. Attempts to compute these quantities have run
into a controversy in the literature \cite{r-HK}\cite{r-HM}\cite{RSZ4}. We
will now clarify that at the root of the controversy is the need to
introduce nonassociativity of the star product to extract nontrivial results
from VSFT.

The equation of motion for $T$ at the linearized level is%
\begin{equation}
T_{1}=\Xi\ast T_{1}+T_{1}\ast\Xi.   \label{ST}
\end{equation}
In Appendix-2, it is shown that the general solution to Eq.(\ref{ST}) is

\begin{equation}
T_{1}\left( \bar{x},x_{e},p_{e}\right) =\int d\lambda~\left[ f_{+}\left(
\bar{x},\lambda\right) \left( e^{\xi\left( \frac{1+m}{2}\right) \lambda
}-1\right) \Xi+f_{-}\left( \bar{x},\lambda\right) \left( e^{\xi\left( \frac{%
1-m}{2}\right) \lambda}-1\right) \Xi\right]   \label{T1}
\end{equation}
for any functions $f_{\pm}\left( \bar{x},\lambda\right) .$ Taking any $\bar{x%
}$ dependent solution $T_{1}\left( \bar{x}\right) $ with definite center of
mass momentum $p,$ and replacing it back in the action, one hopes to
identify the mass and the coupling constant for the particle represented by
the solution, with a properly normalized $T_{1}$, as follows
\begin{align}
-\tilde{K}\int d\bar{x}~Tr_{m}\left( T_{1}\ast\left( \frac{1}{2}-\Xi\right)
\ast T_{1}\right) & =\left( p^{2}+\left( mass\right) ^{2}\right)
~f^{2}\left( p\right)  \label{mass} \\
-\tilde{K}\int d\bar{x}~Tr_{m}\left( -\frac{1}{3}T_{1}\ast T_{1}\ast
T_{1}\right) & =\left( coupling\right) ~f^{3}\left( p\right)
\label{coupling}
\end{align}
where $f\left( p\right) $ represents the particle wavefunction in momentum
space. Of course, the left side of Eq.(\ref{mass}) must vanish as long as $%
T_{1}$ is the solution of Eq.(\ref{ST}), which implies $p^{2}+\left(
mass\right) ^{2}=0$. So to identify the normalization $f^{2}\left( p\right) $
as the coefficient of $p^{2}$one works slightly off shell. We will show that
the left hand side of Eq.(\ref{coupling}) exactly vanishes at any $N$ as a
consequence of associativity, therefore the coupling vanishes. So it is
problematic to extract the coupling, $\tilde{K}$ or the $D_{25}$-brane
tension.

To prove this point we give the following arguments. Assuming associativity
is satisfied by the star product, we can always map this problem to a matrix
problem in which $\Xi$ is a projector represented by a diagonal matrix that
contains one entry $1$ and the rest zeros. Then it is easy to see that the
most general solution of Eqs.(\ref{SS},\ref{SS}) for the matrix problem is%
\begin{equation}
\Xi=\left(
\begin{array}{cc}
1 & 0 \\
0 & 0%
\end{array}
\right) ,\;T_{1}=\left(
\begin{array}{cc}
0 & b^{\dagger} \\
b & 0%
\end{array}
\right)   \label{STmatrices}
\end{equation}
where $b$ is a complex column matrix. This solution may also be rewritten as
$T_{1}=i\left[ H,\Xi\right] $ for any hermitian matrix $H.$ For this matrix
solution it is easy to see that Tr$\left( T_{1}\ast T_{1}\ast T_{1}\right)
=0,$ and hence the cubic coupling of the fluctuations vanishes.

The same statements can now be made for any associative star product.
Namely, if $\Xi$ is given by the string field of Eq.(\ref{SSS}), then the
most general solution of Eq.(\ref{ST}), including Eq.(\ref{T1}), must take
the form%
\begin{equation}
T_{1}=i(H\ast\Xi-\Xi\ast H),   \label{THS}
\end{equation}
where $H$ is any string field. This solution can be easily verified%
\begin{align}
\Xi\ast T_{1}+T_{1}\ast\Xi & =i\Xi\ast(H\ast\Xi-\Xi\ast
H)+i(H\ast\Xi-\Xi\ast H)\ast\Xi \\
& =-i\Xi\ast\Xi\ast H+iH\ast\Xi\ast\Xi \\
& =iH\ast\Xi-i\Xi\ast H=T_{1}
\end{align}
Inserting the result into the formulas for mass and coupling we see that the
left hand sides of Eqs.(\ref{mass},\ref{coupling}) vanish by simple algebra
and cyclicity of the trace.

This shows that, independent of any detail of the solution, associativity
leads to a unique consequence, namely that the fluctuations $T_{1}$ are pure
gauge. Indeed, if we define the gauge transformation for $T$ as
\begin{equation}
\Xi+T^{\prime}=U\ast\left( \Xi+T\right) \ast U^{\dagger}
\end{equation}
we see that for the small fluctuation $T_{1}$ we can write the infinitesimal
gauge transformation for $U=\exp_{\ast}\left( iH\right) $ in the form%
\begin{equation}
T_{1}^{\prime}=U\ast\Xi\ast U^{\dagger}-\Xi+\cdots=i(H\ast\Xi-\Xi\ast
H)+\cdots
\end{equation}
which has the form of the general solution.

Are there cohomologically non-trivial solutions $T_{1}$ (i.e. not pure
gauge) as suggested in \cite{r-HK} ? In analyzing their suggestion, we find
that in fact it is not a solution at all, iff we use associativity freely.
One way to see the problem is to examine the quantities $G,H$ related to
mass and coupling, as identified in \cite{r-HM}. We find that these
quantities vanish identically at any $N,\kappa_{e},\kappa_{o}$ when we
insert our explicit expressions for the Neumann coefficients. The same
result was obtained in \cite{r-HM} by using the Neumann coefficient
identities in Eqs.(\ref{cM0}-\ref{cc0}), which now hold also for any $%
N,\kappa_{e},\kappa_{o}.$ This vanishing is a consequence of associativity
pure and simple. The vanishing of $G$ implies that the $T_{1}$ given in \cite%
{r-HK} is not a solution, while the vanishing of $H$ corresponds to the
vanishing of the coupling.

To avoid this outcome, VSFT must have an associativity anomaly, which we
discuss next.

\subsection{Focus on anomalies}

We have shown very generally that there can be no interesting
nonperturbative phenomena in VSFT unless there is an associativity anomaly.
In a previous paper we had argued that the existence of closed strings in
open string field theory is also closely related to an associativity anomaly
\cite{r-BarsMatsuo}. So, how can nonassociativity arise in the theory in
detail? In our previous paper \cite{r-BarsMatsuo} we had argued that the
associativity anomaly of the star product is closely linked to the
associativity anomaly of the matrices $T,R,w,v$ in the infinite $N$ limit.
This was due to the fact that the matrix $T_{eo}$ develops a zero mode in
that limit, $Tv\rightarrow0$ as shown clearly in Eqs.(\ref{blockdiagonal1}).
Hence whenever the inverse of the matrix occurs one must define it
carefully, and watch that sometimes the zero mode is compensated by infinite
sums, thus giving rise to anomalies. In particular the sliver in Eq.(\ref%
{SSS}) does involve the inverse of the matrix $a^{-1}$ and this enters in
the expressions that determine the mass and couplings of the fluctuations.

Also, as it is clear from the computation of $N$-string vertices from MSFT,
the basic ingredients are the $\tilde{m}_{0}$ and vector $W$. All the
Neumann coefficients are written in terms of them. In this sense, in the
MSFT context, the understanding of the associativity anomaly can be reduced
to the study of $\tilde{m}_{0}$ and $W$ in the large $N$ limit.

Since we derived the explicit form of the Neumann coefficients in terms of $%
\tilde{m}_{0}$, we can pinpoint in which combination of the Neumann
coefficients, such an anomaly occurs. Some of the examples are,
\begin{equation}
1+3\mathcal{M}_{0}=\frac{4\tilde{m}_{0}^{2}}{\tilde{m}_{0}^{2}+3},\quad%
\mathcal{M}_{odd}=\frac{4\tilde{m}_{0}}{\tilde{m}_{0}^{2}+3},\quad1-\mathcal{%
M}_{0}^{gh}=\frac{4\hat{m}_{0}^{2}}{3\hat{m}_{0}^{2}+1}.
\end{equation}
All of them have zero-eigenvalue in the large $N$ limit as seen from Eqs.(%
\ref{msquare},\ref{msquare}) and Eq.(\ref{tau}). In this sense, whenever we
try to invert these matrices, we meet the anomaly as will be discussed in
the next subsection. We note that the order of $\tilde{m}_{0}$ in these
expressions coincides with the degree of singularity introduced in \cite%
{r-HM}.

Actually all fluctuations around the D-brane vacuum, that are claimed to be
cohomologically non-trivial in the literature, use the inverse of such
matrices. For example, the tachyon wave function conjectured in \cite{r-HK}
takes the form,
\begin{equation}
|\Phi_{T}\rangle=\exp\left( -\sum_{n\geq1}t_{n}p_{\mu}a_{n}^{\dagger\mu
}\right) |\Phi_{C}\rangle,\quad t=3(1+\mathcal{T})(1+3\mathcal{M}_{0})^{-1}%
\mathcal{V}_{0}\,\,.
\end{equation}
$|\Phi_{C}\rangle$ is a classical solution which describes the D-brane.
Also, the pure ghost BRST charge itself takes the \textquotedblleft
singular\textquotedblright\ form \cite{r-HK}\cite{rsz3}\cite{r-Kishimoto},
\begin{equation}
\mathcal{Q}=c_{0}+\sum_{n\geq1}f_{n}(c_{n}+(-1)^{n}c_{n}^{\dagger}),\qquad
f=(1-\mathcal{M}_{0}^{gh})^{-1}\mathcal{V}_{0}^{gh}\,\,.
\end{equation}
By now, it is quite well-known that such a singular vectors are related to
the midpoint \cite{rsz3}\cite{r-Okuyama}\cite{r-BarsMatsuo}.

So far, the nonperturbative effects from VSFT are based on such singular
states. For example, in recent literature \cite{r-HM} the computation of the
tachyon mass and coupling of Eq.(\ref{mass},\ref{coupling}) were performed
based on such states. If we try to keep associativity everywhere it would
give only a trivial value for such physical quantities independent of the
details of a solution $T_{1}$ as discussed in the previous subsection.

The authors of \cite{r-HM} suggested a regularization scheme that introduces
a \textquotedblleft twist anomaly\textquotedblright\ to obtain nontrivial
values. In our language their proposal is equivalent to a slight shift of
the eigenvalues of $\tilde{m}_{0}$ in such a way that in Eq.(\ref{m2}) the
eigenvalues in the upper block are different than those in the lower block.
While this prescription gave the correct nonzero value for $G,$ it produced
the wrong result for $H$. Since we have argued in the previous paragraph
that a cutoff consistent with associativity cannot alter the conclusion, a
twist anomaly must be equivalent to nonassociativity. But breaking
associativity also breaks the gauge invariance of the theory, and this is
likely to be the reason for obtaining the wrong value of $H.$

Nevertheless some of the arguments in \cite{r-HM} seem to point in the right
direction. The precise way in which the anomaly could occur is in the
definition of the inverse of the infinite matrix $\tilde{m}_{0}^{-2}$. This
general issue should now be investigated in a systematic way by using our
consistent techniques, which tie together all the places where the zero
eigenvalue $\tau=0$ could occur in the large $N$ limit, and concentrating on
the proper definition of $\left( \tilde{m}_{0}^{-2}\right) ^{1/2}$. This
definition has to introduce an associativity anomaly, but the anomaly should
be gentle enough to keep sufficient gauge symmetry intact.

Thus, the emergence of closed strings in open string field theory, as well
as the nontrivial values of the masses or couplings of fluctuations in VSFT,
all need the same source of associativity anomaly that resides in the
definition of $\left( \tilde{m}_{0}^{-2}\right) ^{1/2}$. It is crucial to
weigh the desirability of the anomaly versus the gauge invariance of the
theory. In the next subsection we identify the source of the anomaly.

Before closing this section, let us mention that there is a different
proposal for the resolution of the controversy \cite{okawa} for computing
the D-brane tension on the one hand and mass and coupling of the
fluctuations on the other hand. In this case the approach is based on BSFT
which uses conformal field theory techniques. So far it has not been
possible to translate this proposal to the context of algebraic techniques
used in previous investigations \cite{r-HM} or in the present paper. It must
be emphasized however, that in the proposal of \cite{okawa} there is no $%
T_{1}$ that solves the equations of motion in Eq.(\ref{ST}), and this is a
way of avoiding the pure gauge configuration. Until this proposal is
understood in the algebraic approach and the problems encountered above
solved in all languages, there remains a cloud in our understanding. We
believe that a key here is the associativity anomaly.

\subsection{Origin of associativity anomaly in MSFT}

One approach is to define the inverse of potentially singular matrices at
finite $N$ when they are not singular, perform all the computations, and set
$N$ to infinity at the end. However, as we have seen, even at finite $N$ the
left hand sides of Eqs.(\ref{mass},\ref{coupling}) vanish. So this means
that associativity itself must be broken in some other subtle way, and this
must be incorporated as part of the principles of setting up VSFT. For this
reason, we re-examine the \textquotedblleft bare\textquotedblright\ infinite
matrices directly in the following discussion.

Let us first describe the characteristic behavior of $m_{0}^{2}$ in the
limit, (recall $\tilde{m}_{0}^{2}=Km_{0}^{2}K^{-1}$ and $m_{0}=M_{0}\sigma$
is given in Eq.(\ref{M0}))
\begin{equation}
m_{0}^{2}=\left(
\begin{array}{cc}
\kappa_{e}Z & 0 \\
0 & Z\kappa_{e}%
\end{array}
\right) \equiv\left(
\begin{array}{cc}
\bar{\Gamma} & 0 \\
0 & \Gamma%
\end{array}
\right) \,\,.   \label{m02}
\end{equation}
$Z$ was given in Eq.(\ref{Z}-\ref{Znum}) and $\Gamma$ was discussed in Eqs.(%
\ref{gamma}-\ref{gt}). Recall that this matrix occurs in the computation of
the wedge states as in Eq.(\ref{msms}) and it needs to be inverted to define
the sliver state as in Eq.(\ref{msliver}). We would like to re-examine this
operation in the large $N$ limit to identify a source of associativity
anomaly.

By using the large $N$ version of the properties (\ref{tr}),(\ref{rtv}) and (%
\ref{tt}), we have the properties of the infinite matrices
\begin{equation}
T\bar{T}=1,\;\;\bar{T}T=1-v\bar{v},\;\;Tv=0,\;\;\bar{v}v=1,\;\;v=\bar{T}w.
\end{equation}
In the same way we can derive the large $N$ equations as the limits of Eqs.(%
\ref{gbarg}-\ref{u})
\begin{equation}
\Gamma\bar{\Gamma}=1\,,\quad\bar{\Gamma}\Gamma=1-u\,\bar{u},\quad\Gamma
u=0\,\,,\;\;\bar{u}u=1,\;\;u=\bar{\Gamma}w.
\end{equation}
If one worries about the associativity of the product involved in the
definition of $\Gamma$, one may use the following explicit form of the
matrix elements of $\Gamma$ and $u$ to prove these identities directly
\begin{equation}
\Gamma_{e,e^{\prime}}=Z_{e,e^{\prime}}e^{\prime},\qquad\bar{\Gamma }%
_{e,e^{\prime}}=eZ_{e,e^{\prime}},\qquad u_{e}=\frac{e}{\sqrt{2}}Z_{e,0}
\end{equation}
where $e,e^{\prime}$ are non-negative even indices as before, but to define
the last formula we formally extended the expression in Eq.(\ref{zpsi}) for $%
Z_{e,e^{\prime}}$ to include $e^{\prime}=0$.

We see that the matrix $\Gamma$ defines a \textit{shift operator} in the
Hilbert space $\mathcal{H}^{even}$ which appears typically in the solution
generating technique \cite{HKL}. To see it more explicitly, we introduce a
new basis,
\begin{equation*}
e_{0}\equiv u,\quad e_{1}\equiv\bar{\Gamma}u,\quad e_{2}\equiv\bar{\Gamma }%
\,e_{1},\cdots.
\end{equation*}
It is easy to see that $\Gamma$ and $\bar{\Gamma}$ are the shift operators
on this basis,
\begin{equation*}
\bar{\Gamma}\,e_{n}=e_{n+1},\qquad\Gamma\,e_{n+1}=e_{n},\quad(n>0),\qquad
\Gamma e_{0}=0,\qquad\bar{e}_{n}\cdot e_{m}=\delta_{nm}.
\end{equation*}

We note that we have a close analogy between $T$ and $\Gamma.$ One
difference is that while the operator $T$ interpolates between the different
Hilbert spaces $\mathcal{H}^{odd}$ and $\mathcal{H}^{even}$, $\Gamma$ is
defined within the same Hilbert space $\mathcal{H}^{even}$. In this sense,
the eigenstate equation is well-defined for $\Gamma$. Obviously, it has one
zero eigenstate $u$. As in our previous paper \cite{r-BarsMatsuo}, it causes
the associativity anomaly,
\begin{align}
& (\Gamma\bar{\Gamma}\,)\,w\ =\ w,\quad versus\quad\Gamma\,(\bar{\Gamma }%
\,w)=\Gamma u=0\,\,,  \notag \\
& (\Upsilon\,\Gamma)\,u=u,\quad versus\quad\Upsilon\,(\Gamma\,u)=0\,\,,\;etc.
\end{align}
where the inverse $\Upsilon$ was defined in (\ref{Ginv}) with $\Upsilon
\Gamma=1$. It is interesting that just like $R$ exists, $\Upsilon$ also
exists in the infinite $N$ limit.

Thus, while $\Gamma$ has a zero eigenstate in the large $N$ limit, its
transpose $\bar{\Gamma}$ has no zero eigenstate. The other nonzero
eigenvalues are shared by $\Gamma$ and $\bar{\Gamma}$. This is seen from Eq.(%
\ref{gt}) where $\bar{\Gamma}$ was diagonalized%
\begin{equation}
\bar{\Gamma}=\left( \kappa_{e}\right) ^{1/2}V^{e}\left( \tau\right) ^{2}\bar{%
V}^{e}\left( \kappa_{e}\right) ^{-1/2},   \label{m2}
\end{equation}
and we saw that $\tau_{k}$ became the continuous function $\tau\left(
k\right) =-\tanh\left( \pi k/4\right) $ at large $N.$ Thus, every nonzero
eigenvalue of $m_{0}^{2}$ comes in pairs at $k\neq0.$ On the other hand we
have just argued that at $k=0$ there is only one eigenstate. This asymmetry
which occurs in the large $N$ limit is related to the associativity anomaly.

To summarize, the associativity anomaly occurs when we try to invert the
matrices which contain the zero eigenvalue in the continuous spectrum. It is
curious that inverses exist explicitly such as $R$ and $\Upsilon$ as we have
seen. Now we see that for $m_{0}^{2}$ of Eq.(\ref{m02}), at infinite $N$ the
left inverse is different than the right inverse%
\begin{equation}
\left( m_{0}^{-2}\right) _{L}=\left(
\begin{array}{cc}
\Gamma & 0 \\
0 & \Upsilon%
\end{array}
\right) ,\;\;\left( m_{0}^{-2}\right) _{R}=\left(
\begin{array}{cc}
\bar{\Upsilon} & 0 \\
0 & \bar{\Gamma}%
\end{array}
\right)
\end{equation}
This is a characteristic structure of the nonassociative algebra. {}From
these we may define $\tilde{m}_{0}^{-2}=Km_{0}^{-2}K^{-1}$ and then try to
find a proper definition of $\left( \tilde{m}_{0}^{-2}\right) ^{1/2}$. It is
already evident that associativity is not trivial and that in this way an
associativity anomaly is likely to be introduced. The gauge symmetry needs
to be analyzed carefully and then the nonperturbative quantities in VSFT
could be extracted. Since this was not the focus of this paper, we will
discuss these issues in future work.

\section{Summary and Outlook}

We have formulated MSFT and showed that it is an easier framework for
performing computations in string field theory. We introduced the monoid
algebra as a tool that facilitates computations. The expressions we obtained
for various quantities, such as wedge, sliver, projectors, Neumann
coefficients, vertices both perturbative and nonperturbative, etc. are
generalizations that are valid for any frequencies $\kappa_{e},\kappa_{o}$
for any number of string oscillators $2N$. Our analytic expressions are not
only more general, but also considerably simpler than those in the existing
literature, while agreeing with them whenever they are available in the
large $N$ limit.

MSFT, taken with a specific cutoff procedure that guarantees associativity,
is apparently a consistent field theory of open strings. The appearance of
closed strings in open string field theory require an associativity anomaly
\cite{r-HS}-\cite{r-Strominger2}, \cite{r-BarsMatsuo}.

In the context of VSFT we have shown that there has to be an associativity
anomaly in order to recover certain nonperturbative results. Quantities such
as masses and couplings for fluctuations around a D-brane, the D-brane
tension, as well as closed strings, all depend on the same source of
nonassociativity. Using the MSFT framework, we have identified in detail the
possible source of the anomaly, namely the zero eigenvalue of $\tilde{m}_{0}$
at $N\rightarrow\infty,$ and the corresponding nontrivial definition of $%
\left( \tilde{m}_{0}^{2}\right) ^{-1/2}.$ This zero eigenvalue is introduced
by the sliver field which defines the D25 brane. The breaking of
associativity by the definition of $\left( \tilde{m}_{0}^{2}\right) ^{-1/2}$
has to be sufficiently gentle as to maintain enough gauge symmetry. If this
can be successfully accomplished then MSFT will be useful to compute certain
nonperturbative quantities in the VSFT scenario. We have left this task to
future work.

In the setup of MSFT we have used an equal number of even and odd
oscillators. When $N\rightarrow\infty$ it is impossible to say that they are
equal in number. Therefore, it would be interesting to investigate the case
of $N$ even oscillators and $N+1$ odd oscillators in the cutoff theory and
then send $N$ to infinity at the end. In this case $T_{eo}$ is a rectangular
matrix at finite $N$ and does not have an inverse. It could be that this
would be an approach for incorporating the associativity anomaly while still
having a cutoff $N.$


The Moyal star formulation for fermionic ghosts should be possible. Although
we have already included bosonized ghosts in our formulation, we expect to
learn more about the ghost sector and simplify computations that involve the
fermionic ghosts (such as the BRST operator) more directly.

The Moyal formulation allows us to consider the system one mode at a time in
the formalism of noncommutative geometry that is formally the same as the
quantization of the relativistic spinless particle in phase space a la
Wigner-Weyl-Moyal (although the star in MSFT does not follow from quantum
mechanics, as emphasized at the end of section IIA). This makes it tempting
to consider the spinning particle and the superparticle with this method of
quantization and inventing the corresponding noncommutative geometry in
superspace. Perhaps this would be an approach to building the supersymmetric
string field theory that has been a challenge so far.

The same temptation applies to building the generalization of 2T-physics to
string field theory \cite{twoT}. The particle version of 2T-physics field
theory uses precisely the same noncommutative geometry approach (which does
follow from quantum mechanics) and is therefore a formalism that is directly
related to our current formulation of MSFT. In fact, the hint for
introducing the Moyal star product in string field theory by one of the
authors, came directly from the formalism of 2T-physics field theory.
Reversing the process, now one can try to find the string field theory
version of 2T-physics by incorporating the features of 2T field theory \cite%
{twoT}.

Since we have introduced arbitrary frequencies $\kappa_{e},\kappa_{o}$ it
may be possible to apply our more general string field theory formalism to
circumstances where string backgrounds alter the frequencies, such as a
quadratic \textquotedblleft mass\textquotedblright\ term in the sigma model.
One such case that has arisen recently is called the \textquotedblleft
pp-wave\textquotedblright. It would be of interest to investigate this and
similar cases in our string field theory context.

Generalizations of the Moyal product are known in noncommutative geometry
\cite{kontsevich}. If such generalizations of the star product are used in
our setup of string field theory, one wonders whether this corresponds to
strings in various nontrivial backgrounds. This question can be investigated
by computing $n$-point vertices using our methods with a given star products
and comparing them to vertices computed in conformal field theory.

We also note that when the NS-NS two-form field $B$ has a nontrivial
curvature $dB\neq0$, Kontsevich's star product becomes inevitably
non-associative \cite{rCS},
\begin{equation}
(f\star g)\star h-f\star(g\star h)\sim C_{ijk}\partial_{i}f\,\partial
_{j}g\,\partial_{k}h,\qquad C=dB\,\,.
\end{equation}
This is an indication that nontrivial closed string physics can be recovered
when associativity is broken. One related issue is that in the presence of
nontrivial $dB$, the gauge symmetry of Born-Infeld theory gets modified
because of the coupling $\mathcal{F}=F+B$. In this sense, the conventional
gauge transformation is affected by the gauge symmetry of the $B$ field.
This is the origin of the appearance of twisted $K$-theory. It is tempting
to imagine that the breaking of the gauge symmetry of Witten type SFT in the
presence of the associativity anomaly is directly related to the coupling to
the closed string degree of freedom. In this sense, a generalization of \cite%
{rCS} to the full string variables (like we did for the flat space) should
be quite interesting.

These and other investigations are underway and they will be reported in
future publications.

\begin{center}
\noindent{\large \textbf{Acknowledgments}}
\end{center}

I.B. is supported in part by a DOE grant DE-FG03-84ER40168. Y.M. is
supported in part by Grant-in-Aid (\# 13640267) from the Ministry of
Education, Science, Sports and Culture of Japan. Both Y.M. and I.B. thank
the NSF and the JSPS for making possible the collaboration between Tokyo
University and USC through the collaborative grants, NSF-9724831 and JSPS
(US-Japan cooperative science program). I.B. is grateful to University of
Tokyo and Y.M. is grateful to the Caltech-USC Center for hospitality.

We would like thank, T. Kawano, I. Kishimoto and Y. Okawa for helpful
discussions.

\bigskip 

\section{Appendix}

\subsection{Derivation of Monoid algebra}

We first note that for any function $A_{1}$, $A_{2}$, the star product acts
as, (using $\partial_{i}$ to mean $\partial/\partial\xi^{i},$ and
suppressing the midpoint insertion)
\begin{align}
A_{1}\left( \xi\right) \ast A_{2}\left( \xi\right) & =\,e^{\frac{1}{2}%
\partial\sigma\partial^{\prime}}A_{1}\left( \xi\right) A_{2}\left(
\xi^{\prime}\right) |_{\xi^{\prime}=\xi} \\
& =A_{1}\left( \xi+\frac{1}{2}\sigma\partial^{\prime}\right) A_{2}\left(
\xi^{\prime}\right) |_{\xi^{\prime}=\xi} \\
& =A_{2}\left( \xi^{\prime}-\frac{1}{2}\sigma\partial\right) A_{1}\left(
\xi\right) |_{\xi^{\prime}=\xi}\,.
\end{align}
We apply this formula for the product of the two elements in the Monoid
algebra,
\begin{align}
A_{12}\left( \xi\right) & =A_{1}\left( \xi\right) \ast A_{2}\left( \xi\right)
\\
& =\mathcal{N}_{1}e^{-\xi M_{1}\xi-\xi\lambda_{1}}\ast\mathcal{N}_{2}e^{-\xi
M_{2}\xi-\xi\lambda_{2}} \\
& =\mathcal{N}_{1}\mathcal{N}_{2}e^{-\left( \xi+\frac{1}{2}\sigma
\partial^{\prime}\right) ^{T}M_{1}\left( \xi+\frac{1}{2}\sigma
\partial^{\prime}\right) }\,e^{-\left( \xi+\frac{1}{2}\sigma\partial
^{\prime}\right)
^{T}\lambda_{1}}e^{-\xi^{\prime}M_{2}\xi^{\prime}-\xi^{\prime}\lambda_{2}} \\
& =\mathcal{N}_{1}\mathcal{N}_{2}e^{\frac{1}{4}\lambda_{2}M_{2}^{-1}%
\lambda_{2}}e^{-\xi M_{1}\xi-\xi\lambda_{1}} \\
& \times e^{\partial^{\prime}\left( \sigma M_{1}\xi+\frac{1}{2}\sigma
\lambda_{1}\right) }e^{\frac{1}{4}\partial^{^{\prime}}\sigma
M_{1}\sigma\partial^{\prime}}e^{-\left( \xi^{\prime}+\frac{1}{2}%
M_{2}^{-1}\lambda_{2}\right) ^{T}M_{2}\left( \xi^{\prime}+\frac{1}{2}%
M_{2}^{-1}\lambda_{2}\right) }
\end{align}
To perform the derivatives we use the basic relations
\begin{align}
e^{\frac{1}{4}\partial A\partial}e^{-\left( \xi+u\right) B\left(
\xi+u\right) } & =\frac{e^{-\left( \xi+u\right) \left( A+B^{-1}\right)
^{-1}\left( \xi+u\right) }}{\left( \det\left( 1+BA\right) \right) ^{d/2}}, \\
e^{\partial v}f\left( \xi\right) & =f\left( \xi+v\right) ,
\end{align}
for any constant vectors $u,v$ and constant matrices $A,B.$ Notice that the
dimension $d$ appears in the power of the determinant. Then we get
\begin{align}
A_{12} & =\frac{N_{1}N_{2}e^{\frac{1}{4}\lambda_{2}M_{2}^{-1}%
\lambda_{2}}e^{-\xi M_{1}\xi-\xi\lambda_{1}}}{\left( \det\left(
1+M_{2}\sigma M_{1}\sigma\right) \right) ^{d/2}} \\
& \times e^{\partial^{\prime}\left( \sigma M_{1}\xi+\frac{1}{2}\sigma
\lambda_{1}\right) }e^{-\left( \xi^{\prime}+\frac{1}{2}M_{2}^{-1}\lambda
_{2}\right) ^{T}\left( M_{2}^{-1}+\sigma M_{1}\sigma\right) ^{-1}\left(
\xi^{\prime}+\frac{1}{2}M_{2}^{-1}\lambda_{2}\right) } \\
& =\frac{N_{1}N_{2}e^{\frac{1}{4}\lambda_{2}M_{2}^{-1}\lambda_{2}}e^{-\xi
M_{1}\xi-\xi\lambda_{1}}}{\left( \det\left( 1+M_{2}\sigma M_{1}\sigma\right)
\right) ^{d/2}} \\
& \times e^{-\left( \xi+\sigma M_{1}\xi+\frac{1}{2}\sigma\lambda_{1}+\frac{1%
}{2}M_{2}^{-1}\lambda_{2}\right) ^{T}\left( M_{2}^{-1}+\sigma
M_{1}\sigma\right) ^{-1}\left( \xi+\sigma M_{1}\xi+\frac{1}{2}\sigma
\lambda_{1}+\frac{1}{2}M_{2}^{-1}\lambda_{2}\right) } \\
& =\frac{N_{1}N_{2}e^{\frac{1}{4}\lambda_{2}M_{2}^{-1}\lambda_{2}}e^{-\xi
M_{1}\xi-\xi\lambda_{1}}}{\left( \det\left( 1+M_{2}\sigma M_{1}\sigma\right)
\right) ^{d/2}} \\
& \times e^{-\left( \xi+\sigma M_{1}\xi+\frac{1}{2}\sigma\lambda_{1}+\frac{1%
}{2}M_{2}^{-1}\lambda_{2}\right) ^{T}\left( M_{2}^{-1}+\sigma
M_{1}\sigma\right) ^{-1}\left( \xi+\sigma M_{1}\xi+\frac{1}{2}\sigma
\lambda_{1}+\frac{1}{2}M_{2}^{-1}\lambda_{2}\right) } \\
& =N_{12}e^{-\xi M_{12}\xi-\xi\lambda_{12}}
\end{align}
In the last expression the coefficients of the quadratic, linear, and zeroth
order terms in $\xi,$ have been collected to the form
\begin{align}
M_{12} & =M_{1}+\left( 1-M_{1}\sigma\right) \left( M_{2}^{-1}+\sigma
M_{1}\sigma\right) ^{-1}\left( 1+\sigma M_{1}\right) \\
\lambda_{12} & =\lambda_{1}+\left( 1-M_{1}\sigma\right) \left(
M_{2}^{-1}+\sigma M_{1}\sigma\right) ^{-1}\left(
\sigma\lambda_{1}+M_{2}^{-1}\lambda_{2}\right) \\
N_{12} & =\frac{N_{1}N_{2}e^{\frac{1}{4}\lambda_{2}M_{2}^{-1}\lambda_{2}}}{%
\left( \det\left( 1+M_{2}\sigma M_{1}\sigma\right) \right) ^{d/2}}e^{-\frac{1%
}{4}\left( \sigma\lambda_{1}+M_{2}^{-1}\lambda_{2}\right) ^{T}\left(
M_{2}^{-1}+\sigma M_{1}\sigma\right) ^{-1}\left( \sigma
\lambda_{1}+M_{2}^{-1}\lambda_{2}\right) }.
\end{align}
These can be rewritten in a form that make explicit the symmetry under $%
\left( 1\longleftrightarrow2\right) $ and $\sigma\longleftrightarrow -\sigma$
\begin{align}
M_{12} & =M_{1}+\left( 1-M_{1}\sigma\right) \left( M_{2}^{-1}+\sigma
M_{1}\sigma\right) ^{-1}\left( 1+\sigma M_{1}\right) ,  \label{M121} \\
& =M_{2}+\left( 1+M_{2}\sigma\right) \left( M_{1}^{-1}+\sigma
M_{2}\sigma\right) ^{-1}\left( 1-\sigma M_{2}\right) ,  \label{M122} \\
& =\left( M_{1}+M_{2}\sigma M_{1}\right) \left( 1+\sigma M_{2}\sigma
M_{1}\right) ^{-1}+\left( M_{2}-M_{1}\sigma M_{2}\right) \left( 1+\sigma
M_{1}\sigma M_{2}\right) ^{-1},   \label{M12}
\end{align}
and%
\begin{align}
\lambda_{12} & =\lambda_{1}+\left( 1-M_{1}\sigma\right) \left(
M_{2}^{-1}+\sigma M_{1}\sigma\right) ^{-1}\left(
\sigma\lambda_{1}+M_{2}^{-1}\lambda_{2}\right) , \\
& =\lambda_{2}+\left( 1+M_{2}\sigma\right) \left( M_{1}^{-1}+\sigma
M_{2}\sigma\right) ^{-1}\left( -\sigma\lambda_{2}+M_{1}^{-1}\lambda
_{1}\right) , \\
& =\left( 1-M_{1}\sigma\right) \left( 1+M_{2}\sigma M_{1}\sigma\right)
^{-1}\lambda_{2}+\left( 1+M_{2}\sigma\right) \left( 1+M_{1}\sigma
M_{2}\sigma\right) ^{-1}\lambda_{1},   \label{la12}
\end{align}
and%
\begin{equation}
N_{12}=N_{1}N_{2}\left( \det\left( 1+M_{2}\sigma M_{1}\sigma\right) \right)
^{-d/2}e^{\frac{1}{4}\lambda_{a}K_{ab}\lambda_{b}},
\end{equation}
$\allowbreak$where the matrices $\left( K_{ab}\right) _{ij}$ with $a$ or $b$%
=1,2, are the coefficients of $\lambda_{a}^{i}\lambda_{b}^{j}$
\begin{align}
K_{11} & =\left( \sigma^{-1}M_{2}^{-1}\sigma^{-1}+M_{1}\right) ^{-1},\quad
K_{12}=\left( \sigma^{-1}+M_{2}\sigma M_{1}\right) ^{-1}, \\
K_{21} & =-\left( \sigma+M_{1}\sigma M_{2}\right) ^{-1},\quad K_{22}=\left(
M_{2}+\sigma^{-1}M_{1}^{-1}\sigma^{-1}\right) ^{-1}.
\end{align}
By fairly complex matrix manipulations they can be reassembled to the form%
\begin{equation}
\lambda_{a}K_{ab}\lambda_{b}=\left( \lambda_{1}+\lambda_{2}\right) \left(
M_{1}+M_{2}\right) ^{-1}\left( \lambda_{1}+\lambda_{2}\right) -\left(
\lambda_{12}\right) ^{T}\left( M_{12}\right) ^{-1}\lambda_{12},
\end{equation}
and, we can also show that
\begin{equation}
\det\left( M_{12}\right) \det\left( 1+M_{2}\sigma M_{1}\sigma\right)
=\det\left( \left( M_{1}+M_{2}\right) \right) .
\end{equation}
Thus, we can write $N_{12}$ in the following form which will be useful later
\begin{align}
N_{12} & =N_{1}N_{2}\left( \frac{\det\left( 2M_{12}\sigma\right) }{%
\det\left( 2\left( M_{1}+M_{2}\right) \sigma\right) }\right) ^{d/2}\,e^{%
\frac{1}{4}\left( \lambda_{1}+\lambda_{2}\right) \left( M_{1}+M_{2}\right)
^{-1}\left( \lambda_{1}+\lambda_{2}\right) -\frac{1}{4}\left(
\lambda_{12}\right) ^{T}\left( M_{12}\right) ^{-1}\lambda_{12}},  \label{N12}
\\
& =T_{12}\det\left( 2M_{12}\sigma\right) ^{d/2}\exp\left( -\frac{1}{4}\left(
\lambda_{12}\right) ^{T}\left( M_{12}\right) ^{-1}\lambda _{12}\right)
\end{align}
where
\begin{equation}
T_{12}=\frac{N_{1}N_{2}}{\det\left( 2\left( M_{1}+M_{2}\right) \sigma\right)
^{d/2}}\exp\left( \frac{1}{4}\left( \lambda_{1}+\lambda _{2}\right)
^{T}\left( M_{1}+M_{2}\right) ^{-1}\left( \lambda_{1}+\lambda_{2}\right)
\right)
\end{equation}
is the trace of $A_{12}.$ Therefore, we have shown that%
\begin{equation}
A_{12}=\,T_{12}\det\left( 2M_{12}\sigma\right) ^{d/2}\exp\left( -\left( \xi+%
\frac{1}{2}M_{12}^{-1}\lambda_{12}\right) ^{T}M_{12}\left( \xi+\frac {1}{2}%
M_{12}^{-1}\lambda_{12}\right) \right) .
\end{equation}
The results of this appendix were given in \cite{witmoy} and are used in the
text.

\subsection{General solution of fluctuations around a projector}

Let us consider any projector $\Xi$ of the form of Eq.(\ref{SSS}) for any
matrix $m$ that satisfies $m^{2}=1.$ We want to give here the solution that
is analogous to the matrix solution for $T_{1}$ given in Eq.(\ref{STmatrices}%
). Namely, we expect the analog of a row and a column, which we will denote
as $T_{\pm}.$ Two particular solutions for $T_{\pm}$ have the following
explicit form%
\begin{equation}
T_{\pm}=\mathcal{N}_{\pm}e^{ik\cdot\bar{x}}\det\left( 2^{d/2}\right)
\exp\left( -\xi m\sigma^{-1}\xi\right) ~\left( e^{\xi\frac{1\pm m}{2}%
\lambda}-1\right)
\end{equation}
where $\bar{x}^{\mu}$ is the midpoint. The factor $e^{ik\cdot\bar{x}}$ is
insensitive to the star product, so we can ignore it in the following
argument.

To verify that these $T_{\pm}$ are solutions we use the product formula for
monoid elements $A_{i}=\mathcal{N}_{i}e^{-\xi m\sigma^{-1}\xi-\xi\lambda_{i}}
$ with identical $m^{\prime}$s, which simplifies to the following form when $%
m^{2}=1$
\begin{align}
A_{1}\ast A_{2} & =\frac{\mathcal{N}_{1}\mathcal{N}_{2}e^{K_{12}/4}}{%
\det\left( 2^{d/2}\right) }e^{-\xi m\sigma^{-1}\xi-\xi\lambda_{12}} \\
\lambda_{12} & =\frac{1+m}{2}\lambda_{1}+\frac{1-m}{2}\lambda_{2} \\
K_{12} & =\frac{1}{2}\bar{\lambda}_{1}\sigma m\lambda_{1}+\frac{1}{2}\bar{%
\lambda}_{2}\sigma m\lambda_{2}+\frac{1}{2}\left( \bar{\lambda}%
_{1}\sigma\lambda_{2}-\bar{\lambda}_{2}\sigma\lambda_{1}\right)
\end{align}
Using this formula we compute $\Xi\ast T_{\pm}$ (with $\lambda_{1}=0,$ and $%
\lambda_{2}=0,\frac{1\pm m}{2}\lambda$ as needed). We find
\begin{equation}
\Xi\ast T_{\pm}=\mathcal{N}_{\pm}e^{ik\cdot\bar{x}}\det\left( 2^{d/2}\right)
\exp\left( -\xi m\sigma^{-1}\xi\right) \left( e^{\frac{1}{32}\lambda\left(
1\pm m^{T}\right) \sigma m\left( 1\pm m\right) \lambda}e^{\xi\left( \frac{1-m%
}{2}\right) \left( \frac{1\pm m}{2}\right) \lambda}-1\right)
\end{equation}
Similarly we compute $T_{\pm}\ast\Xi$ (with $\lambda_{1}=0,\frac{1\pm m}{2}%
\lambda~$as needed$,$ and $\lambda_{2}=0$). We find%
\begin{equation}
T_{\pm}\ast\Xi=\mathcal{N}_{\pm}e^{ik\cdot\bar{x}}\det\left( 2^{d/2}\right)
\exp\left( -\xi m\sigma^{-1}\xi\right) \left( e^{\frac{1}{32}\lambda\left(
1\pm m^{T}\right) \sigma m\left( 1\pm m\right) \lambda_{+}}e^{\xi\left(
\frac{1+m}{2}\right) \left( \frac{1\pm m}{2}\right) \lambda_{+}}-1\right)
\end{equation}
Now, taking into account that $m^{2}=1$ and $m^{T}=-\sigma m\sigma^{-1}$
(symmetric $M=m\sigma^{-1}$), we see that these expressions simplify to%
\begin{equation}
e^{\frac{1}{32}\lambda_{\pm}\left( 1\pm m^{T}\right) \sigma m\left( 1\pm
m\right) \lambda_{\pm}}=e^{0}=1.
\end{equation}
Therefore%
\begin{align}
\Xi\ast T_{+} & =0,\;\Xi\ast T_{-}=T_{-} \\
T_{+}\ast\Xi & =T_{+},\;T_{-}\ast\Xi=0.
\end{align}
as expected from columns and rows if the $T_{\pm}$ were matrices. So we find%
\begin{equation}
\Xi\ast T_{\pm}+T_{\pm}\ast\Xi=T_{\pm},
\end{equation}
which shows that we have indeed a solution for any $\lambda$ and any
coefficients $\mathcal{N}_{\pm}e^{ik\cdot\bar{x}}.$

Using these properties of $T_{\pm}$ a more general solution is constructed
as a superposition of the form%
\begin{equation}
T\left( \bar{x},x_{e},p_{e}\right) =\int d\lambda~\left[ f_{+}\left( \bar{x}%
,\lambda\right) \left( e^{\xi\left( \frac{1+m}{2}\right) \lambda }-1\right)
\Xi+f_{-}\left( \bar{x},\lambda\right) \left( e^{\xi\left( \frac{1-m}{2}%
\right) \lambda}-1\right) \Xi\right]   \label{tach}
\end{equation}
for any functions $f_{\pm}\left( \bar{x},\lambda\right) .$


\newpage

\begin{thebibliography}{99}
\bibitem{Witten} E. Witten, ``Noncommutative geometry and String Field
Theory'', Nucl. Phys. B268 (1986) 253.

\bibitem{r-RSZ} L.~Rastelli, A.~Sen and B.~Zwiebach, ``String field theory
around the tachyon vacuum'', arXiv:hep-th/0012251;\newline
``Classical solutions in string field theory around the tachyon vacuum'',
arXiv:hep-th/0102112;\newline
JHEP\textbf{0111} (2001) 035 [arXiv:hep-th/0105058].

\bibitem{r-RSZ2} L.~Rastelli, A.~Sen and B.~Zwiebach, JHEP \textbf{0111}
(2001) 045 [arXiv:hep-th/0105168].

\bibitem{r-GT} D.~J.~Gross and W.~Taylor, \textquotedblleft Split string
field theory I\textquotedblright, JHEP \textbf{0108} (2001) 009
[arXiv:hep-th/0105059];\newline
\textquotedblleft Split string field theory. II\textquotedblright, JHEP
\textbf{0108} (2001) 010 [arXiv:hep-th/0106036].


\bibitem{r-KO} T.~Kawano and K.~Okuyama, \textquotedblleft Open string
fields as matrices,\textquotedblright\ JHEP \textbf{0106} (2001) 061
[arXiv:hep-th/0105129].

\bibitem{witmoy} I.~Bars, ``Map of Witten's * to Moyal's'', Phys.\ Lett.\ B
\textbf{517} (2001) 436 [arXiv:hep-th/0106157].

\bibitem{r-Matsuo} Y.~Matsuo, ``BCFT and sliver state,'' Phys.\ Lett.\ B
\textbf{513} (2001) 195 [arXiv:hep-th/0105175];\newline
Phys.\ Lett.\ B \textbf{514} (2001) 407 [arXiv:hep-th/0106027];\newline
Mod.\ Phys.\ Lett.\ A \textbf{16} (2001) 1811 [arXiv:hep-th/0107007].

\bibitem{r-Bars2} 
I.~Bars, \textquotedblleft Nonperturbative effects of extreme localization
in noncommutative geometry\textquotedblright, arXiv:hep-th/0109132.

\bibitem{r-Kluson} J.~Kluson, ``Proposal for background independent
Berkovits' superstring field theory,'' JHEP \textbf{0107} (2001) 039
[arXiv:hep-th/0106107].

\bibitem{r-FO} K.~Furuuchi and K.~Okuyama, ``Comma vertex and string field
algebra,'' JHEP \textbf{0109} (2001) 035 [arXiv:hep-th/0107101].

\bibitem{r-HK} H.~Hata and T.~Kawano, ``Open string states around a
classical solution in vacuum string field theory,'' JHEP \textbf{0111}
(2001) 038 [arXiv:hep-th/0108150].

\bibitem{r-Kishimoto} I.~Kishimoto, ``Some properties of string field
algebra,'' JHEP \textbf{0112} (2001) 007 [arXiv:hep-th/0110124].

\bibitem{GRSZ} D.~Gaiotto, L.~Rastelli, A.~Sen and B.~Zwiebach, ``Ghost
structure and closed strings in vacuum string field theory,''
arXiv:hep-th/0111129.

\bibitem{r-HI} A.~Hashimoto and N.~Itzhaki, ``Observables of string field
theory,'' arXiv:hep-th/0111092.

\bibitem{r-HM} H.~Hata and S.~Moriyama, \textquotedblleft Observables as
twist anomaly in vacuum string field theory,\textquotedblright\
arXiv:hep-th/0111034. H.~Hata, S.~Moriyama and S. Teraguchi,
\textquotedblleft Exact results on twist anomaly\textquotedblright,
arXiv:hep-th/0201117.

\bibitem{RSZ4} L. Rastelli, A. Sen, and B.~Zwiebach, \textquotedblleft A
note on a proposal for the tachyon state in VSFT", arXiv:hep-th/0111153.

\bibitem{r-MT} G.~Moore and W.~Taylor, ``The singular geometry of the
sliver'', arXiv:hep-th/0111069.

\bibitem{-rRSZ2} L.~Rastelli, A.~Sen and B.~Zwiebach, ``Star algebra
spectroscopy,'' arXiv:hep-th/0111281.

\bibitem{r-Okuyama} K.~Okuyama, \textquotedblleft Siegel gauge in vacuum
string field theory,\textquotedblright\ arXiv:hep-th/0111087;\newline
\textquotedblleft Ghost kinetic operator of vacuum string field
theory,\textquotedblright\ arXiv:hep-th/0201015;\newline
\textquotedblleft Ratio of Tensions from Vacuum String Field
Theory,\textquotedblright\ arXiv:hep-th/0201136.

\bibitem{furuuchiokuyama} K.\symbol{126}Furuuchi and K.\symbol{126}Okuyama,
\textquotedblleft Comma vertex and string field algebra,\textquotedblright\
JHEP \{\TEXTsymbol{\backslash}bf 0109\}, 035 (2001), [arXiv:hep-th/0107101].

\bibitem{r-KOh} I.~Kishimoto and K.~Ohmori, ``CFT description of identity
string field: Toward derivation of the VSFT action,'' arXiv:hep-th/0112169.

\bibitem{r-AGM} I.~Y.~Arefeva, A.~A.~Giryavets and P.~B.~Medvedev,
\textquotedblleft NS matter sliver,\textquotedblright\ arXiv:hep-th/0112214.

\bibitem{marinoschiappa} M.\symbol{126}Marino and R.\symbol{126}Schiappa,
\textquotedblleft Towards vacuum superstring field theory: The
supersliver,\textquotedblright\ arXiv:hep-th/0112231.

\bibitem{r-BarsMatsuo} I. Bars and Y. Matsuo, `` Associativity Anomaly in
String Field Theory'', arXiv:hep-th/0202030.

\bibitem{DLMZ} M. Douglas, H. Liu, G. Moore and B. Zwiebach,
\textquotedblleft Open String Star as a Continuous Moyal
Product\textquotedblright, arXiv:hep-th/0202087.

\bibitem{belov} D.\symbol{126}M.\symbol{126}Belov, \textquotedblleft
Diagonal representation of open string star and Moyal
product,\textquotedblright\ arXiv:hep-th/0204164.

\bibitem{arafeva} I.\symbol{126}Y.\symbol{126}Arefeva and A.\symbol{126}A.%
\symbol{126}Giryavets, \textquotedblleft Open superstring star as a
continuous Moyal product,\textquotedblright\ arXiv:hep-th/0204239.

\bibitem{rsz3} D. Gaiotto, L. Rastelli, A. Sen, and B.~Zwiebach,
\textquotedblleft Patterns in string field theory
solutions\textquotedblright,\ arXiv:hep-th/0201159; \textquotedblleft Star
algebra projectors\textquotedblright, arXiv:hep-th//0202151.

\bibitem{potting} R.\symbol{126}Potting and J.\symbol{126}Raeymaekers,
\textquotedblleft Ghost sector of vacuum string field theory and the
projection equation,\textquotedblright\ arXiv:hep-th/0204172.

\bibitem{chu} C.\symbol{126}S.\symbol{126}Chu, P.\symbol{126}M.\symbol{126}%
Ho and F.\symbol{126}L.\symbol{126}Lin, \textquotedblleft Cubic string field
theory in pp-wave background and background independent Moyal
structure,\textquotedblright\ arXiv:hep-th/0205218.

\bibitem{GJ} D.~J.~Gross and A.~Jevicki,
Nucl.\ Phys.\ B \textbf{283} (1987) 1; 
Nucl.\ Phys.\ B \textbf{287} (1987) 225.

\bibitem{giddings} S. Giddings, Nucl. Phys. \textbf{B278}, (1986) 242.

\bibitem{r-Bordes} H.~M.~Chan, J.~Bordes, S.~T.~Tsou and L.~Nellen,
Phys.\ Rev.\ D \textbf{40} (1989) 2620;\newline
J.~Bordes, H.~M.~Chan, L.~Nellen and S.~T.~Tsou,
Nucl.\ Phys.\ B \textbf{351} (1991) 441;\newline
J.~Bordes, A.~Abdurrahman and F.~Anton,
Phys.\ Rev.\ D \textbf{49} (1994) 2966 [arXiv:hep-th/9306029];\newline
A.~Abdurrahman, F.~Anton and J.~Bordes,
Nucl.\ Phys.\ B \textbf{411} (1994) 693;\newline
A.~Abdurrahman and J.~Bordes,
Phys.\ Rev.\ D \textbf{58} (1998) 086003.

\bibitem{r-Moyal} J. Moyal, Proc. Camb. Phil. Soc. \textbf{45 }(1949)\textbf{%
\ }99.

\bibitem{r-DouglasNekrasov} For a recent review and extensive references
see: M. Douglas and N. Nekrasov, \textquotedblleft Noncommutative Field
Theory\textquotedblright, hep-th/0106048.

\bibitem{okawa} Y.~Okawa, ``Open string states and D-brane tension from
vacuum string field theory,'' arXiv:hep-th/0204012.

\bibitem{r-HS} G.~T.~Horowitz and A.~Strominger, ``Translations As Inner
Derivations And Associativity Anomalies In Open String Field Theory,''
Phys.\ Lett.\ B \textbf{185} (1987) 45.

\bibitem{r-Strominger} A.~Strominger,
Phys.\ Rev.\ Lett.\ \textbf{58} (1987) 629;\newline
Z.~Qiu and A.~Strominger,
Phys.\ Rev.\ D \textbf{36} (1987) 1794.

\bibitem{r-Strominger2} A.~Strominger, 
Nucl.\ Phys.\ B \textbf{294} (1987) 93.

\bibitem{wigner} E.~Wigner, Phys. Rev. \textbf{40} (1932) 749--759;\newline
For recent studies, see for example T.~Curtright, T.~Uematsu and
C.~K.~Zachos, 
J.\ Math.\ Phys.\ \textbf{42} (2001) 2396 [arXiv:hep-th/0011137] and
references therein.

\bibitem{HKL} J.~A.~Harvey, P.~Kraus and F.~Larsen,
JHEP \textbf{0012} (2000) 024 [arXiv:hep-th/0010060].

\bibitem{twoT} I. Bars, \textquotedblleft$U_{\ast}(1,1)$ noncommutative
gauge theory as the foundation of 2T-physics in field theory", [
arXiv:hep-th/0106013 ] Phys. Rev. \textbf{D64} (2001) 126001;
\textquotedblleft2T-physics 2001" , arXiv:hep-th/0106021.

\bibitem{kontsevich} M.~Kontsevich, ``Deformation quantization of Poisson
manifolds, I,'' arXiv:q-alg/9709040.

\bibitem{rCS} L.~Cornalba and R.~Schiappa,
Commun.\ Math.\ Phys.\ \textbf{225} (2002) 33 [arXiv:hep-th/0101219].
\end{thebibliography}
\end{document}